\documentclass[aps,prd,twocolumn,citeautoscript,superscriptaddress,groupedaddress,nofootinbib]{revtex4} 
\usepackage{graphicx}  
\usepackage{dcolumn}   
\usepackage{bm}        
\usepackage{amssymb}   
\usepackage[caption=false]{subfig}
\usepackage{amsmath}
\usepackage{longtable}     
\usepackage[hyperfootnotes=true]{hyperref}
\hypersetup{
 colorlinks=true,
 citecolor=blue,
 linkcolor=blue,
 urlcolor=blue}

\usepackage{color}
\definecolor{darkblue}{rgb}{0,0,0.5}
\definecolor{darkgreen}{rgb}{0.1,0.4,0.1}
\definecolor{orange}{rgb}{0.99,0.4,0}

\DeclareRobustCommand{\Sec}[1]{Sec.~\ref{#1}}

\DeclareRobustCommand{\Figs}[2]{Figs.~\ref{#1} and \ref{#2}}

\usepackage{amsmath}
\usepackage{amssymb}
\usepackage{comment}

\newcommand{\R}{\mathbb{R}}

\newcommand{\Zp}{$Z^\prime$}
\newcommand{\Higgs}{$H$}

\newcommand{\Hbb}{$H\rightarrow b\overline{b}$}
\newcommand{\gbb}{$g\rightarrow b\overline{b}$}

\begin{document}

\title{Automating the Construction of Jet Observables with Machine Learning}

\author{Kaustuv Datta}
\affiliation{Department of Physics, ETH Z\"{u}rich, 8093 Z\"{u}rich, Switzerland}
\email{kdatta@ethz.ch}
\author{Andrew Larkoski}
\affiliation{Physics Department, Reed College, Portland, OR 97202, USA}
\email{larkoski@reed.edu}
\author{Benjamin Nachman}
\affiliation{Physics Division, Lawrence Berkeley National Laboratory, Berkeley, CA 94720, USA}
\email{bpnachman@lbl.gov}

\begin{abstract}
Machine-learning assisted jet substructure tagging techniques have the potential to significantly improve searches for new particles and Standard Model measurements in hadronic final states.  Techniques with simple analytic forms are particularly useful for establishing robustness and gaining physical insight.  We introduce a procedure to automate the construction of a large class of observables that are chosen to completely specify $M$-body phase space.  The procedure is validated on the task of distinguishing $H\rightarrow b\bar{b}$ from $g\rightarrow b\bar{b}$, where $M=3$ and previous brute-force approaches to construct an optimal product observable for the $M$-body phase space have established the baseline performance. We then use the new method to design tailored observables for the boosted $Z'$ search, where $M=4$ and brute-force methods are intractable.  The new classifiers outperform standard $2$-prong tagging observables, illustrating the power of the new optimization method for improving searches and measurement at the LHC and beyond.

\end{abstract}

\maketitle
\section{\label{sec:intro}Introduction}

Effective identification of hadronic decays of boosted heavy particles like the top quark or $W$, $Z$ and Higgs (\Higgs) bosons is essential for analyses at the Large Hadron Collider (LHC). Jet substructure observables that identify specific discriminating information in the radiation pattern of jets originating from different particles are now necessary, both in the search for new physics and precision Standard Model (SM) measurements. As a result, there is an extensive literature developing observables and techniques for identifying boosted topologies to increase the efficacy of LHC analyses probing extreme regions of phase space~\cite{Larkoski:2017jix,Asquith:2018igt}.

Modern machine learning (ML) methods have emerged as useful tools for automating the creation of optimal observables for classification.  These methods are particularly powerful for high-dimensional, low-level inputs such as fixed-length sets of four-vectors~\cite{Pearkes:2017hku}, variable-length sets of four-vectors~\cite{Komiske:2018cqr}, physics-inspired bases~\cite{Komiske:2017aww,Erdmann:2018shi,Datta:2017rhs,Datta:2017lxt,Butter:2017cot}, images~\cite{Cogan:2014oua,Almeida:2015jua,deOliveira:2015xxd,Komiske:2016rsd,Barnard:2016qma,Kasieczka:2017nvn,Dreyer:2018nbf,Lin:2018cin,Fraser:2018ieu,Chien:2018dfn,Macaluso:2018tck}, sequences~\cite{Guest:2016iqz,Egan:2017ojy,Andreassen:2018apy,Fraser:2018ieu}, trees~\cite{Cheng:2017rdo,Louppe:2017ipp}, and graphs~\cite{henrion}.  Some deep learning-based tagging schemes have already been demonstrated using collider data as well as with full detector simulations for top quark tagging~\cite{Aaboud:2018psm,CMS-DP-2017-049}, boson tagging~\cite{Aaboud:2018psm,CMS-DP-2018-046}, quark/gluon tagging~\cite{ATL-PHYS-PUB-2017-017,CMS-DP-2017-027}, and $b$-jet tagging~\cite{ATL-PHYS-PUB-2017-003,ATL-PHYS-PUB-2017-013,CMS-DP-2018-058,Sirunyan:2017ezt}.  In addition to improving classification performance, ML techniques may also be able to make jet tagging more independent from simulation and robust to differences between simulation and data as well as between sideband and signal regions~\cite{Metodiev:2018ftz,Komiske:2018oaa,Metodiev:2017vrx,Dery:2017fap,Cohen:2017exh,Louppe:2016ylz,Shimmin:2017mfk,ATL-PHYS-PUB-2018-014}.  These and related techniques have also been proposed as more model-agnostic approaches to new particle searches~\cite{Collins:2018epr,Collins:2019jip,Heimel:2018mkt,Farina:2018fyg,Hajer:2018kqm}.

One of the key challenges with ML taggers is to identify what information the machine is using for classification.  Understanding the origin of discrimination can lead to robustness when taggers are applied outside of the region they were trained, can result in new theoretical insight for other applications, and may produce new simple observables that capture most of the information.  While there are many proposals for ML metacognition~\cite{Cohen:2017exh,deOliveira:2015xxd,Lin:2018cin,Komiske:2017aww,Komiske:2018cqr,Datta:2017lxt,Datta:2017rhs}, one particularly powerful approach is to identify simple product observables that capture most of the information from an ML algorithm trained on the full phase space~\cite{Datta:2017lxt}.  This approach results in analytically tractable observables that can capture nearly all of the power of a more complicated algorithm, but are also very robust and insightful.  One of the most challenging aspects of the approach presented in Ref.~\cite{Datta:2017lxt} is the fitting process for picking the optimal simple product observable.

%

In this paper, we describe a new procedure based on ML for automating the feature extraction originally presented in Ref.~\cite{Datta:2017lxt}.  This method is applied to derive an optimal product observable for discriminating \Hbb~vs. \gbb~ and the outcome is compared to the result of Ref.~\cite{Datta:2017lxt} which used a brute force approach.  Having validated the method, a new classifier is developed to distinguish a $Z'$ from generic quark and gluon jets.  The phase space scan required in this later tagging task is too big for the brute force approach and therefore the automated method is required to find the optimal tagger.  The resulting classifier has a simple form and is competitive with a tagger using high-dimensional, low-level inputs.  In addition to Ref.~\cite{Shimmin:2017mfk}, this is the only other study of the dependence on the mass of the new boson, which is timely given new searches for light boosted bosons~\cite{Sirunyan:2017dnz,Sirunyan:2017nvi,Aaboud:2018zba}.

This paper is organized as follows.  The method for constructing product observables is described in Sec.~\ref{sec:prodobs} and the machine learning approaches are detailed in Sec.~\ref{sec:deeplearn}.  Results for both the Higgs and $Z'$ classification tasks are presented in Sec.~\ref{sec:results}.  The paper ends with conclusions and future outlook in Sec.~\ref{sec:concl}.

\section{\label{sec:prodobs}$N$-subjettiness Product Observables }

The information about the kinematic phase space of $M$-subjets in a jet is resolved with a set of $(3M-4)$ $N$-subjettiness~\cite{Stewart:2010tn,Thaler:2010tr,Thaler:2011gf} observables.  By increasing $M$, one can identify the number of subjets required to saturate the classification performance based on the spanning set of $N$-subjettiness observables~\cite{Datta:2017rhs}:

\begin{align}
\nonumber
\left\{
\tau_1^{(0.5)},\tau_1^{(1)},\tau_1^{(2)},...,\tau_{M-2}^{(0.5)},\tau_{M-2}^{(1)},\tau_{M-2}^{(2)},\tau_{M-1}^{(1)},\tau_{M-1}^{(2)}
\right\}\,,
\end{align}

\noindent where

\begin{align}
\tau_N^{(\beta)}=\frac{1}{\sum_{i\in\text{jet}} p_\text{T,$i$}R^\beta}\sum_{i\in\text{jet}} p_\text{T,$i$}\min_\text{axes $j$}(\Delta R_{j,i})^\beta,
\end{align}

\noindent for some choice of $N$ axes within the jet; $R$ is the jet radius parameter, and $(\Delta R)^2=(\Delta\phi)^2+(\Delta\eta)^2$. Given the minimal $M$, one can posit an ansatz\footnote{The product form may not be flexible enough to capture the full discrimination power.  We find that it can capture a significant portion of the classification performance, but Appendix~\ref{app:betaML_comp} indicates that further information can be useful.} for a simple product observable that captures most of the information contained in a neural network trained on the entire spanning set:

\begin{align}
\label{eq:betaml}
\beta_{M}^\text{ML}= \left(\tau_1^{(0.5)}\right)^{a}\left(\tau_1^{(1)}\right)^{b}\left(\tau_1^{(2)}\right)^{c}\left(\tau_2^{(1)}\right)^{d}\cdots.
\end{align}

\noindent For distinguishing \Hbb~vs. \gbb~ jets, Ref.~\cite{Datta:2017lxt} showed that the useful information for classification is saturated by $M=3$ and $\beta_3^\text{ML}$ has nearly the same tagging performance as the full $3$-body phase space.  The parameters $a,b,c,d,e$ that specify $\beta_3^\text{ML}$ were identified by randomly scanning the five-dimensional phase space and exploiting minimal correlations between some of the parameters.  This becomes intractable when the optimal $M$ is bigger than $3$.

In this paper, we explore methods to overcome the difficulties of extending this procedure to higher dimensions.  In one approach, we replace the random sampling segment of the procedure with a combination of neural networks carrying out regression from the parameter space to the distributions of the product observable for individual jets.  Off-the-shelf minimization routines can then be used to optimize any metric of the classifier performance.  A complementary and simpler approach is to directly use the form in Eq.~\ref{eq:betaml} in the machine learning optimization, where the learnable parameters are the exponents $\{a,b,c,...\}$.  Further details are described in the next sections.


\section{\label{sec:deeplearn} Machine Learning\newline Implementation}

\subsection{\label{subsec:dataset} Dataset}



Proton-proton collisions with $Z'\rightarrow\text{hadrons}$, $H\rightarrow b\bar{b}$, and generic quark and gluon jets (QCD) at $\sqrt{s}=13$~TeV are generated using \textsc{Pythia}~8.226~\cite{Sjostrand:2006za,Sjostrand:2014zea}.  For the $H\rightarrow b\bar{b}$ case, the background is enriched in $g\rightarrow b\bar{b}$ as in Ref.~\cite{Alwall:2014hca} by generating the gluon splitting matrix element in MadGraph 5 v2.5.4~\cite{Alwall:2014hca}.  All detector-stable particles excluding neutrinos and muons are clustered into jets using the anti-$k_t$ algorithm~\cite{Cacciari:2008gp} with $R=0.8$ as implemented in Fastjet~\cite{Cacciari:2011ma}.  Jets are groomed by reclustering the constituents using the Cambridge-Aachen algorithm~\cite{Dokshitzer:1997in,Wobisch:1998wt} and applying the soft drop algorithm~\cite{Larkoski:2014wba} with $\beta=0$ and $z_\text{cut}=0.1$ (equivalent to modified mass drop tagging or mMDT~\cite{Dasgupta:2013ihk}).  The $N$-subjettiness observables are computed using the axes that minimize $\tau_N^{(\beta)}$, using the exclusive $k_t$ algorithm~\cite{Catani:1993hr,Ellis:1993tq} with standard $E$-scheme recombination \cite{Blazey:2000qt}. For comparison with other state-of-the-art two-prong tagging techniques, the $D_2$~\cite{Larkoski:2014gra}, $N_2$~\cite{Moult:2016cvt} observables, and $\tau_{21}^{(\beta)}$ with winner-take-all (WTA) recombination \cite{Bertolini:2013iqa,Larkoski:2014uqa,Larkoski:2014bia}, are also computed from the jet constituents.

\subsection{\label{subsec:regopt} Construction of optimized product observables}

Using the approach followed in Ref.~\cite{Datta:2017lxt}, the point of saturation of discrimination power is first identified using a deep neural network (DNN) classifier. For \Zp~vs. QCD and \Hbb~vs.~\gbb~ discrimination, we note that discrimination power saturates at 4-body (8-dimensional) and 3-body phase space (5-dimensional), respectively. Then it is simple to form the product observable from the elements of the $M$-body basis corresponding to saturation. 

We examine two approaches for finding the optimal product observable.  The first approach follows a similar method as the brute-force algorithm.  Neural networks approximate signal and background probability distributions conditioned on the parameters $\{a,b,c,...\}$ and then any automated optimization procedure can be used to identify the best exponents. For each task, the product observable is calculated for 25,000 signal and background jets for different values of the parameters [$a-e$] (\Hbb) or [$a-h$] (\Zp), in the range $[-5,5]$. These distributions are then stored to generate training sets for the neural networks used to carry out regression from the parameter space to the calculation of $\beta_{M}^{\text{ML}}$ with those exponents.  

\begin{figure}[h!]
\centering
\includegraphics[width=0.45\textwidth]{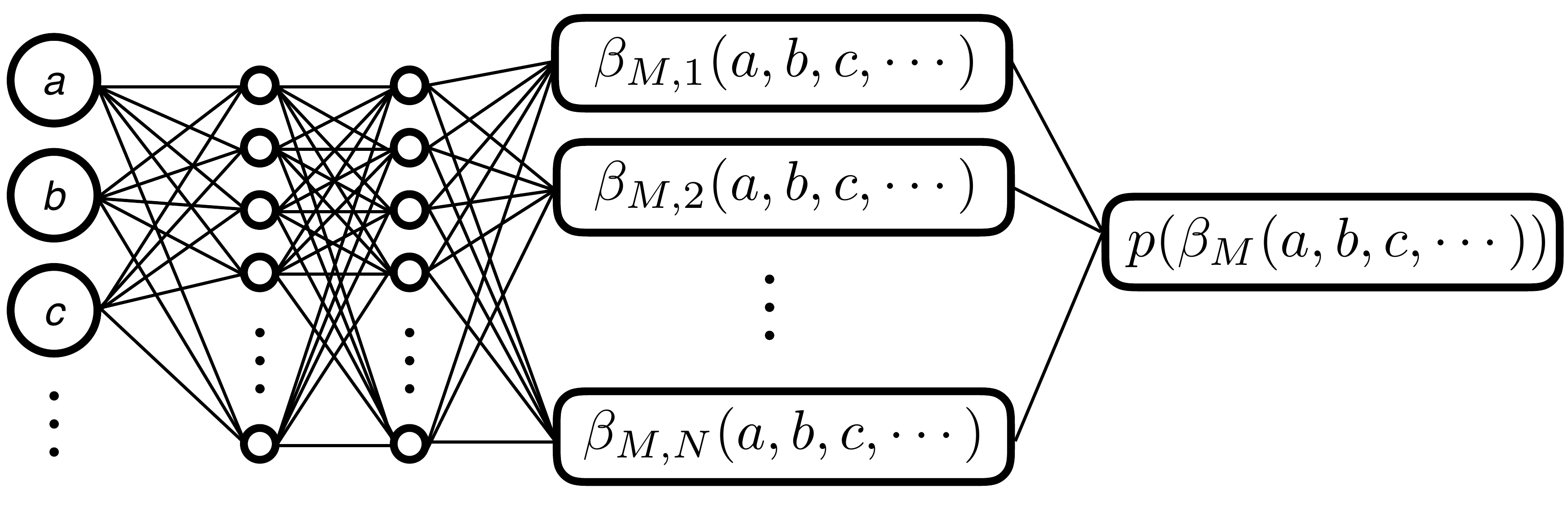}
\caption{A schematic diagram of the network architecture used to produce the probability distribution of $\beta_M$ for a given set of input parameters $\{a,b,c,...\}$.  In this case, $N=25,000$.}
\label{fig:schematic}
\end{figure}

While there are multiple possibilities for learning the probability distribution of $\beta_M$ given $\{a,b,c,...\}$, such as generative adversarial networks~\cite{NIPS2014_5423} and variational autoencoders~\cite{DBLP:journals/corr/KingmaW13,Rezende:2014:SBA:3044805.3045035}, the method that we found works well for the product observables is illustrated in Fig.~\ref{fig:schematic}.  The network takes as input 5 (Higgs) or 8 ($Z'$) inputs and outputs 25,000 numbers, which represent a dataset that is the same size as the training data, but with the specified parameter values $\{a,b,c,...\}$.  From these 25,000 values, the probability distribution of $\beta$ is formed for signal and background and the one-dimensional likelihood ratio is constructed for optimizing the classifier performance.  Variations on this setup are possible, such as (significantly) reducing the number of points needed to specify the probability distributions, but this approach was found to be robust to perturbations in initialization and network architecture.  For this paper, it was found that the network did not work well with fewer than 25k example jets per parameter point.  For each network, 250k (450k) parameter points were used for training in the $Z'$ and ungroomed Higgs (groomed Higgs) case.  In only the groomed Higgs case, a single network was trained for signal and background with a 1/0 switch added to the input.  Separate networks were trained for signal and background in the $Z'$ and ungroomed Higgs cases. To reduce the effects of numerical instability on the training of these networks, we train on samples after taking the natural logarithm of the 25k measured values of the product observables.



Aside from the use (or not) of the switch input, both the \Hbb~and \Zp~ tasks use simple fully-connected neural networks with two hidden layers.
The input layer is followed by a dense layer with either 250 or 500 nodes, then another dense layer with 100 or 250 nodes, followed by an output layer with 25,000 nodes using a linear activation.  The number of nodes in the hidden layers were bigger for the $Z'$ case with grooming compared with the Higgs case or the ungroomed $Z'$ case.

We use leaky rectified linear units (Leaky ReLU) as the activations for the hidden layers. The networks were compiled with a mean squared error loss function (on the penultimate layer shown in Fig.~\ref{fig:schematic}, not on $p(\beta_M)$ directly), using Adam optimization \cite{DBLP:journals/corr/KingmaB14}. The regression networks were each trained for $\sim10,000$ epochs. All deep learning tasks were carried out with the Keras~\cite{chollet2015keras} deep learning libraries, using the TensorFlow \cite{tensorflow2015-whitepaper} backend.

Given the set of $25,000$ values of the $\beta_M$ observable for a given set of parameters, it is straightforward to use these networks in an optimality scan. For this purpose, we use SciPy's~\cite{scipy} basin-hopping \cite{doi:10.1021/jp970984n} global minima finder using the non-linear, derivative free COBYLA (Constrained Optimization BY Linear Approximation) \cite{Powell1994} minimizer to scan over local minima. In the optimization, the networks are used to predict background and signal distributions for a given set of parameters. The 1-dimensional binned likelihood distributions\footnote{In principle, one can estimate the AUC without binning, but it was found that there was not a significant sensitivity to the choice of binning.} of the observable, constructed from the network outputs, was then used to calculate the area under the ROC curve (AUC) to estimate the discrimination power, where (1-AUC) was explicitly chosen as the metric for the basin-hopping minimization.   Appendix~\ref{sec:crosscheck} illustrates that the regression networks can be used to accurately model the dependence of the AUC as a function of the parameters.  The observable selected using this procedure will be denoted $\beta_{3,H\rightarrow b\bar{b}}^{\mathrm{ML}}$ in the next sections.

We also note that the space of possible inputs is degenerate since a monotonic function of an observable has the same discrimination power as the original observable. However, due to the finite binning required to calculate the AUC's from the likelihood distributions, and statistical fluctuations in a given data sample, the observables do not have precisely the same power as monotonic functions of themselves. The issue of degeneracies is not explicitly dealt with in the minimization procedure, but if the networks are adequately trained over the input space, it is sufficient to locate any one `global' minimum among local minima of similar depth, using basin-hopping or any other global minimizer. 

A second approach to optimizing $\{a,b,c,...\}$ directly uses Eq.~\ref{eq:betaml}.  The product form can be used directly as a tunable function for predicting signal/background with tunable parameters $\{a,b,c,...\}$.  This is a more direct way of identifying the optimal solution without explicitly modeling the probability distributions.  Optimizing a generic function is possible with methods like stochastic gradient decent, but the product observable is amenable to a significant simplification\footnote{We thank Eric Metodiev for this insightful observation.}.  In particular, two classifiers that are monotonic transformations of each other result in the same classification performance.  By taking the logarithm of Eq.~\ref{eq:betaml}, one can transform the problem into linear regression\footnote{Linear regression was proven to be sufficient for all IRC safe observables Ref.~\cite{Komiske:2017aww}, however our results need not be IRC safe.} where the inputs are $\log(\tau)$ and the coefficients are the exponents.  This approach uses the mean squared error loss to identify $\{a,b,c,...\}$.    The observable selected using this procedure will be denoted $\hat{\beta}_{3,H\rightarrow b\bar{b}}^{\mathrm{ML}}$ in the next sections.

In the limit of infinite data and an arbitrarily flexible neural network, both the ensemble learning and linear regression approaches should achieve the same performance.  The latter is significantly easier to train, but the complex approach may provide additional benefits because by providing access to the probability distributions, one can optimize any performance metric directly.  This includes batch-level losses like the AUC, false positive rate at a fixed true-positive rate, etc.  The mean squared error loss should be sufficient to optimize all of these metrics, but maybe prevented from reaching the desired optimum due to limited training statistics.  In practice, we do not find this to be the case with the setup presented here, but the structure may be useful for related tasks in the future.


\section{\label{sec:results} Results}

In this section, we present the new observables obtained for the different classification tasks for the ungroomed \Zp~samples (the groomed case is in Appendix~\ref{subsec:Groomed_ZvQCD}). For closure, we first demonstrate that this new procedure produces an observable for ungroomed \Hbb~discrimination with the same performance as the $\beta_3$ observable proposed in Ref.~\cite{Datta:2017lxt} (the groomed case in Appendix~\ref{subsec:H2bb_sd}). Then we extend the procedure to higher $M$-body phase space by applying it to \Zp~ discrimination for three values of $m_{Z^\prime}$, and propose new observables for those classification tasks. 

\begin{figure*}[t!]
	\centering
	\begin{minipage}{\textwidth}
		\centering
		\subfloat[]
		{
			\includegraphics[width=0.45\textwidth]{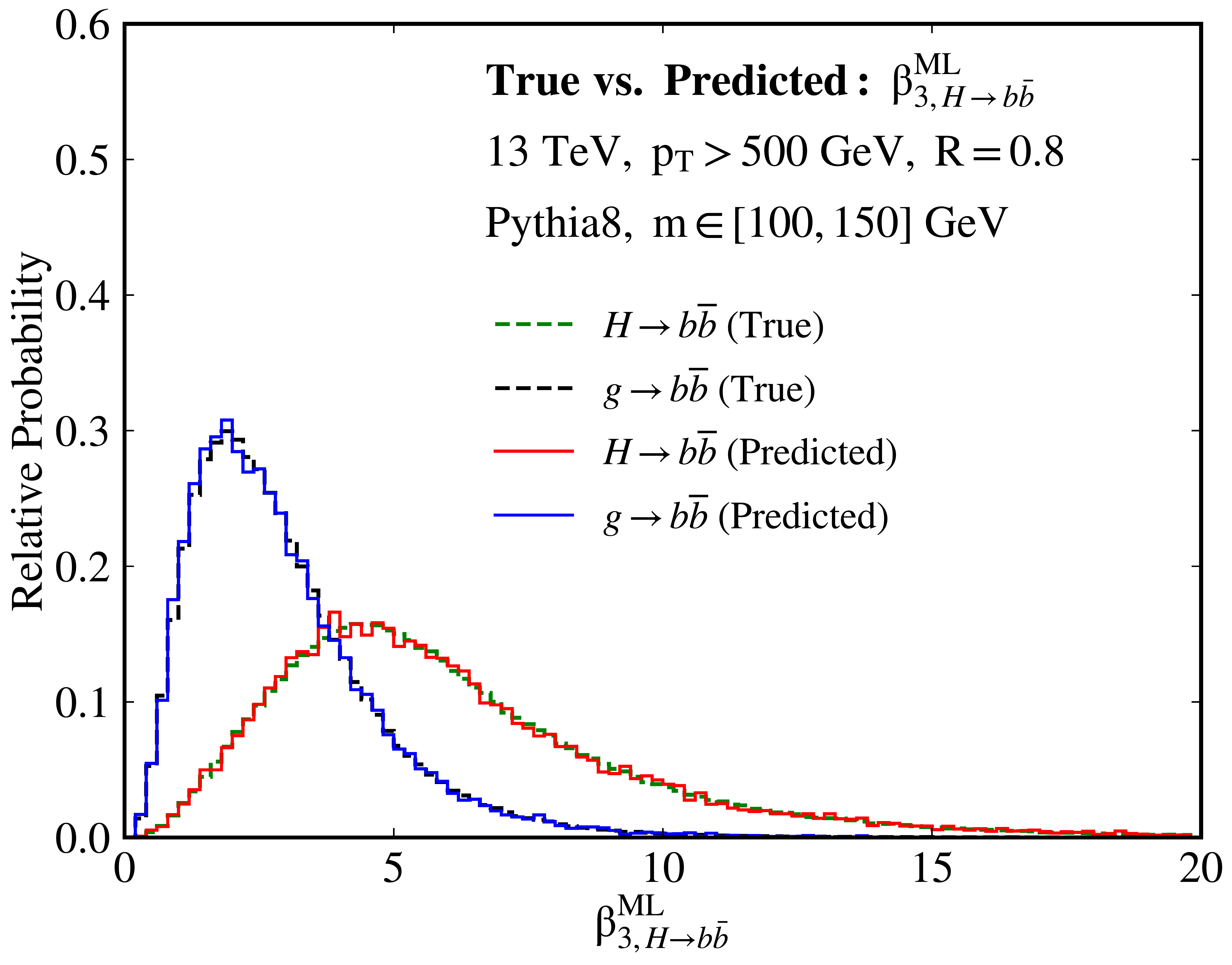}
			\label{fig:H2bb_ungroomed_obsdist_ML}
		}
		\subfloat[]
		{
			\includegraphics[width=0.45\textwidth]{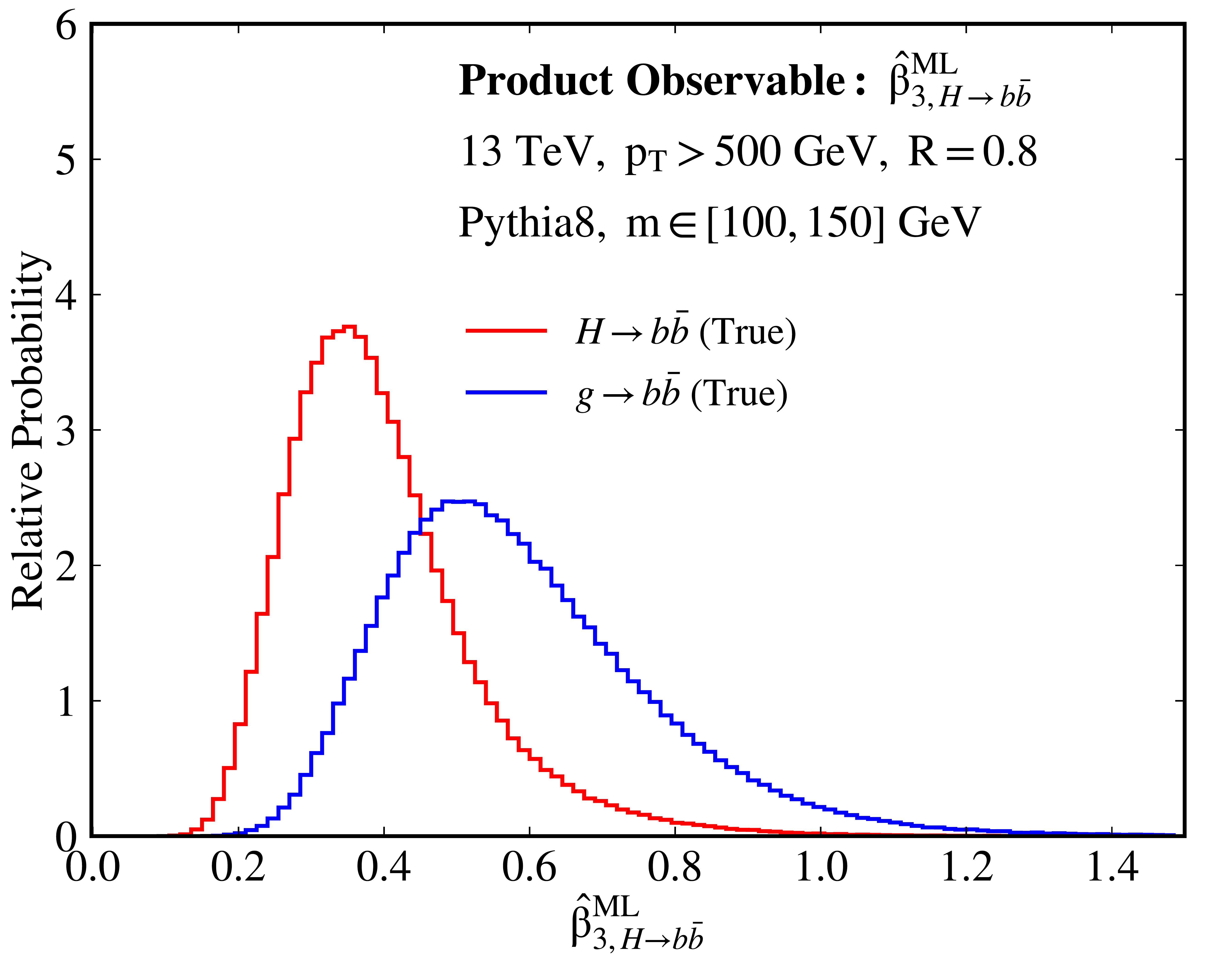}
			\label{fig:H2bb_ungroomed_obsdist_MSE}
		}\\
		\subfloat[]
		{
			\includegraphics[width=0.45\textwidth]{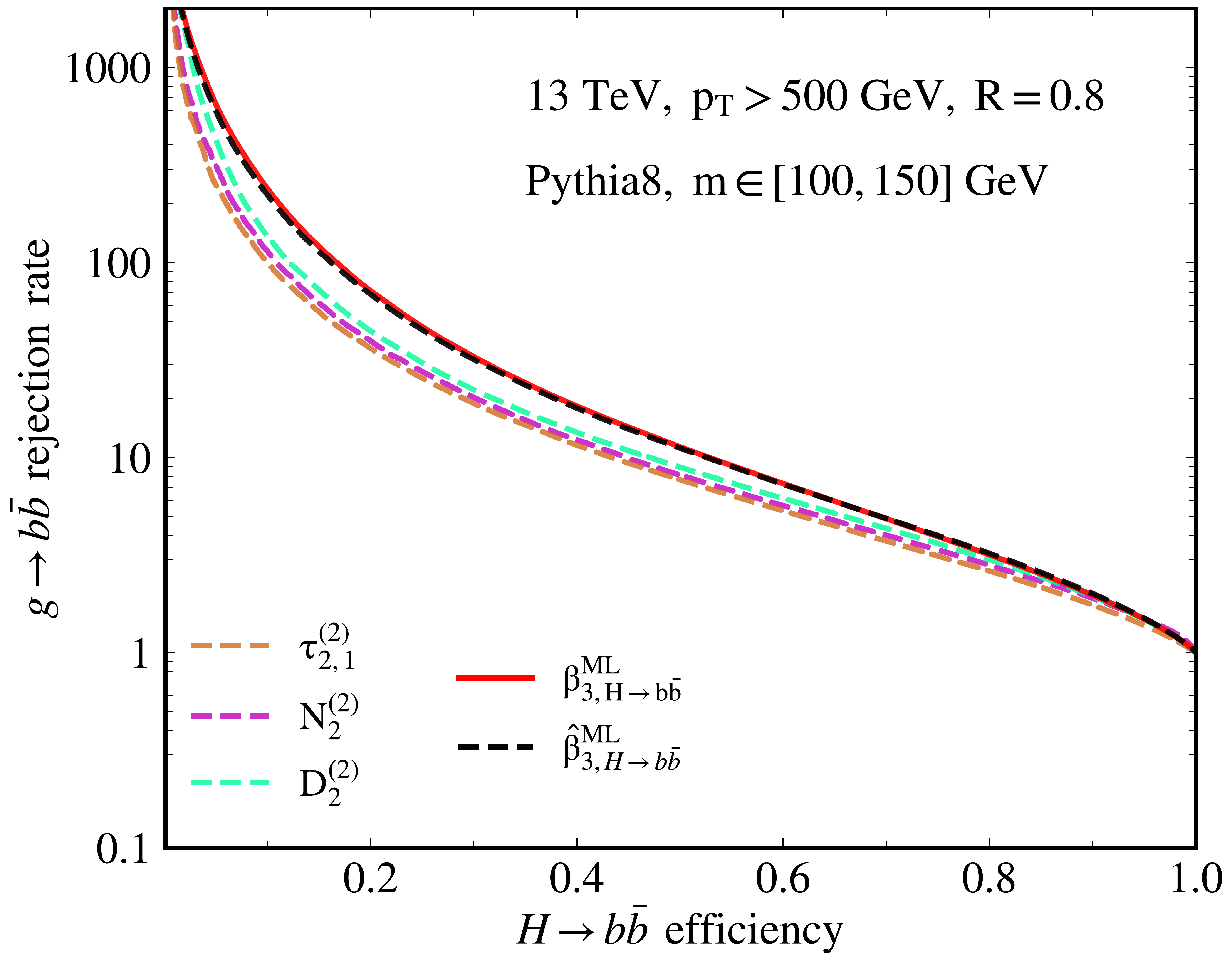}
			\label{fig:H2bb_ungroomed_obscomp_ROC}
		}
		\subfloat[]
		{
			\includegraphics[width=0.45\textwidth]{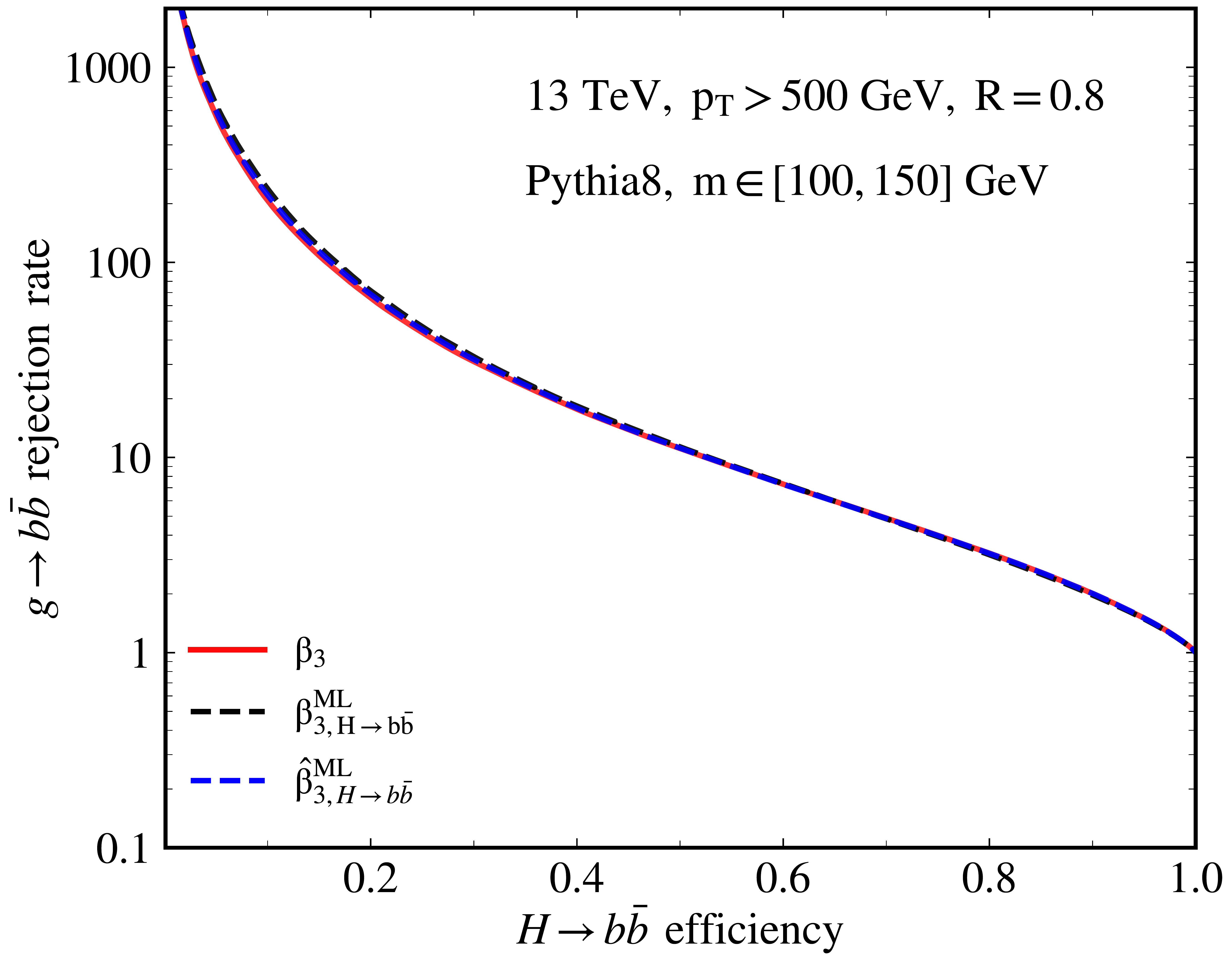}
			\label{fig:H2bb_ungroomed_oldvsnew_ROC}
		}
		\caption{(a): Comparison of the probability density function of the new $\beta_{3,H\rightarrow b\bar{b}}^{ML}$ observables for ungroomed \Hbb~discrimination, using $\sim 500,000$ signal and background samples, and the distributions of the regression DNN prediction. The  distributions are rescaled by a constant for the sake of visual comparison. (b) Probability densities of $\hat{\beta}_{3,H\rightarrow b\bar{b}}^{ML}$ obtained via linear regression. (c): Comparison of discrimination power of $\beta_{3,H\rightarrow b\bar{b}}^{ML}$ and $\hat{\beta}_{3,H\rightarrow b\bar{b}}^{ML}$ to standard observables. (d): Comparison of $\beta_{3,H\rightarrow b\bar{b}}^{ML}$ and $\hat{\beta}_{3,H\rightarrow b\bar{b}}^{ML}$ to $\beta_{3}$ proposed in Ref.~\cite{Datta:2017lxt}; we note that three observables provide essentially the same discrimination power. }
		\label{fig:H2bb_ungroomed_ROC}
	\end{minipage}
\end{figure*}

\subsection{\label{subsec:H2bb}Ungroomed \Hbb~vs.~\gbb~discrimination}

Utilizing the result that discrimination power for ungroomed \Hbb~vs.~\gbb~discrimination saturates at 3-body phase space, we use the procedures proposed in the previous section to find the optimal product observable. The final values for the parameters $\{a,...,e\}$ obtained through the optimization are presented in Table~\ref{tab:ungroomed_Hbb}, along with those obtained in the previous study.  Interestingly, the exponents with the ensemble method are nearly the same for $a$, $b$, $d$, and $e$, but slightly different for $c$.  For the regression method, the exponents are nearly the same as the ensemble method up to a constant factor (approximately $-2$) for $c$, $d$, and $e$, but not for $a$ and $b$.  These results indicate the presence of multiple observables with comparable performance.

\begin{table}[h]
	\begin{center}
		\caption{\label{tab:ungroomed_Hbb}Summary of parameters for the product observables for ungroomed \Hbb~discrimination as proposed in Ref.~\cite{Datta:2017lxt} and as constructed via the procedures presented in this work (\Figs{fig:H2bb_ungroomed_obsdist_ML}{fig:H2bb_ungroomed_obsdist_MSE}).}
		\resizebox{\columnwidth}{!}{
			\begin{ruledtabular}
				\begin{tabular}{ccccccc}
					Observable&$a$&$b$&$c$&$d$&$e$&AUC\\
					
					\hline\vspace{0.1cm}
					$\beta_{3}$   & 2.0 & 0.0 & 0.0 & 0.5 & -1.0 & 0.823\\

					$\beta_{3,H\rightarrow b\bar{b}}^{\mathrm{ML}}$ & 1.87 & -0.02 & -0.14 & 0.66 & -0.98 & 0.823\\ 
					
					$\hat{\beta}_{3,H\rightarrow b\bar{b}}^{\mathrm{ML}}$ & -0.11 & -0.58 & 0.09 & -0.25 &  0.51 & 0.824\\

				\end{tabular}
		\end{ruledtabular}}
	\end{center}
\end{table}

In Fig.~\ref{fig:H2bb_ungroomed_obsdist_ML}, we plot the distributions of the new observable computed for signal and background, along with the prediction from the ensemble neural network. We note that the network provides a good match to the true distribution, where the latter is also calculated on 10 times more jets. Further, in Fig.~\ref{fig:H2bb_ungroomed_obsdist_MSE} we plot the distributions of the observable obtained via the ML regression method. We then compare the ROC curves for the new observables to $D_2^{(2)}$~\cite{Larkoski:2014gra}, $N_2^{(2)}$~\cite{Moult:2016cvt} observables, and $\tau_{21}^{(2)}$ in Fig.~\ref{fig:H2bb_ungroomed_obscomp_ROC}. 

In addition, we also compare the new observables to $\beta_3$ in Fig.~\ref{fig:H2bb_ungroomed_oldvsnew_ROC} to demonstrate that the three observables have essentially the same discrimination power as expected. Then, this allows us to proceed to applying the procedure on higher dimensional problems.

\subsection{\label{subsec:Ungroomed_ZvQCD}Ungroomed \Zp~vs. $\mathrm{QCD}$}
\begin{figure*}[t]
	\centering
	\begin{minipage}{\textwidth}
		\centering
		\subfloat[$m_{Z'}=50~\mathrm{GeV}$]
		{
			\includegraphics[width=0.33\textwidth]{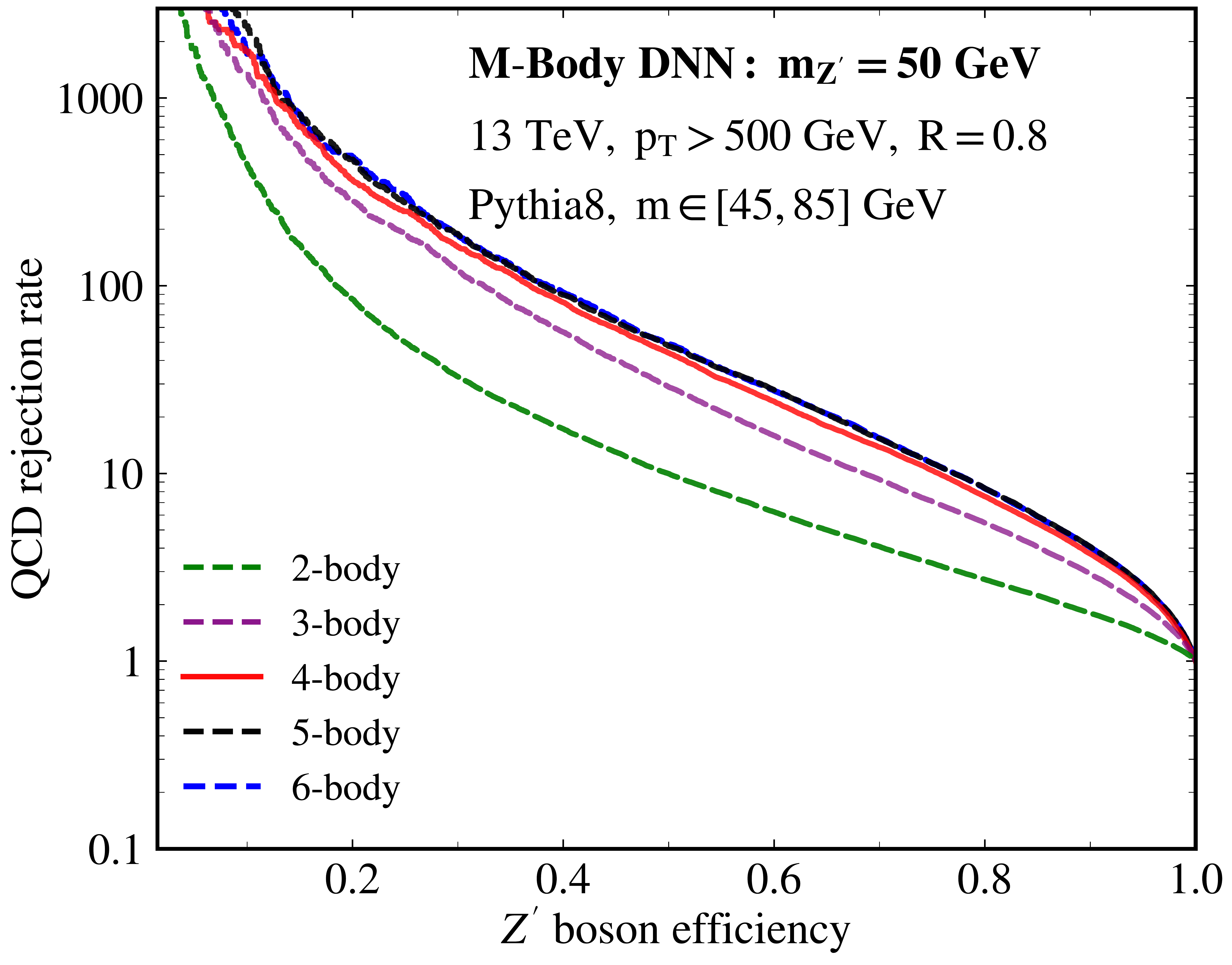}
			\label{fig:satroc50}
		}
		\subfloat[$m_{Z'}=90~\mathrm{GeV}$]
		{
			\includegraphics[width=0.33\textwidth]{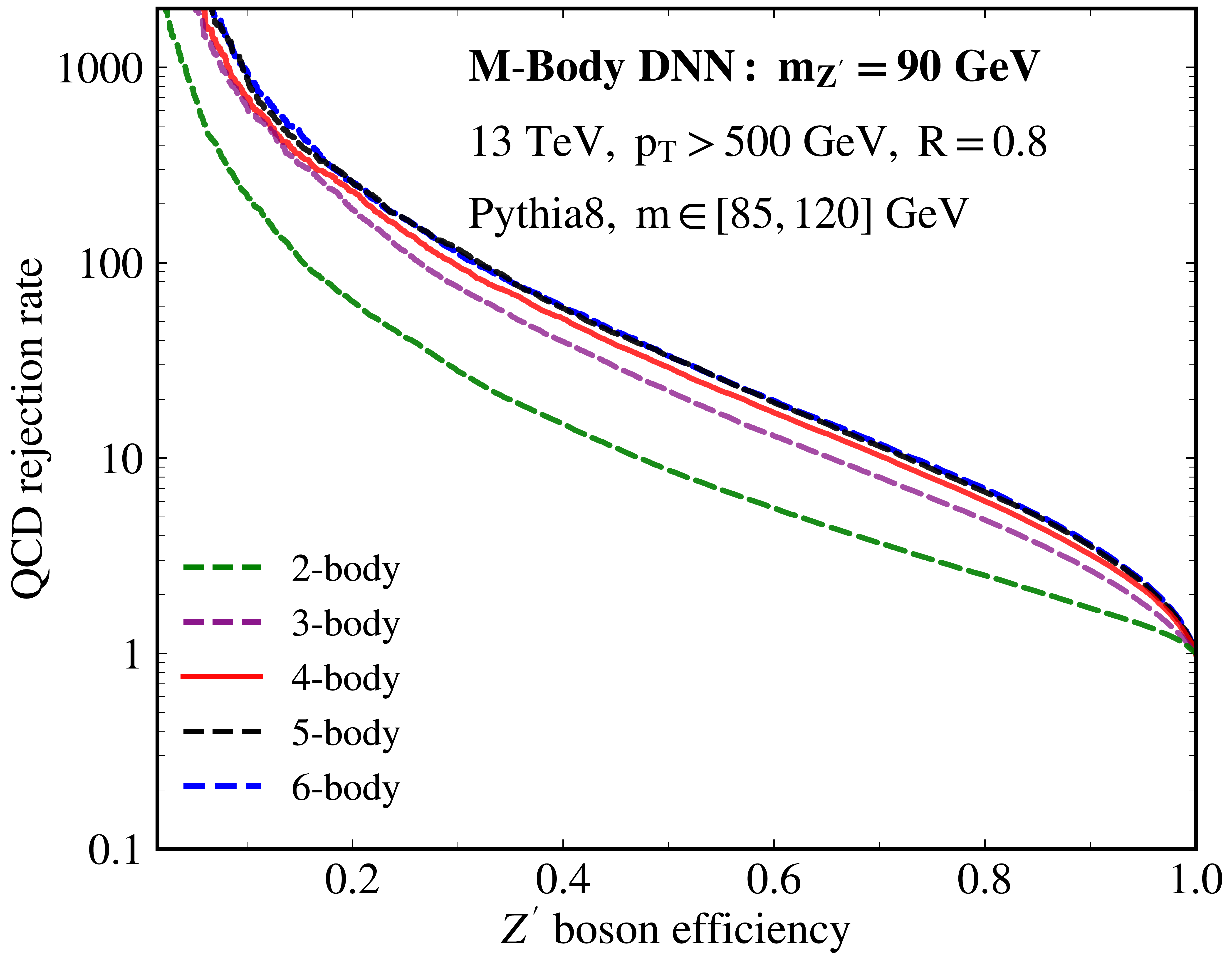}
			\label{fig:satroc90}
		}
		\subfloat[$m_{Z'}=130~\mathrm{GeV}$]
		{
			\includegraphics[width=0.33\textwidth]{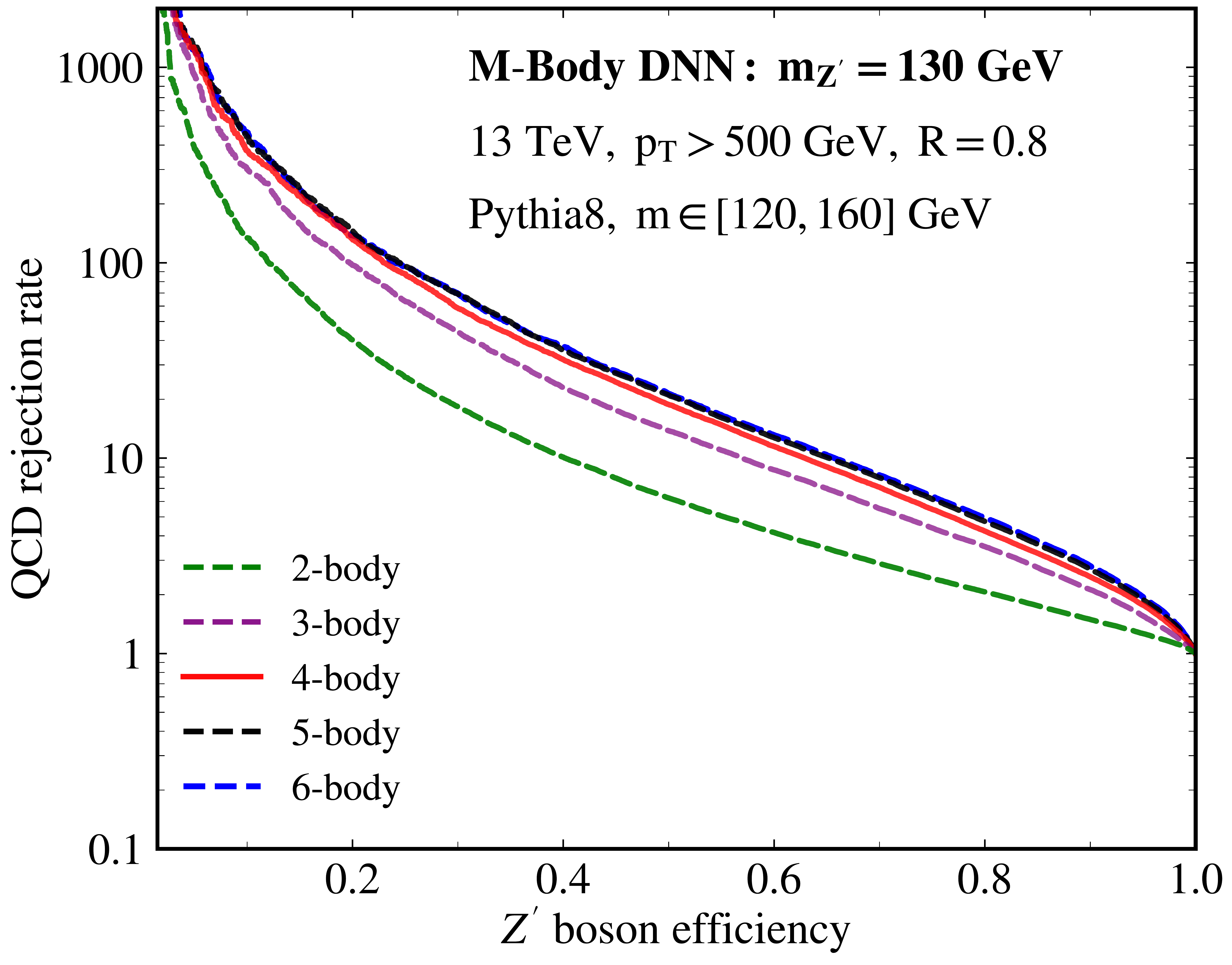}
			\label{fig:satroc130}
		}
		
		\caption{$M$-body discrimination results for ungroomed \Zp~vs.~QCD jets. Discrimination power is effectively saturated at 4-body phase space for each case.}
		\label{fig:mbody_saturation_ungroomedZ}
	\end{minipage}
\end{figure*}
We first train neural network classifiers on the $M$-body $N$-subjettiness bases, to identify the point of saturation of discrimination power for each value of $m_{Z^\prime}$.\footnote{A single neural network architecture, consisting of seven fully connected (five hidden) layers, was utilized for all of the classification tasks. The first four Dense layers consisted of 1000, 1000, 750 and 500 nodes respectively, and were assigned a Dropout \cite{dropout} regularization of 0.2, to prevent over-fitting on training data. The next two Dense layers consisted of 250 nodes with Dropout regularization 0.1, and 100 nodes without Dropout. The input layer and all hidden layers utilized the ReLU activation function \cite{conf/icml/NairH10}, while the output layer, consisting of a single node, used a sigmoid activation. The network was compiled with the binary cross-entropy loss minimization function, using the Adam optimization \cite{DBLP:journals/corr/KingmaB14}. Models were trained with Keras' default EarlyStopping callback, with appropriate patience thresholds, to further negate possible over-fitting.}
The results are presented in Fig.~\ref{fig:mbody_saturation_ungroomedZ}, showing that saturation occurs with the 4-body phase space for each case. 

We then proceed to construct the $\beta_{4, Z^{\prime}}^{ML}$ and  $\hat{\beta}_{4, Z^{\prime}}^{ML}$ product observables with the elements of the 8-dimensional 4-body basis, and run the procedure described in Sec.~\ref{sec:deeplearn} and construct the new observables optimized for \Zp~discrimination at three different values of $m_{Z^\prime}$.

\begin{figure*}[htb]
	\centering
	\begin{minipage}{\textwidth}
		\centering
		
		\subfloat[$m_{Z'}=50~\mathrm{GeV}$]
		{
			\includegraphics[width=0.29\textwidth]{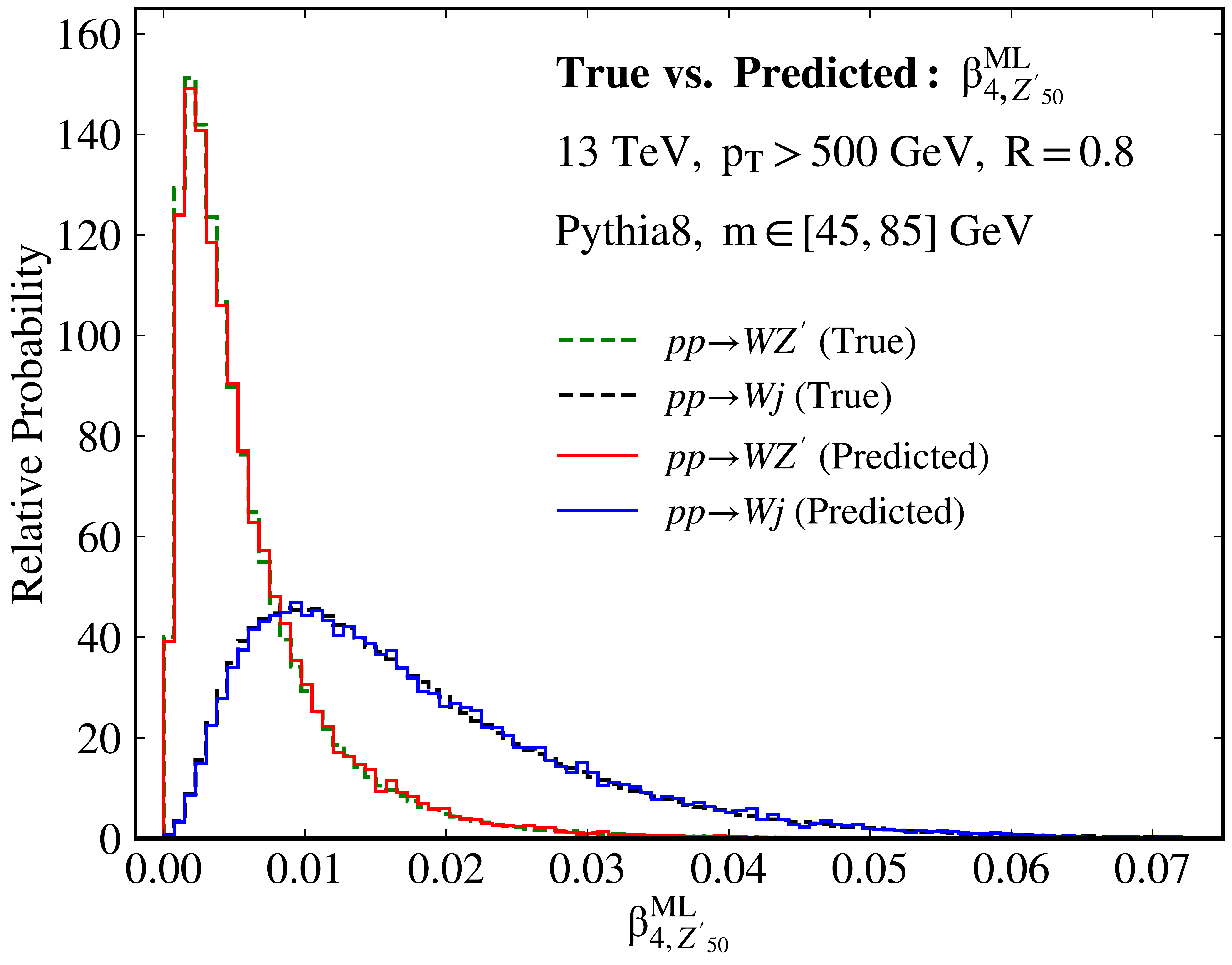}
			\label{fig:obsdist_DNN_50}
		}\hspace{0.7cm}
		\subfloat[$m_{Z'}=90~\mathrm{GeV}$]
		{
			\includegraphics[width=0.29\textwidth]{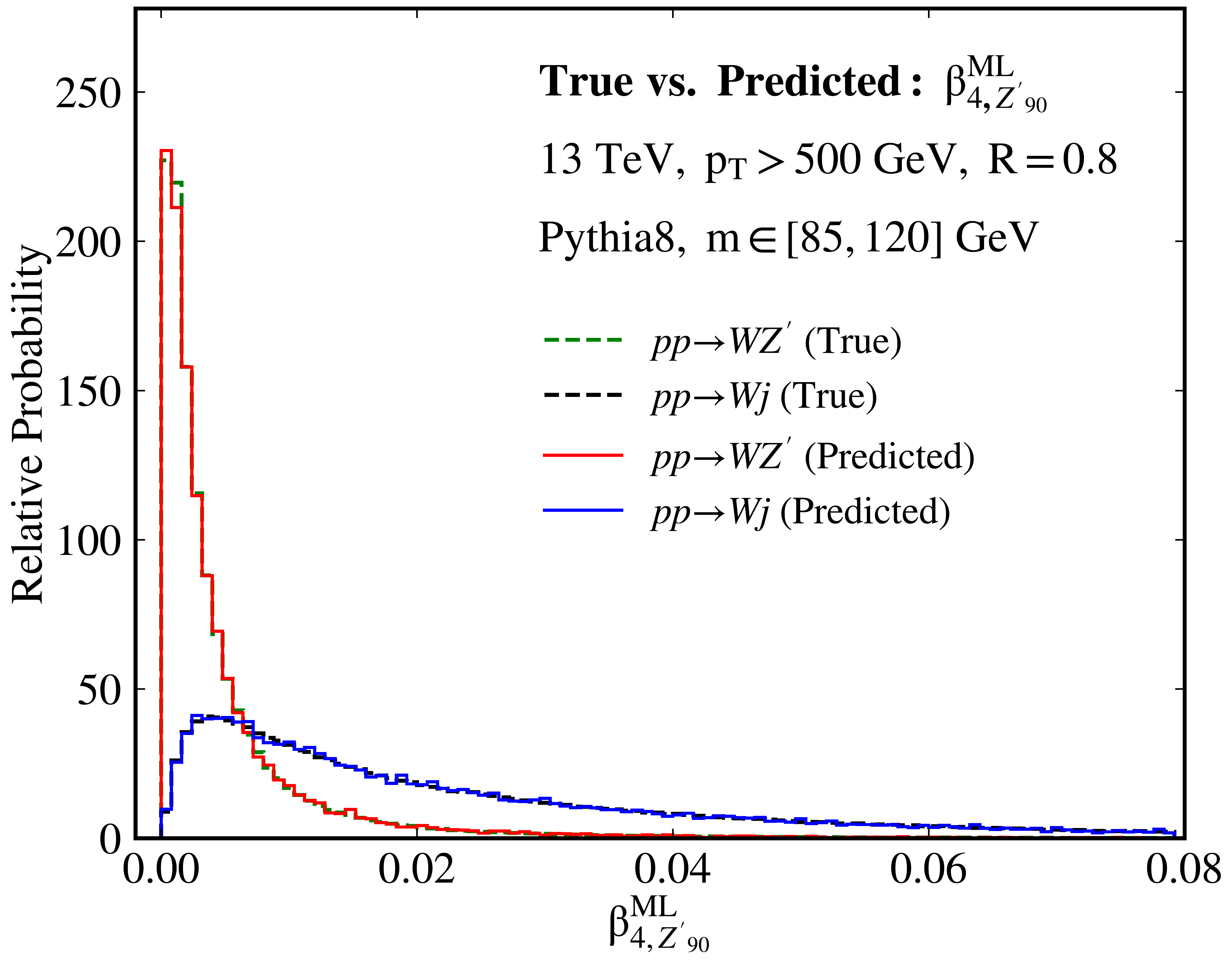}
			\label{fig:obsdist_DNN_90}
		}\hspace{0.7cm}
		\subfloat[$m_{Z'}=130~\mathrm{GeV}$]
		{
			\includegraphics[width=0.29\textwidth]{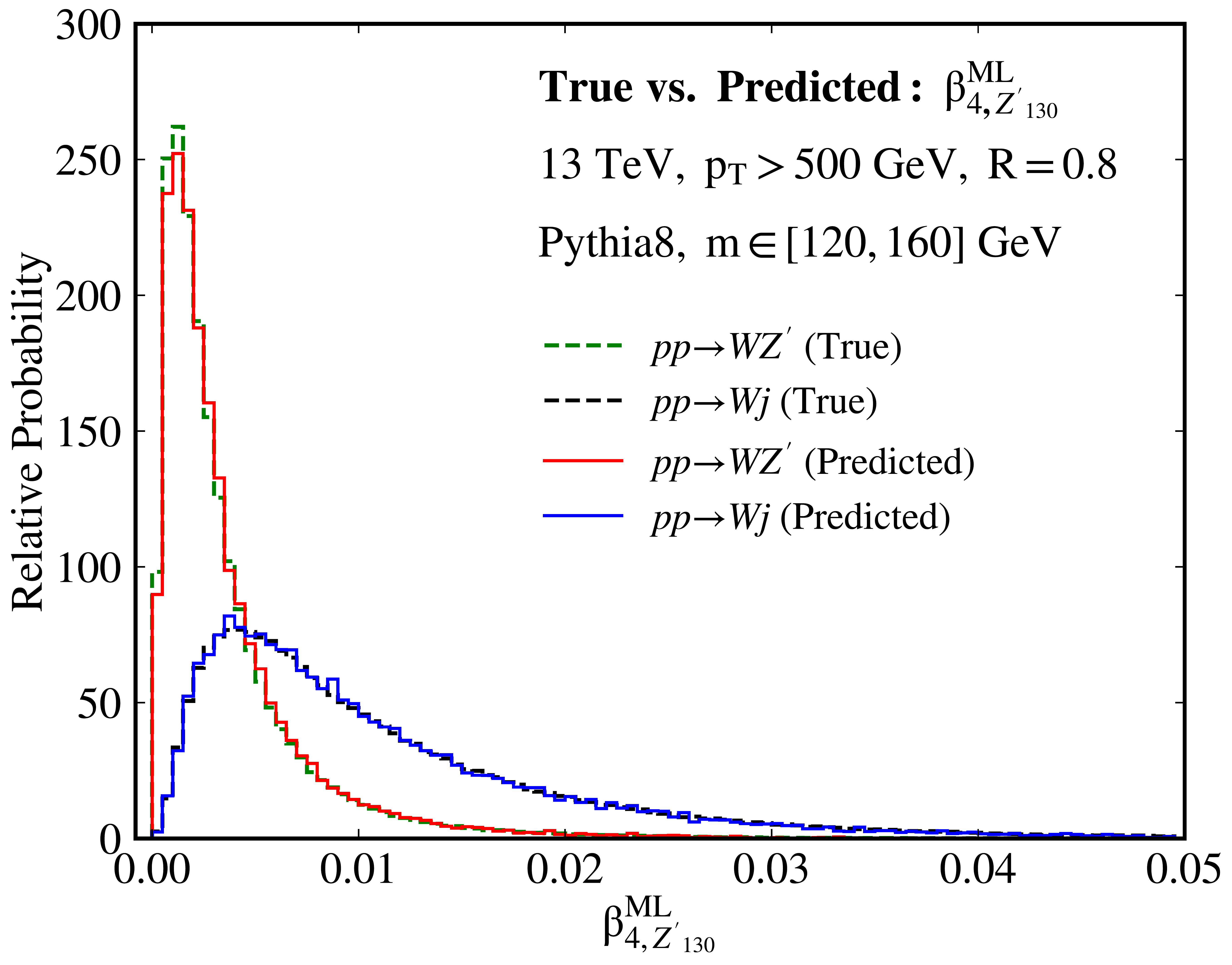}
			\label{fig:obsdist_DNN_130}
		}\\
		\subfloat[$m_{Z'}=50~\mathrm{GeV}$]
		{
			\includegraphics[width=0.29\textwidth]{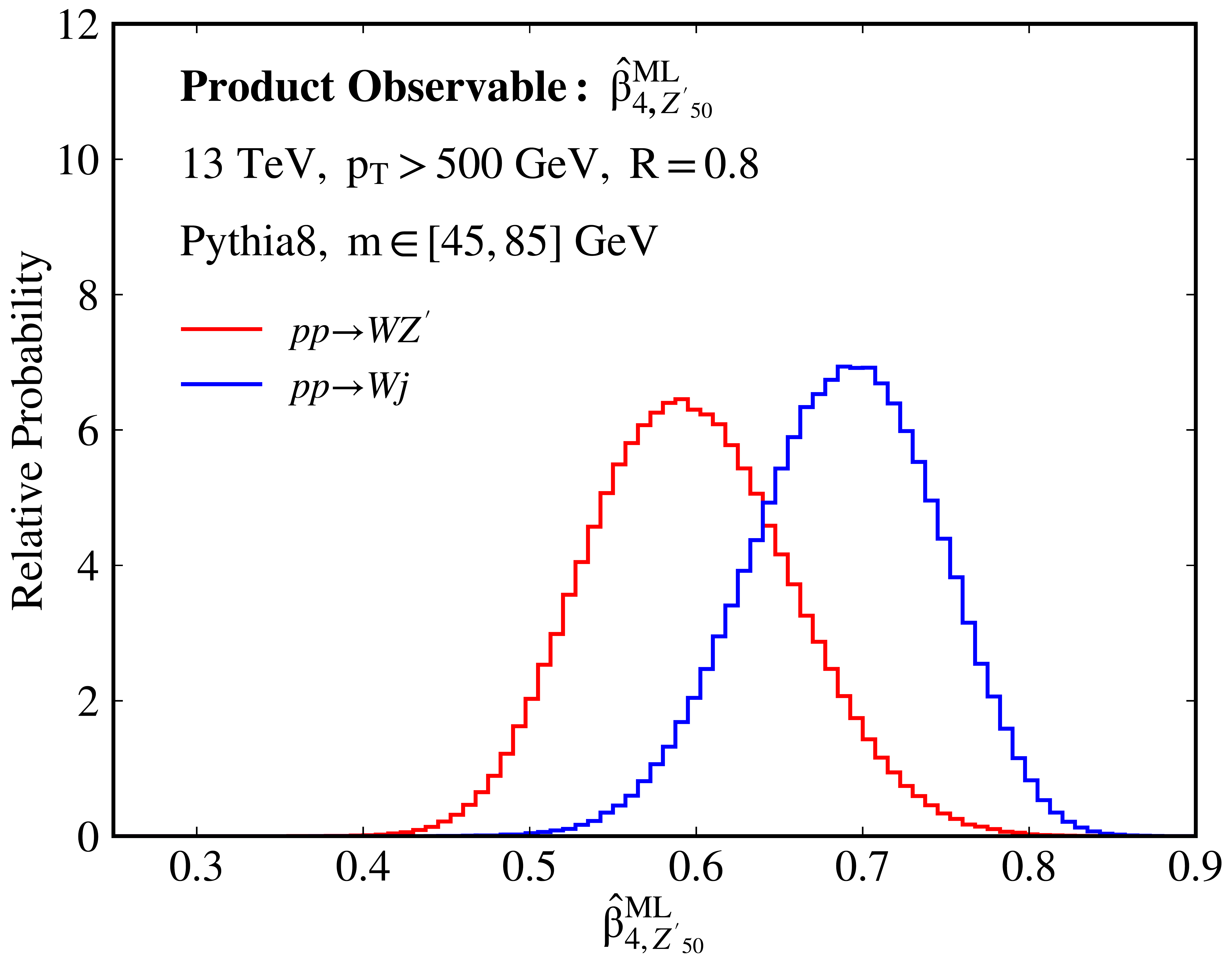}
			\label{fig:obsdist_DNN_50}
		}\hspace{0.7cm}
		\subfloat[$m_{Z'}=90~\mathrm{GeV}$]
		{
			\includegraphics[width=0.29\textwidth]{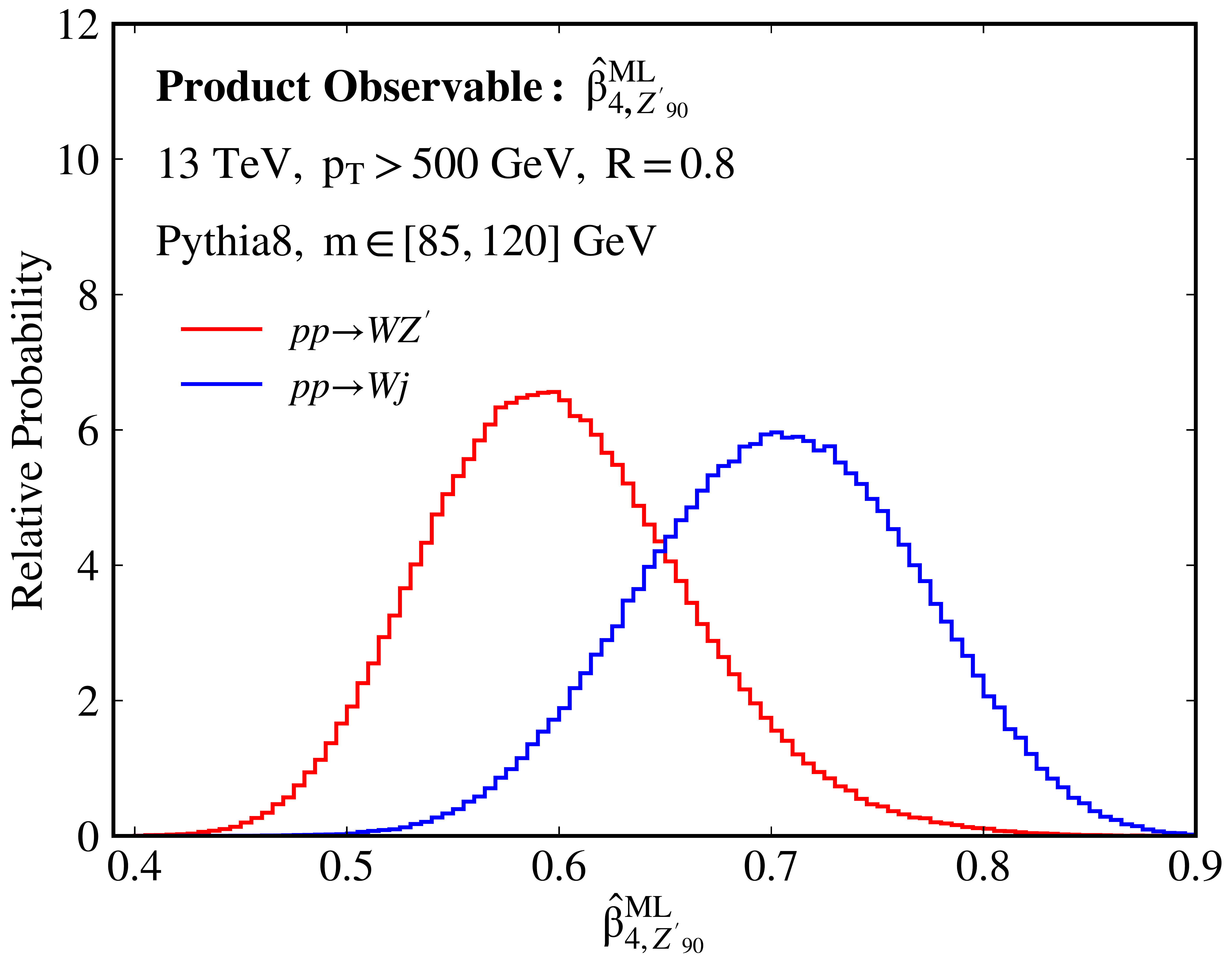}
			\label{fig:obsdist_MSE_DNN_90}
		}\hspace{0.7cm}
		\subfloat[$m_{Z'}=130~\mathrm{GeV}$]
		{
			\includegraphics[width=0.29\textwidth]{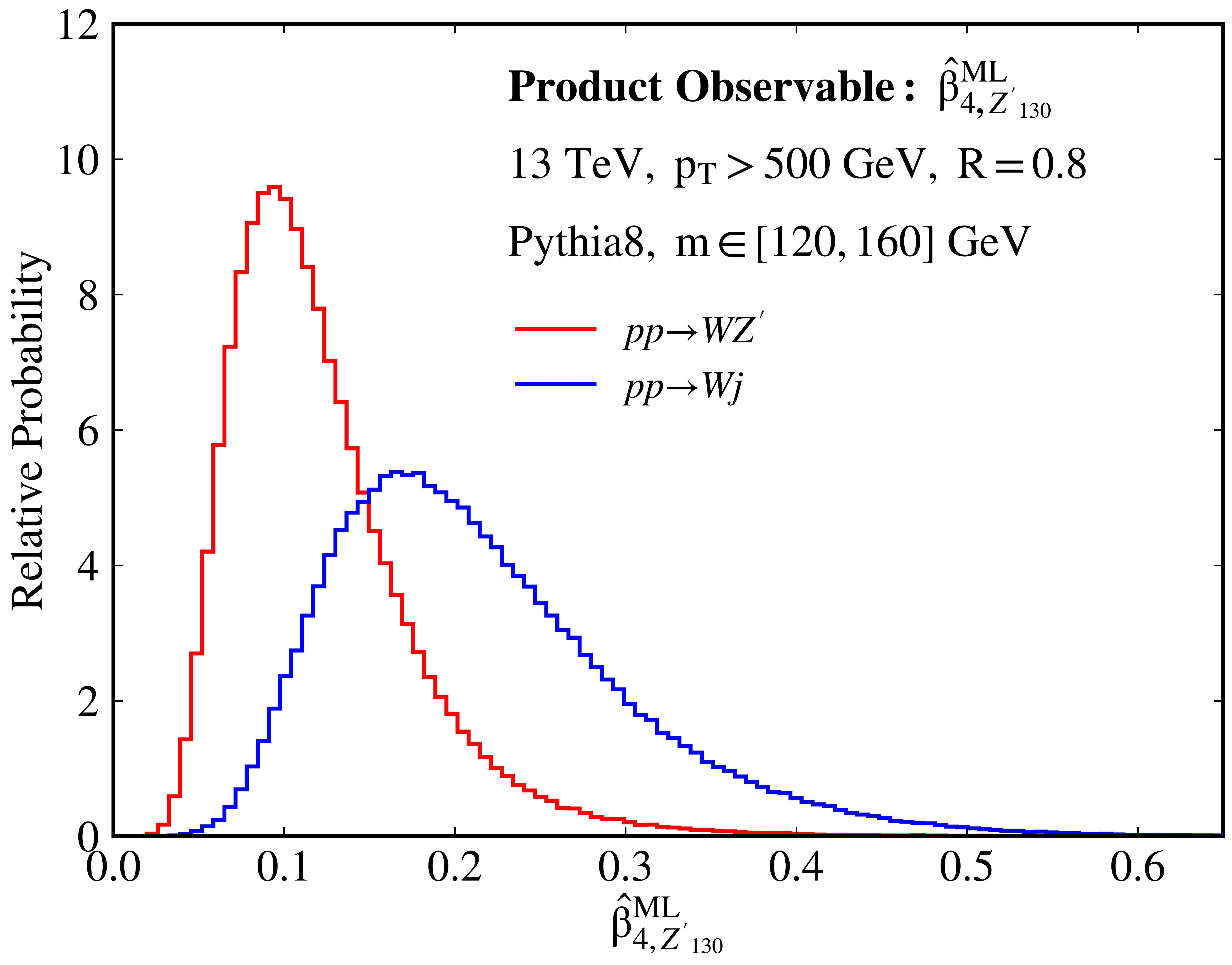}
			\label{fig:obsdist_MSE_DNN_130}
		}\\
		\caption{Top panel [a-c]: Comparison of the probability density function of the new $\beta_{4}^\text{ML}$ observables for ungroomed \Zp~discrimination, calculated for $\sim 500,000$ signal and background samples, and the distributions of the regression DNN predictions of 25,000 observable values. The  distributions are rescaled for the sake of visual comparison. Bottom panel [d-f]: Distributions of the $\hat{\beta}_{4}^\text{ML}$ observables for ungroomed \Zp~discrimination that were obtained via linear regression.}
		\label{fig:distributions_ungroomed}
	\end{minipage}
	
	\begin{minipage}{\textwidth}
		\centering
		
		\subfloat[$m_{Z'}=50~\mathrm{GeV}$]
		{
			\includegraphics[width=0.33\textwidth]{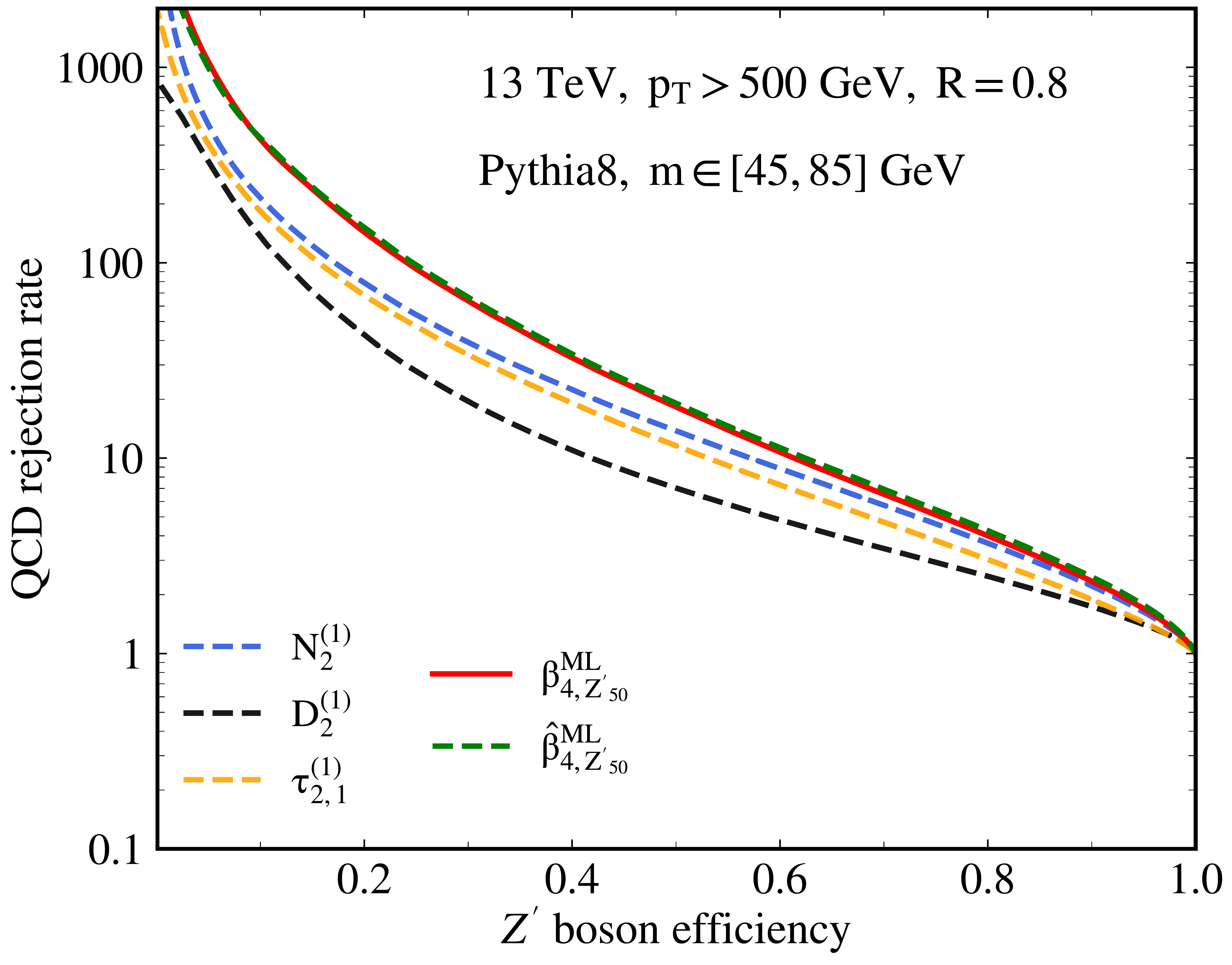}
			\label{fig:obscomp50}
		}
		\subfloat[$m_{Z'}=90~\mathrm{GeV}$]
		{
			\includegraphics[width=0.33\textwidth]{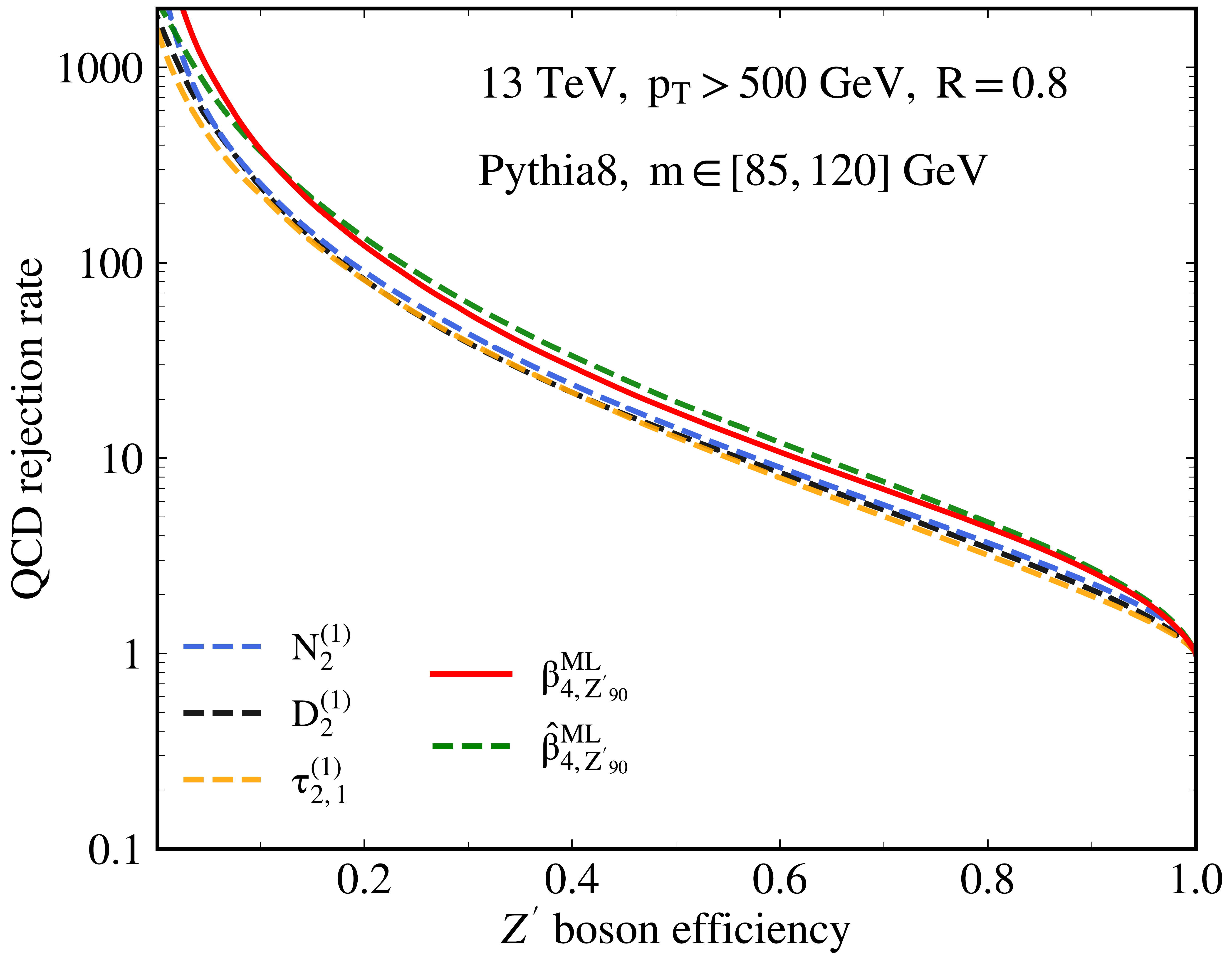}
			\label{fig:obscomp90}
		}		
		\subfloat[$m_{Z'}=130~\mathrm{GeV}$]
		{
			\includegraphics[width=0.33\textwidth]{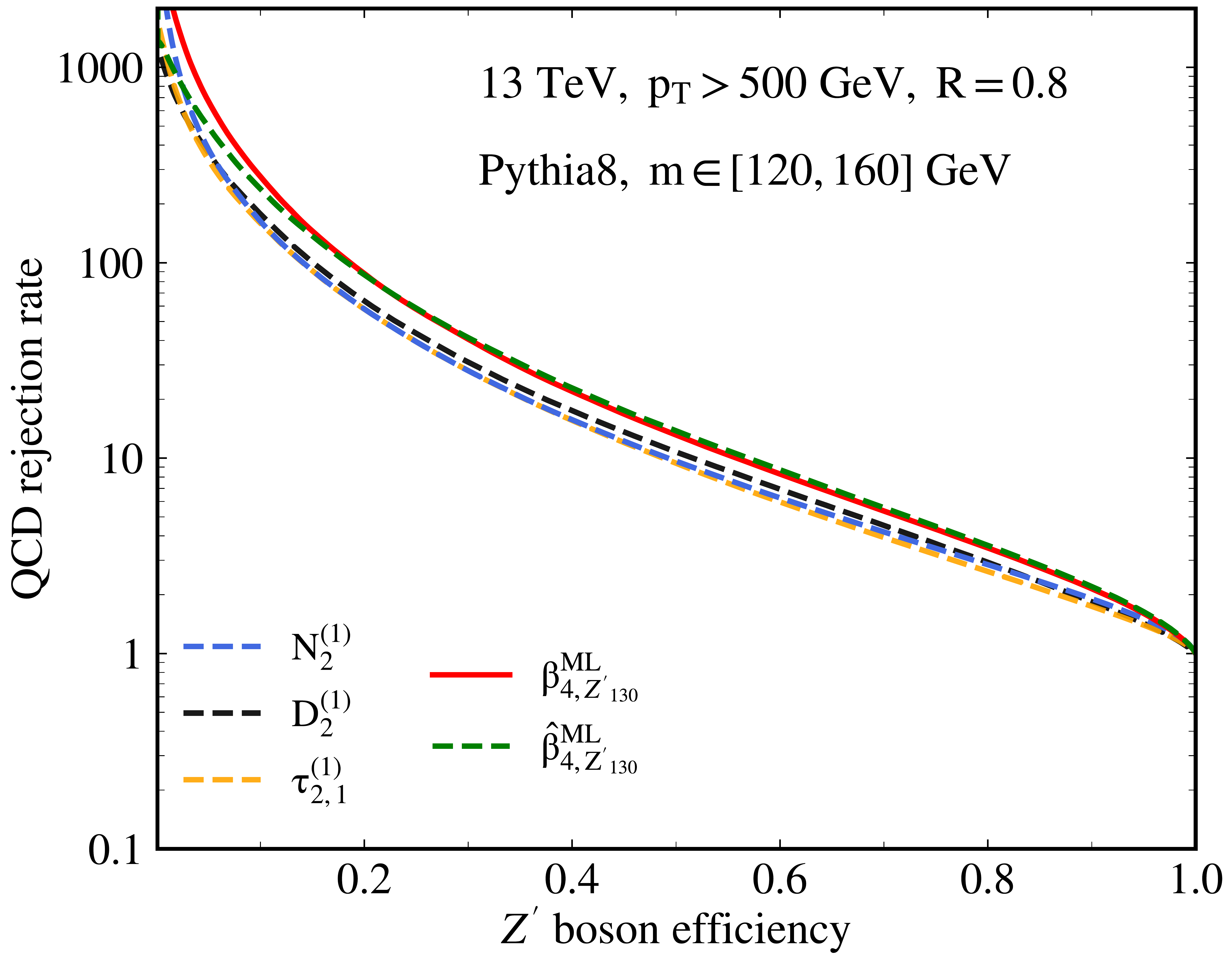}
			\label{fig:obscomp130}
		}\\
		
		\subfloat[$m_{Z'}=50~\mathrm{GeV}$]
		{
			\includegraphics[width=0.33\textwidth]{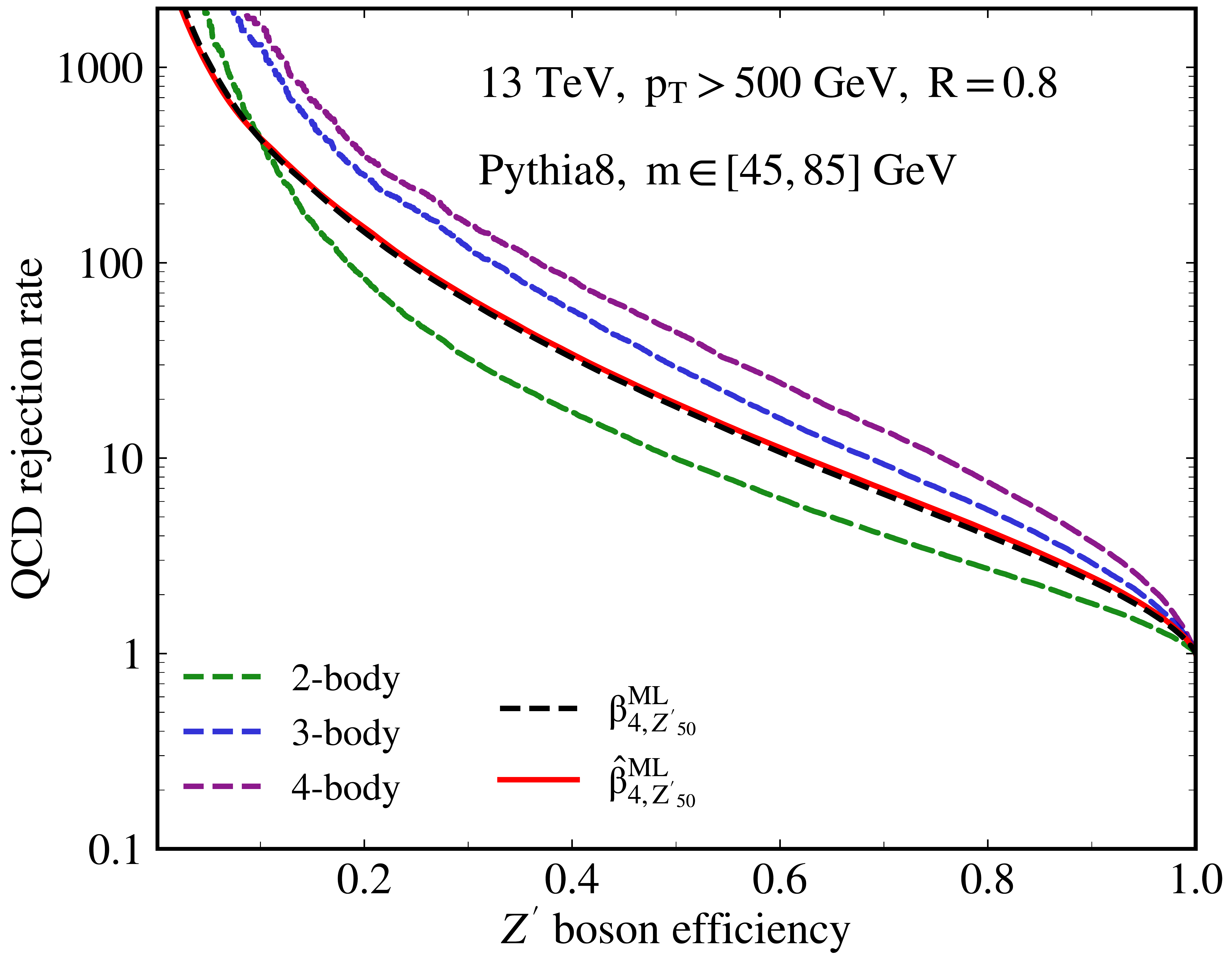}
			\label{fig:dnnobscomp50}
		}
		\subfloat[$m_{Z'}=90~\mathrm{GeV}$]
		{
			\includegraphics[width=0.33\textwidth]{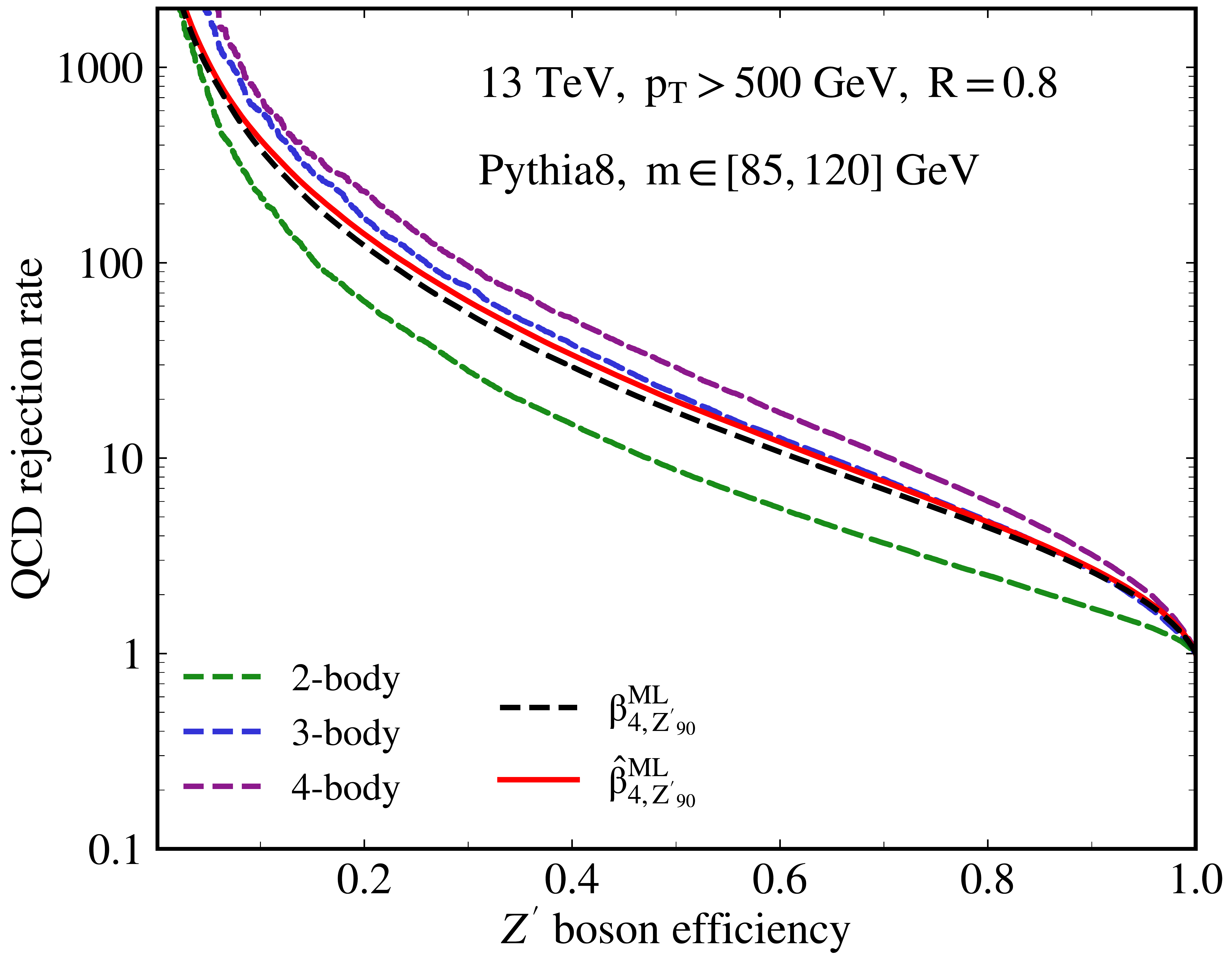}
			\label{fig:dnnobscomp90}
		}		
		\subfloat[$m_{Z'}=130~\mathrm{GeV}$]
		{
			\includegraphics[width=0.33\textwidth]{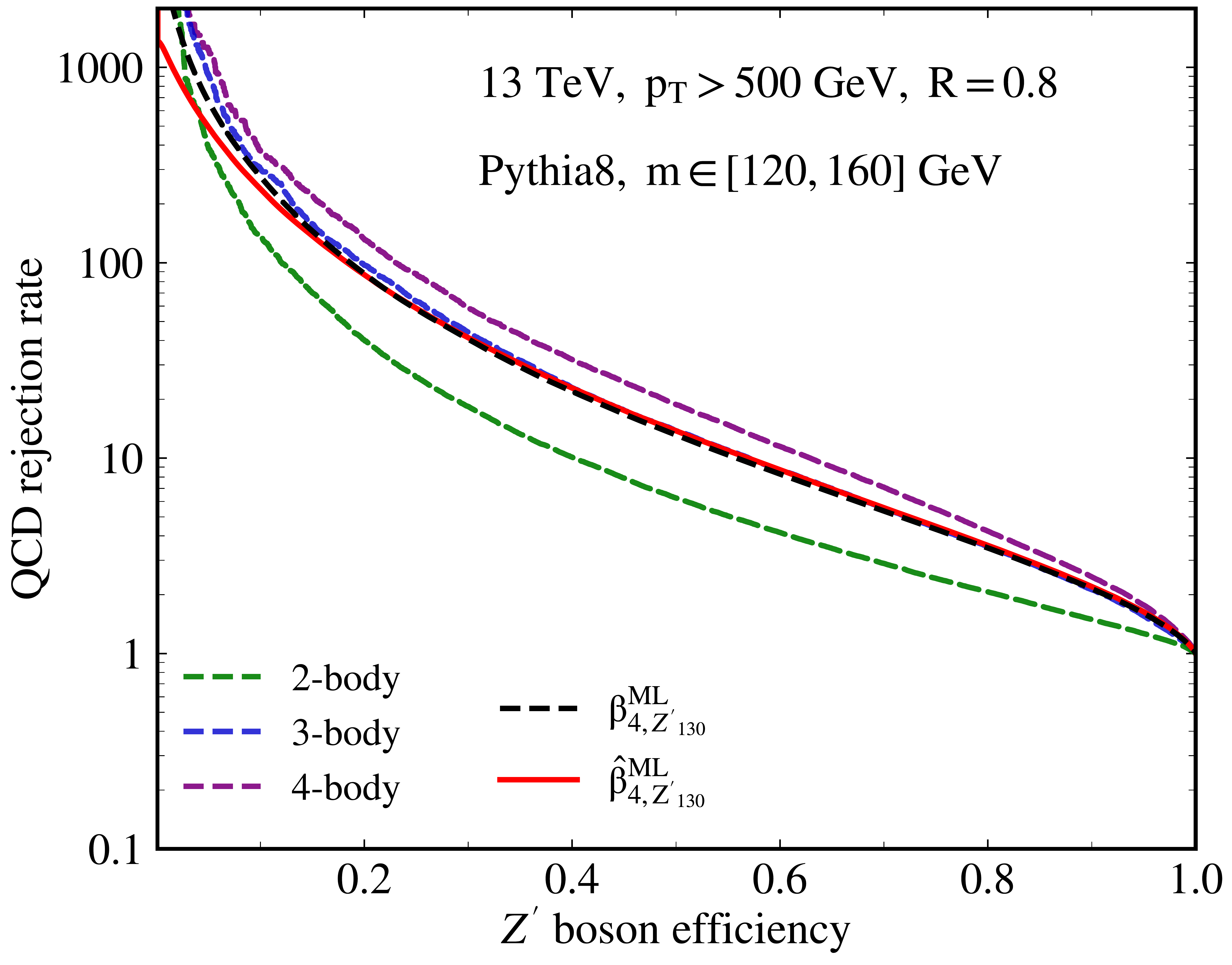}
			\label{fig:dnnobscomp130}
		}

		\caption{Top panel [a-c]: Comparison of discrimination power of $\beta_{4}^\text{ML}$ observables to standard observables; the latter are calculated with an angular exponent of 1, for which they were observed to perform best. Bottom panel [d-f]: Comparison of $\beta_{4}^\text{ML}$ to discrimination power of neural networks trained on the $M$-body observable bases; the observables seem to capture increasing amounts of the discrimination power of the 3- and 4-body neural networks with increasing $m_{Z^\prime}$.}
		\label{fig:ROC_comp_ungroomed}
	\end{minipage}
\end{figure*}

We present the distributions of the new observables for \Zp~ discrimination in Fig.~\ref{fig:distributions_ungroomed} and then compare their discrimination power to standard observables and DNN's trained on the spanning $N$-subjettiness bases in Fig.~\ref{fig:ROC_comp_ungroomed}.  The corresponding values of $\{a,b,c,...,h\}$ and the AUCs are in tables~\ref{tab:ungroomed_Z_ML},~\ref{tab:ungroomed_Z_MSE} and~\ref{tab:ungroomed_Z_AUC}, respectively.  The comparison of the true and predicted distributions in Fig.~\ref{fig:distributions_ungroomed} illustrates the excellent quality of the regression network.  The ROC curves in Fig.~\ref{fig:ROC_comp_ungroomed} show that the learned $\beta^\text{ML}$ and $\hat{\beta}^\text{ML}$ outperform the state-of-the-art single physics-motivated observables (top row), though the product observables do not fully saturate the performance of the DNN trained on the full $4$-body phase space (bottom row).  This suggests that a more flexible form (other than a simple product) is required to build a simple observable to capture more of the classification information.  The product values obtained from the ensemble and regression methods are not simple scaling of each other, though the fact that both have a similar performance suggests that one is a monotonic transformation of the other.

The optimized $\beta^\text{ML}$ and $\hat{\beta}^\text{ML}$ observables are not identical for the different values of $m_{Z'}$ (tables~\ref{tab:ungroomed_Z_ML} and~\ref{tab:ungroomed_Z_MSE}), but it would be interesting to study to what extent the trends are physical or are due to the existence of multiple observables with similar performance.  We leave this study to future work.  However, a first indication that the observables contain similar physical information is studied in Appendix~\ref{app:beta4ML_comp}, where the optimized product for one mass is applied to another mass.  The ROC curves are similar for all three product observables when applied to the same $m_{Z'}$.

\begin{table}[h]
	\begin{center}
		\caption{\label{tab:ungroomed_Z_ML}Summary of parameters for $\beta_{4}^\text{ML}$ for ungroomed \Zp~vs. QCD discrimination at 3 mass points.}
		\resizebox{\columnwidth}{!}{
			\begin{ruledtabular}
				\begin{tabular}{ccccccccc}
					$m_{Z^\prime}$[GeV]&$a$&$b$&$c$&$d$&$e$&$f$&$g$&$h$\\
					\hline	\vspace{0.05cm}	
					50 & 2.72 & -3.78 & 0.63 & -2.77 & 1.54 & 0.20 & 2.36 & -0.28 \\\vspace{0.05cm}	
					90 & 0.90 & -2.87 & 0.18 & -1.78 & -0.72 & 1.79 & 2.48 & -0.44\\
					130 & 1.69 & -2.98 & 0.75 & -0.89 & -0.38 & 0.77 & 1.37 &  0.30 \\ 
				\end{tabular}
		\end{ruledtabular}}
	\end{center}
\end{table}

\begin{table}[h]
	\begin{center}
		\caption{\label{tab:ungroomed_Z_MSE}Summary of parameters for $\hat{\beta}_{4}^\text{ML}$ for ungroomed \Zp~vs. QCD discrimination at 3 mass points.}
		\resizebox{\columnwidth}{!}{
			\begin{ruledtabular}
				\begin{tabular}{ccccccccc}
					$m_{Z^\prime}$[GeV]&$a$&$b$&$c$&$d$&$e$&$f$&$g$&$h$\\
					\hline	\vspace{0.05cm}	
					50 & 1.06 & -1.11 & 0.25 & -0.56 & 0.43 & -0.07 & 0.22 & -0.01 \\\vspace{0.05cm}
					90 & 1.02 & -1.06 & 0.22 & -0.27 & 0.15 & 0.00 & 0.18 & 0.02\\
					130 & -1.09 & -0.43 & 0.25 & -0.97 & 0.37 & 0.12 & 0.60 &  0.19 \\ 
				\end{tabular}
		\end{ruledtabular}}
	\end{center}
\end{table}

\begin{table}[h]
	\begin{center}
		\caption{\label{tab:ungroomed_Z_AUC} Area under the ROC curve (AUC), from Fig.~\ref{fig:ROC_comp_ungroomed}, of the standard observables and the  $\beta_{4}^{\mathrm{ML}}$ observables, optimized for the corresponding signal, for ungroomed \Zp~ vs. QCD discrimination at 3 $m_{Z^\prime}$ points. The ROC curves are calculated using the full datasets, with $\sim$500,000 events passing the mass cut for each value of $m_{Z^\prime}$.}
		\resizebox{\columnwidth}{!}{
			\begin{ruledtabular}
				\begin{tabular}{cccccc}
					$m_{Z^\prime}$[GeV]&$\hat{\beta}_{4}^\text{ML}$&$\beta_{4}^\text{ML}$&$\mathrm{N_2^{(1)}}$&$\mathrm{D_2^{(1)}}$&$\mathrm{\tau_{2,1}^{(1)}}$\\
					\hline				
					50 & 0.864 &0.858 & 0.843 & 0.778 & 0.817 \\
					90 & 0.873 &0.866  & 0.848 & 0.837 & 0.827\\ 
					130 & 0.842 &0.838 & 0.809 & 0.812 & 0.797
				\end{tabular}
		\end{ruledtabular}}
	\end{center}
\end{table}

\section{\label{sec:concl}Conclusions}

This paper has extended the growing literature of machine-learning assisted jet substructure-based tagging in two ways.  First, we have developed a procedure to automatically identify the optimal product observable, using the $N$-subjettiness features as an example.  This is an important innovation because observables with relatively simple analytic forms are robust complements to complex neural network classifiers and prior to this work, there was no efficient way to identify the best coefficients in the product.  Second, we have used this automated framework to identify the optimal product observables for searching for boosted resonances like the $Z$ boson, but with beyond the standard model masses.  Jet substructure has proven to be a powerful toolset for such searches, but until now, there has been few studies of the mass dependence of the optimal observables.

Future extensions of the methods introduced in this paper may be able to simplify the regression procedure, as well as study the connections between different classifiers with similar performance (including the ones connected by monotonic functions).  The power of the method may also be extended by considering other parametric forms besides products.  Classification problems demanding a higher $M$-body phase space are a natural extension of the examples presented here.

As machine learning techniques are used more widely to guide the optimal selection of classifiers, there will be a growing need to simplify and interpret the guidance from the machines.  We have prepared an automated approach to construct optimal observables with simple, analytic forms, which can be used for further theoretical and experimental studies.  This technique will form the basis of multiple extensions in the future to improve classification performance and increase the robustness of searches and measurement at the LHC and beyond.

\section*{Acknolwedgements}

This work was supported by the U.S.~Department of Energy, Office of Science under contract DE-AC02-05CH11231.  We would like to thank  Gregor Kasieczka, Patrick Komiske, Eric Metodiev, and Jesse Thaler for detailed comments on the analysis and manuscript. 





\bibliography{novelobs}

\clearpage

\appendix
\section{Crosscheck for performance of the Regression networks}
\label{sec:crosscheck}
Here we briefly demonstrate that the regression DNNs do actually learn to approximate the mapping from the input parameters of the product observables to their densities, i.e., a mapping from $\R^8\to\R^{25,000}$. We specifically choose the ungroomed 90 GeV case, and choosing values of $\{a,...,h\}$ for the optimal observable, as listed in Table \ref{tab:ungroomed_Z_ML}. 

\begin{figure*}[htb]
	\centering
	\begin{minipage}{\textwidth}
		\centering
		\subfloat[]
		{
			\includegraphics[width=0.25\textwidth]{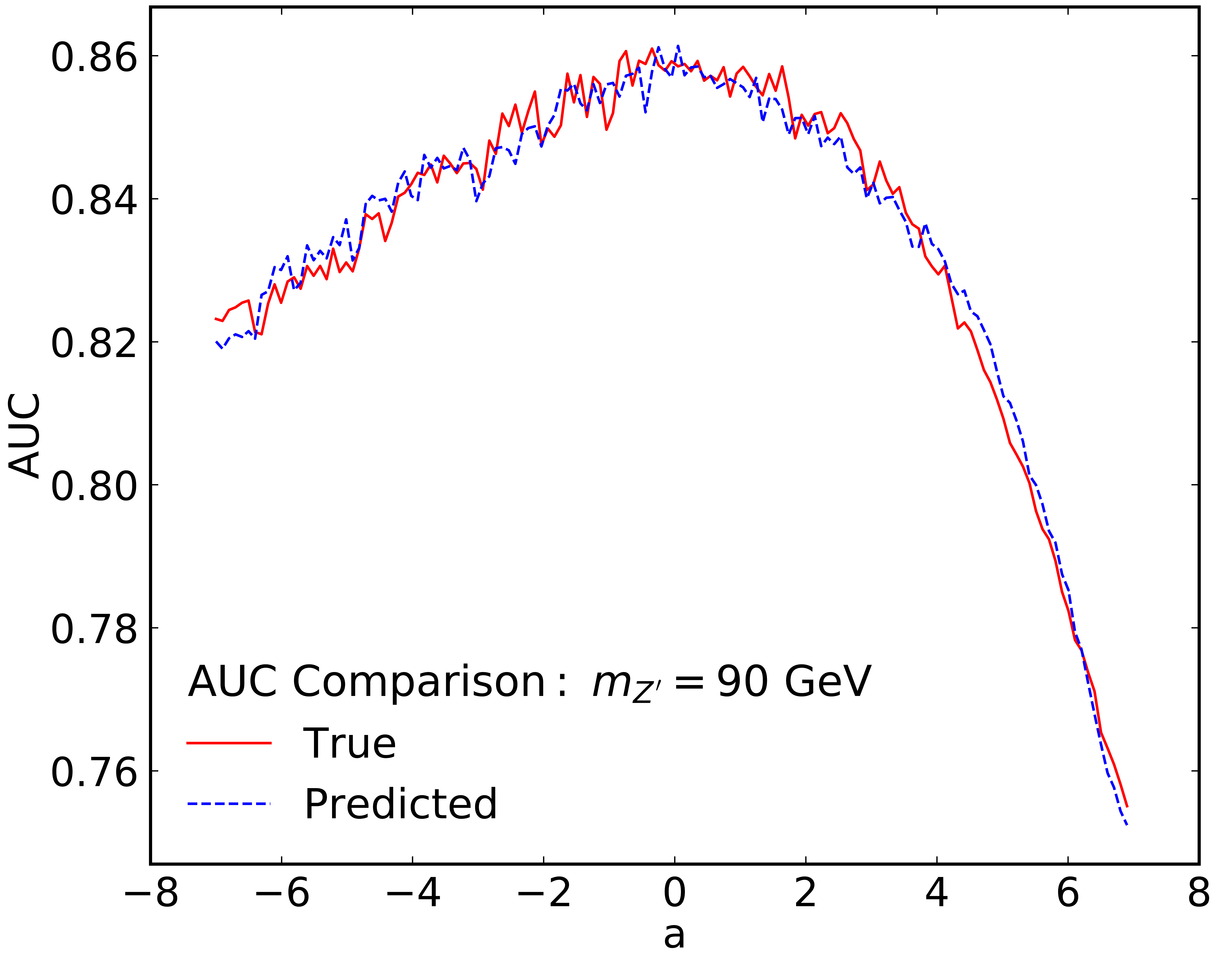}	
		}
		\subfloat[]
		{
			\includegraphics[width=0.25\textwidth]{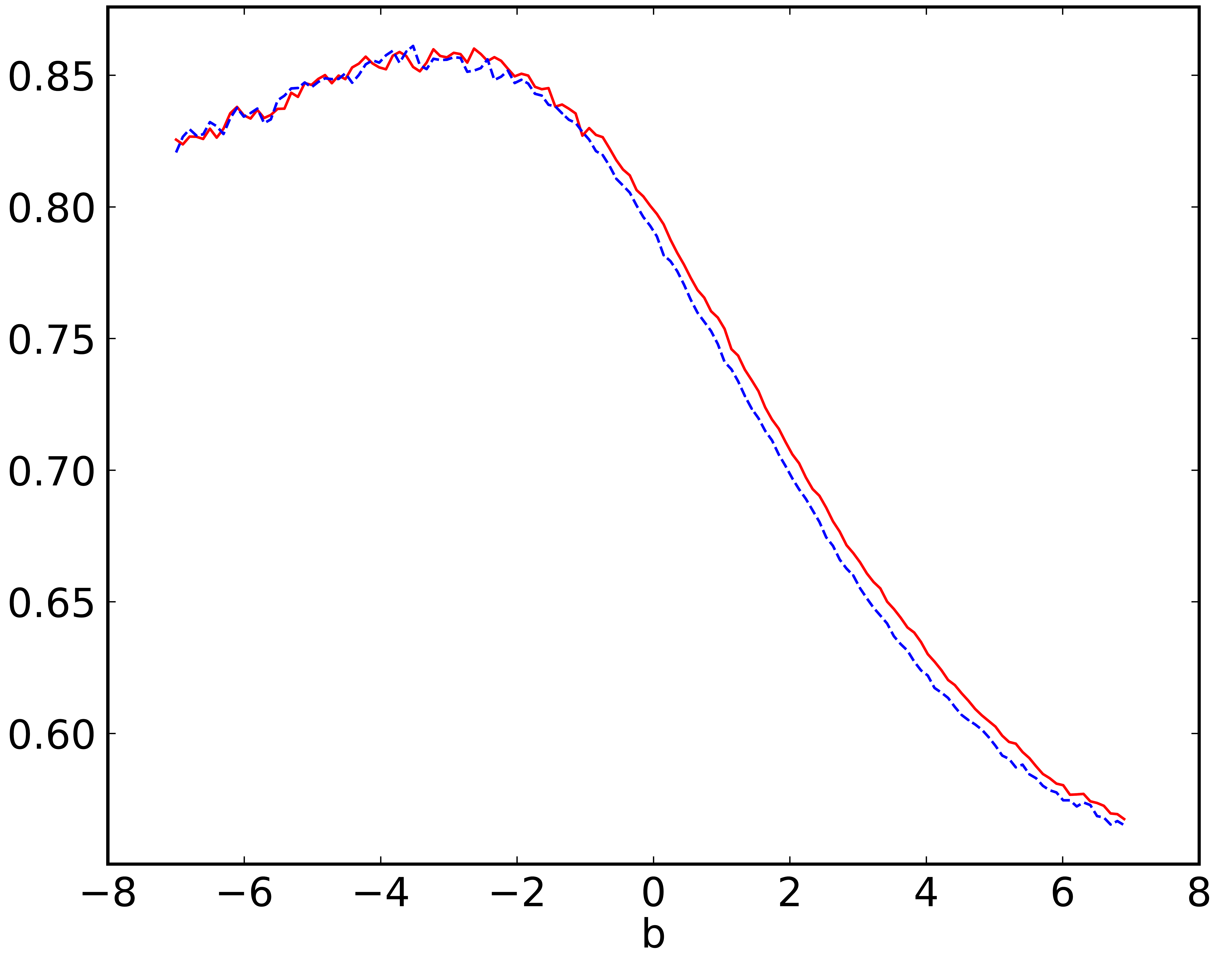}
		}
		\subfloat[]
		{
			\includegraphics[width=0.25\textwidth]{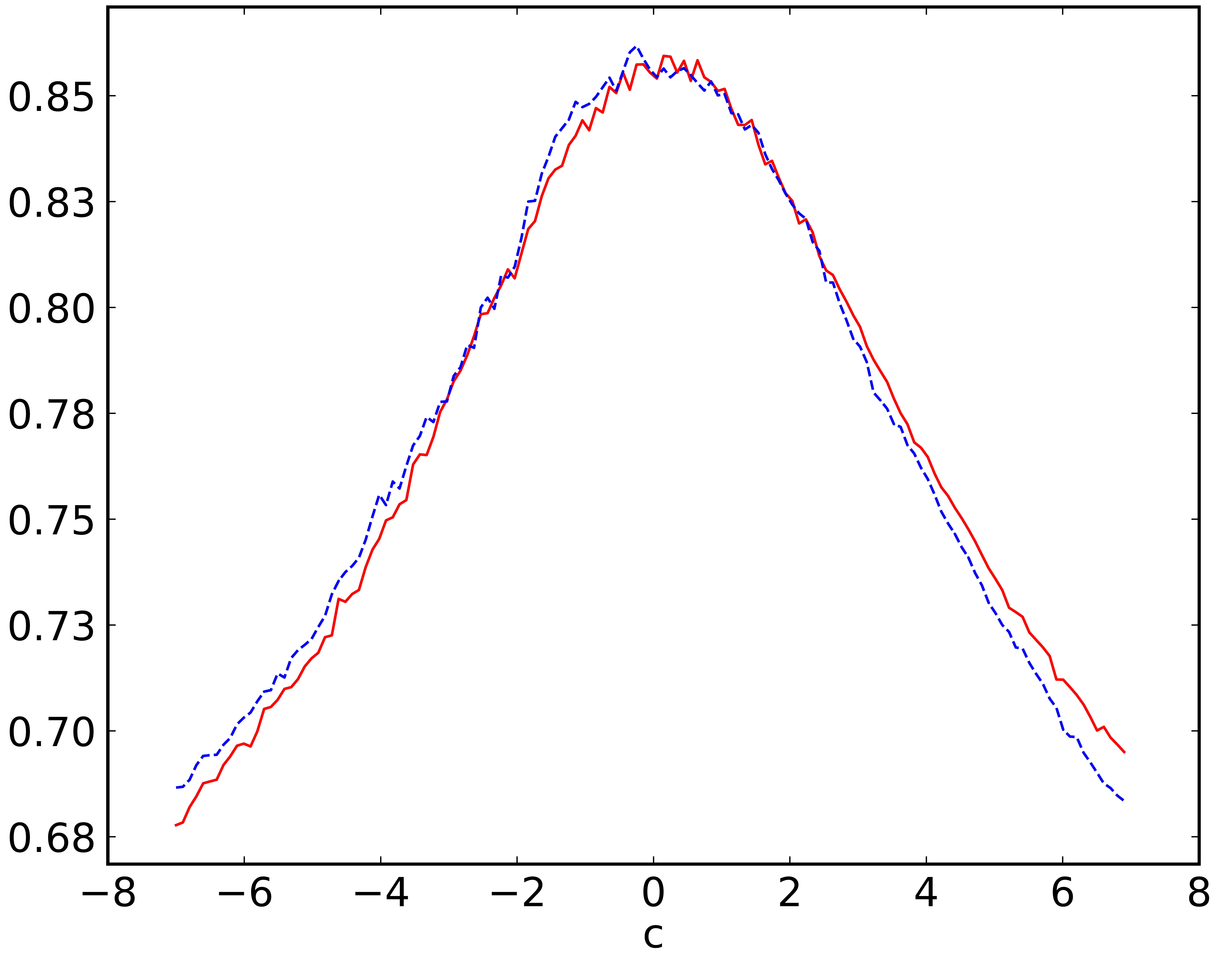}
		}
		\subfloat[]
		{
			\includegraphics[width=0.25\textwidth]{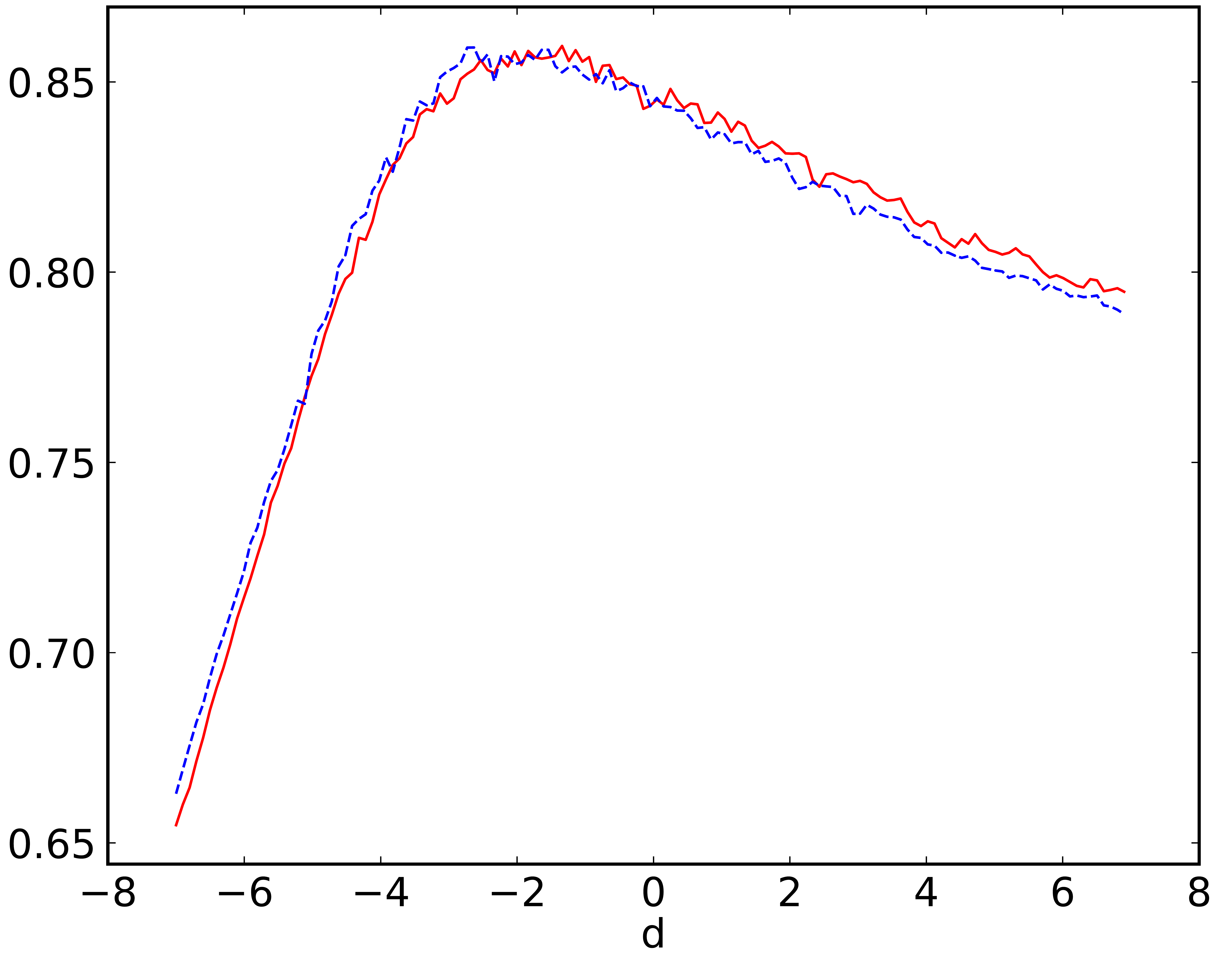}
		}\\
		\subfloat[]
		{
			\includegraphics[width=0.25\textwidth]{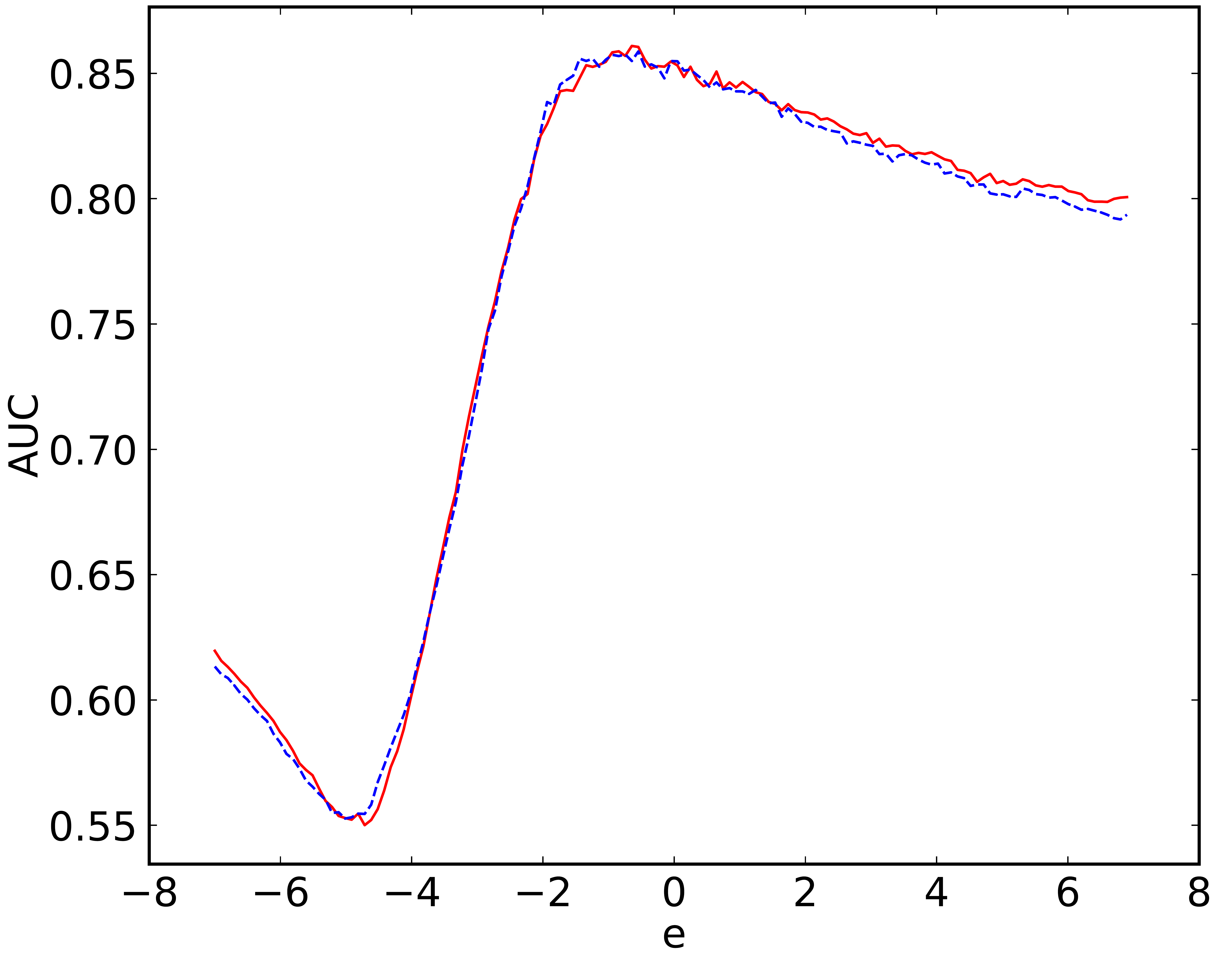}
		}
		\subfloat[]
		{
			\includegraphics[width=0.25\textwidth]{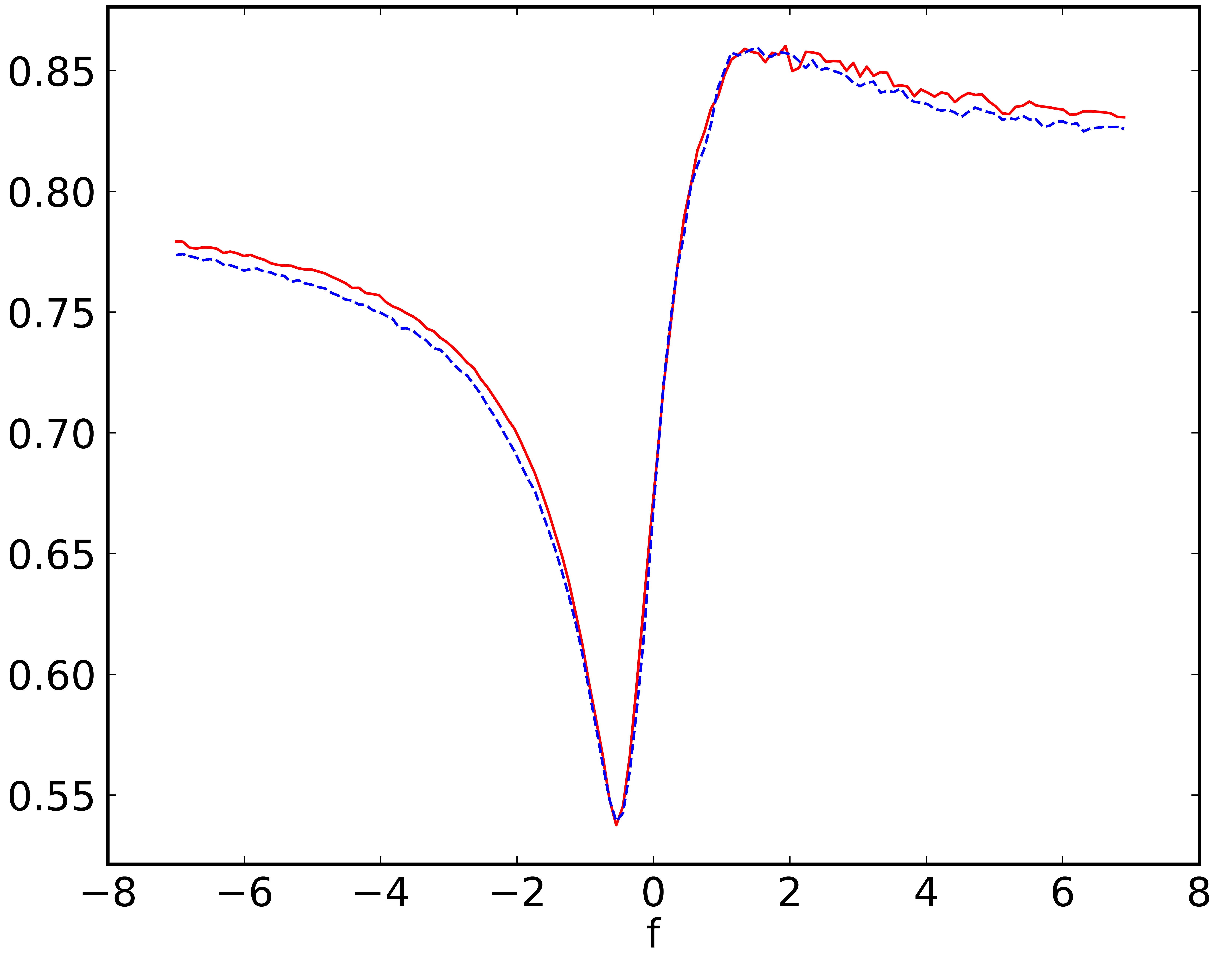}
		}
		\subfloat[]
		{
			\includegraphics[width=0.25\textwidth]{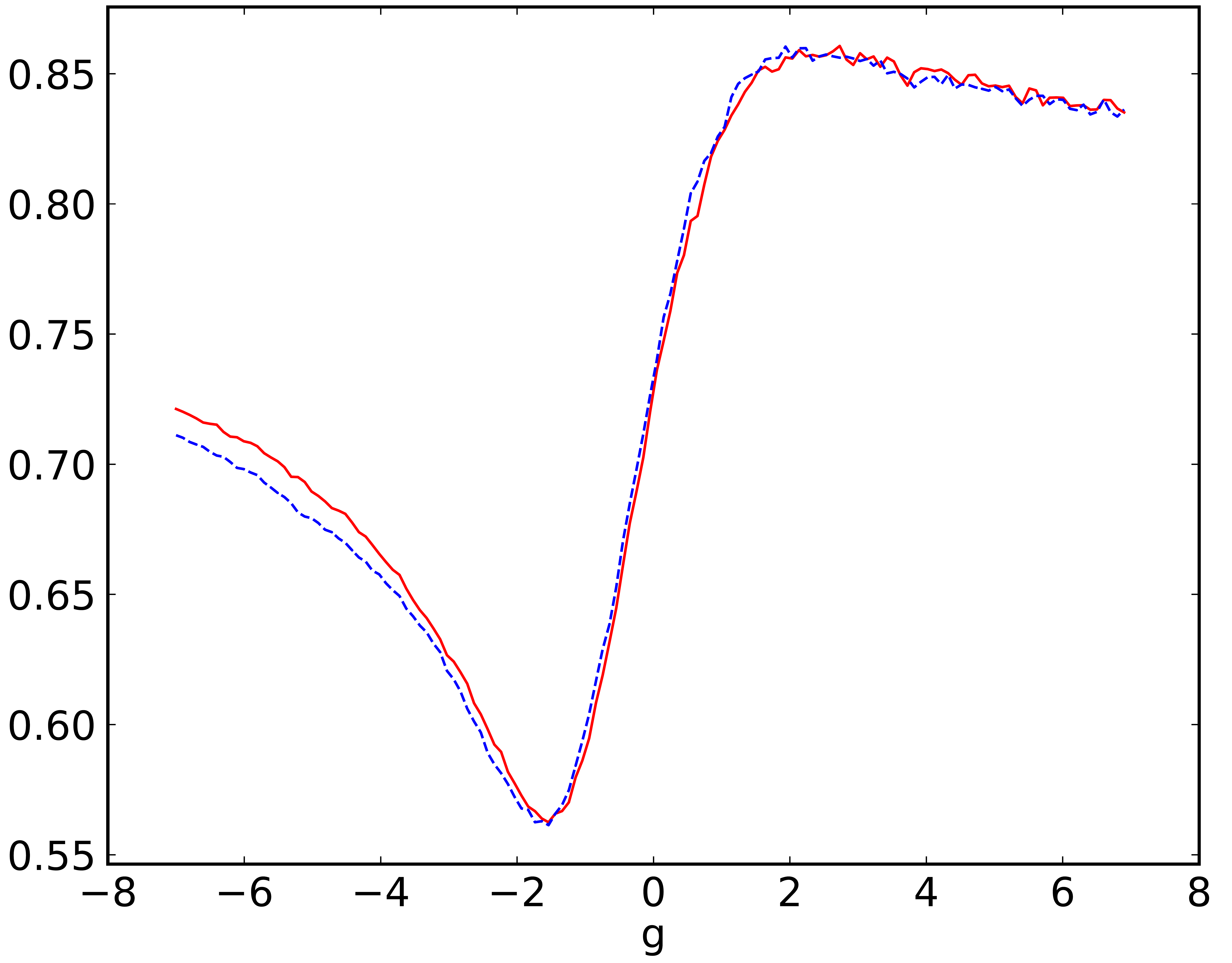}	
		}
		\subfloat[]
		{
			\includegraphics[width=0.25\textwidth]{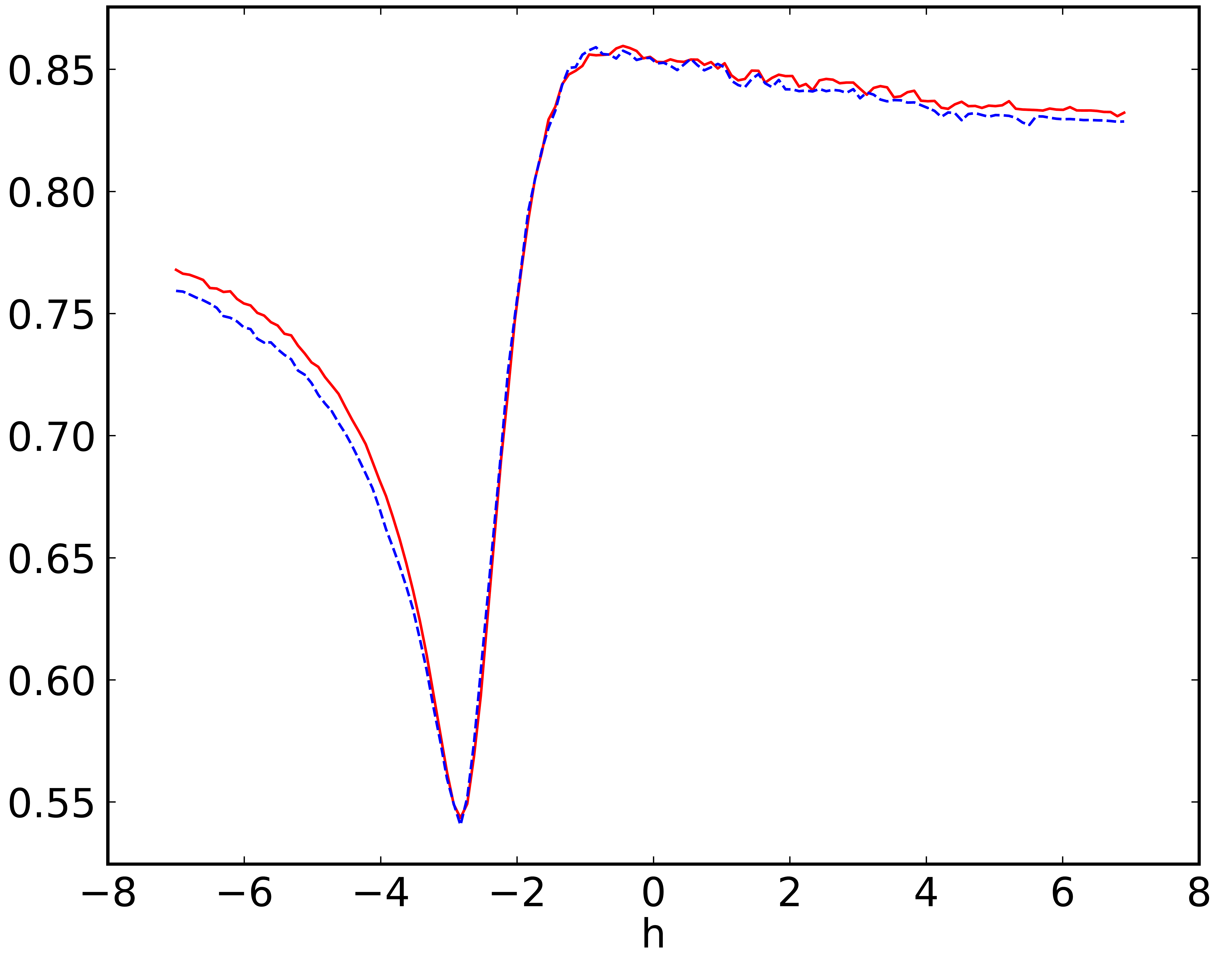}	
		}
		
		\caption{Here we plot results for the calculation of the area under the ROC curve (AUC) from the distributions generated by the signal and background regression DNN's for the ungroomed $m_{Z^\prime}=90~\mathrm{GeV}$ case (blue, dashed) along with the true AUC (red, solid) computed on the same statistics. These are seen to be in good agreement with each other. The results demonstrate the usefulness of the networks to accurately reproduce the change in the signal and background PDFs, represented via the accurate reproduction of the AUCs calculated from them, as a function of the variation of each individual input parameter $\{a,...,h\}$.
			\label{fig:AUC_interpolation}}
	\end{minipage}
\end{figure*}

We then select one of the parameters and vary it between $-7$ and 7 with a step size of 0.1 while keeping the other parameters fixed. This allows us to study how the networks can be used to interpolate AUC's over a range of values around the optimum we locate and, in addition, by going beyond the training range of $[-5,5]$ we also demonstrate that the networks can be used to extrapolate the aforementioned mapping to then still calculate the AUC with a good level of accuracy. The results for this study are shown in Fig.~\ref{fig:AUC_interpolation} and indicate that the regression networks allow to accurately track the trajectories of the AUC in these one-dimensional slices of the parameter space.

\section{\label{subsec:H2bb_sd}Groomed \Hbb~vs.~\gbb~discrimination}

\begin{figure*}[t!]
	\centering
	\begin{minipage}{\textwidth}
		\centering
		\subfloat[]
		{
			\includegraphics[width=0.45\textwidth]{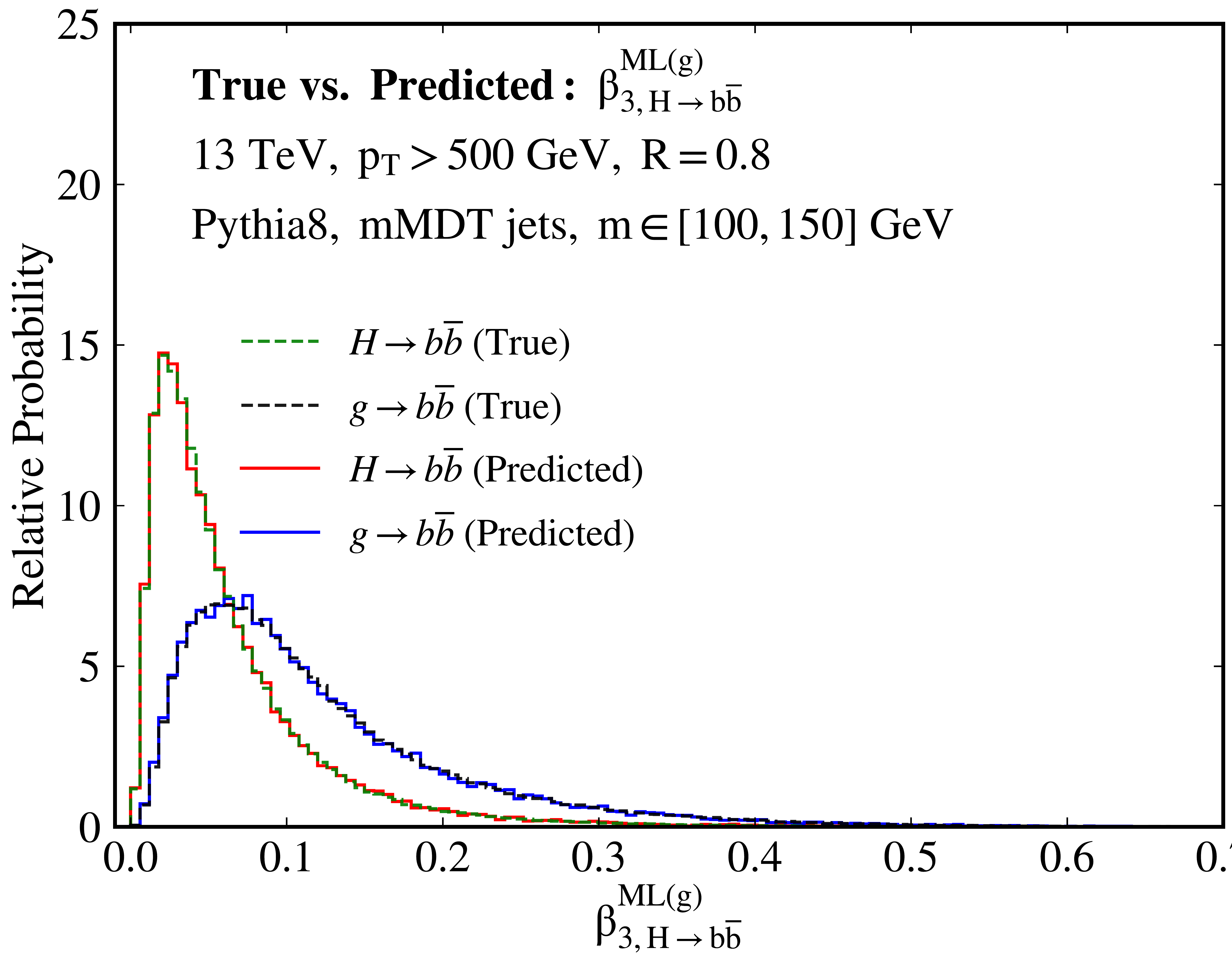}
			\label{fig:H2bb_mmdt_obsdist}
		}
		\subfloat[]
		{
			\includegraphics[width=0.45\textwidth]{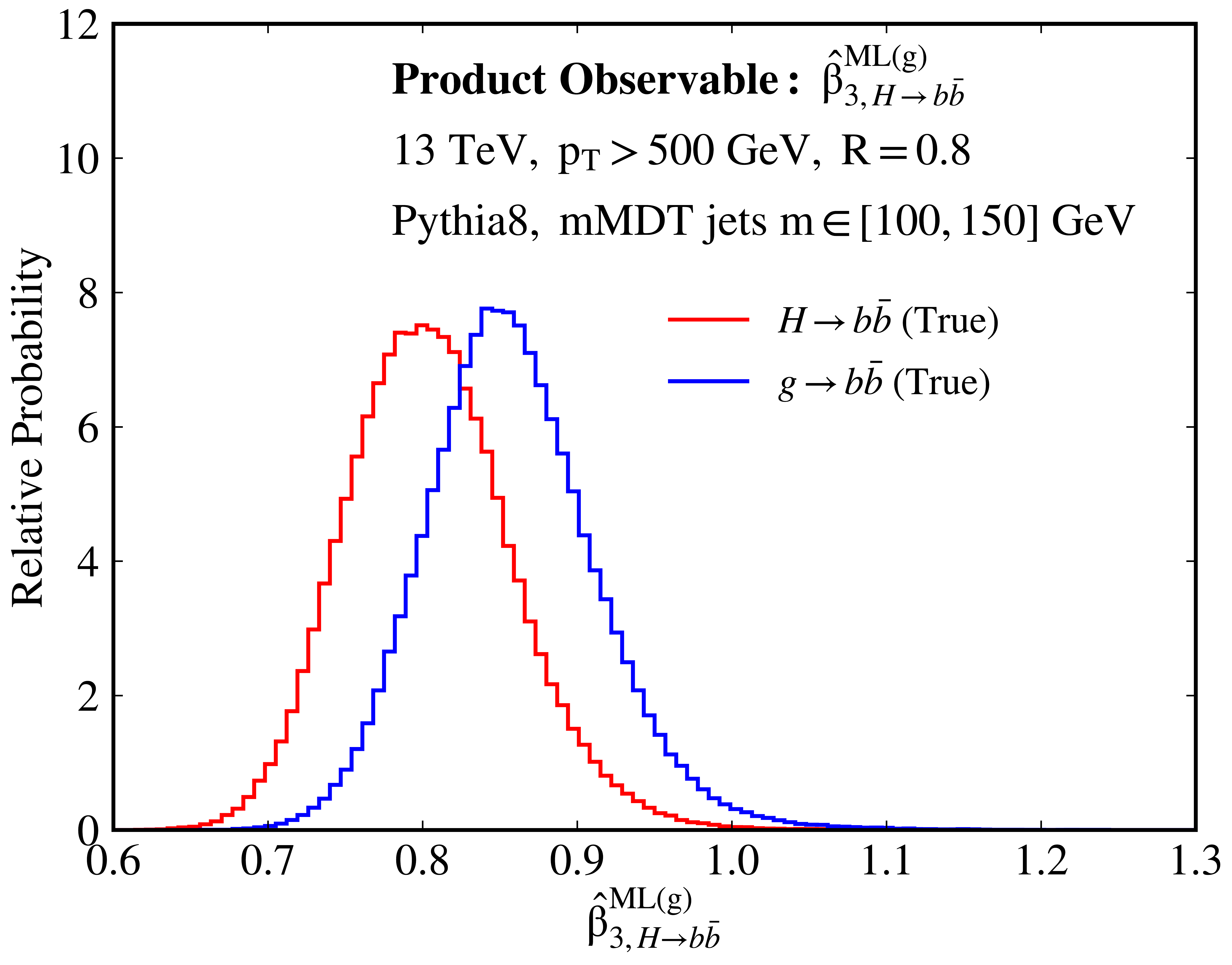}
			\label{fig:H2bb_mmdt_obsdist_MSE}
		}\\
		\subfloat[]
		{
			\includegraphics[width=0.45\textwidth]{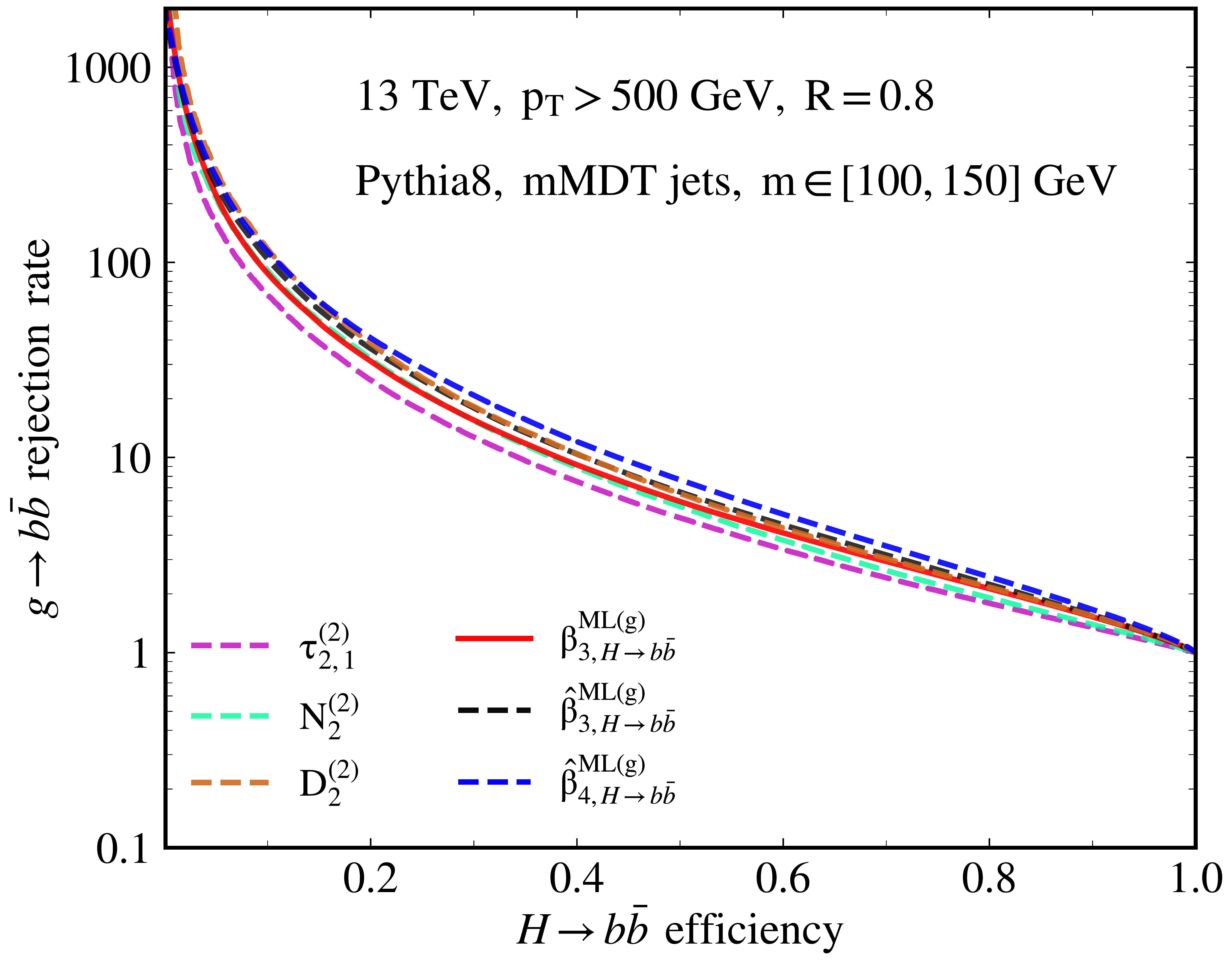}
			\label{fig:H2bb_mmdt_obscomp_ROC}
		}
		\subfloat[]
		{
			\includegraphics[width=0.45\textwidth]{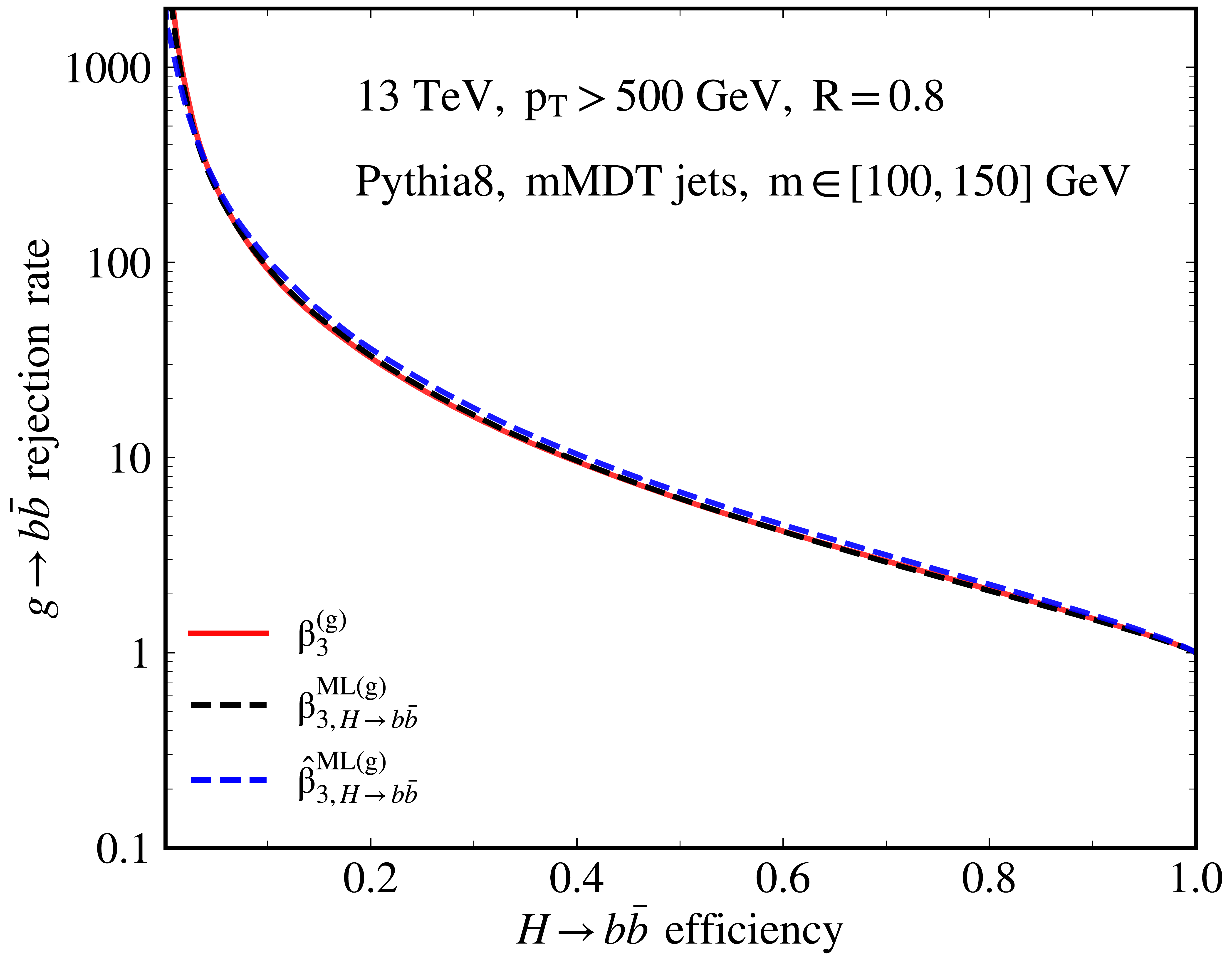}
			\label{fig:H2bb_mmdt_oldvsnew_ROC}
		}
		\caption{
			(a): Comparison of PDFs of $\beta_{3,H\rightarrow b\bar{b}}^{ML(g)}$ for mMDT groomed \Hbb~discrimination, using $\sim 250,000$ signal and background samples, and the distributions of the regression DNN prediction. The distributions are rescaled by a constant for the sake of visual comparison. (b) Probability density distributions of $\hat{\beta}_{3,H\rightarrow b\bar{b}}^{ML(g)}$ obtained via linear regression. (c): Comparison of discrimination power of $\beta_{3,H\rightarrow b\bar{b}}^{ML(g)}$, $\hat{\beta}_{3,H\rightarrow b\bar{b}}^{ML(g)}$ and $\hat{\beta}_{4,H\rightarrow b\bar{b}}^{ML(g)}$ to standard observables, where the 4-body product observable is seen to perform best for groomed \Hbb~discrimination. (d): Comparison of  $\hat{\beta}_{3,H\rightarrow b\bar{b}}^{ML(g)}$ and $\beta_{3,H\rightarrow b\bar{b}}^{ML(g)}$ to $\beta_{3}^{(g)}$ proposed in \cite{Datta:2017lxt}; we note that the latter two 3-body product observables provide essentially the same discrimination power while the 3- and 4-body ones obtained with linear regression outperforms them. }
		\label{fig:H2bb_mmdt_ROC}
	\end{minipage}
\end{figure*}

Utilizing the result that discrimination power for mMDT groomed \Hbb~vs.~\gbb~discrimination saturates at 3-body phase space \cite{Datta:2017lxt}, we use the procedure proposed in the \Sec{sec:deeplearn} to find the optimal product observable. The final values for the parameters $\{a,...,e\}$ obtained through the optimization are presented in Table~\ref{tab:groomed_Hbb}, along with those obtained in the previous study.  Interestingly, the exponents for $\beta_{3,H\rightarrow b\bar{b}}^{\mathrm{ML(g)}}$ are nearly the same for $c$, $d$, and $e$, but are quite different for $a$ and $b$.  The factors $d$ and $e$ are also similar for $\hat{\beta}_{3,H\rightarrow b\bar{b}}^{\mathrm{ML(g)}}$ up to a multiplicative factor.
\begin{table}[h]
	\begin{center}
		\caption{\label{tab:groomed_Hbb}Summary of parameters for the product observables for groomed \Hbb~discrimination as proposed in Ref.~\cite{Datta:2017lxt} and as constructed via the procedure presented in this work (Fig.~\ref{fig:H2bb_mmdt_obsdist}).}
		\resizebox{\columnwidth}{!}{
			\begin{ruledtabular}
				\begin{tabular}{ccccccc}
					Observable&$a$&$b$&$c$&$d$&$e$&AUC\\
					
					\hline\vspace{0.1cm}
					$\beta_{3}^{\mathrm{(g)}}$   & -2.0 & 0.0 & 0.0 & -2.0 &  2.0 & 0.745\\
					
					$\beta_{3,H\rightarrow b\bar{b}}^{\mathrm{ML(g)}}$ & 0.67 & -1.65 & 0.01 & -1.90 &  2.07 & 0.744\\ 

					$\hat{\beta}_{3,H\rightarrow b\bar{b}}^{\mathrm{ML(g)}}$ & -1.54 & 1.01 & -0.17 & -0.15 &  0.16 & 0.758\\ 
					
				\end{tabular}
		\end{ruledtabular}}
	\end{center}
\end{table}

In Fig.~\ref{fig:H2bb_mmdt_obsdist}, we plot the distributions of the new observable computed for signal and background, along with the prediction from the neural network. We note that the network provides a good match to the true distribution, where the latter is also calculated on 10 times more jets. We then compare the ROC curves for the new observable to $D_2^{(2)}$~\cite{Larkoski:2014gra}, $N_2^{(2)}$~\cite{Moult:2016cvt} observables, and $\tau_{21}^{(2)}$ in Fig.~\ref{fig:H2bb_mmdt_obscomp_ROC}.

In addition, we compare the new observable to $\beta_3^{(g)}$ in Fig.~\ref{fig:H2bb_mmdt_oldvsnew_ROC} to demonstrate that both observables have essentially the same discrimination power as expected. Then, this allows us to proceed to applying the procedure on higher dimensional problems. Further, we plot the ROC curve for the 4-body product observable from the linear regression method, noting that it provides the best performance of the observables that have been explored for this problem.~\footnote{Explicitly, the optimal parameter values for  $\hat{\beta}_{4,H\rightarrow b\bar{b}}^{ML(g)}$ are as  follows:
$\{a,...,h\} = \{-2.09, 1.46, -0.31, -0.49, 0.35, 0.03, -0.18, 0.23\}$, and it leads to an AUC of 0.778 in Fig.~\ref{fig:H2bb_mmdt_obscomp_ROC}.}

\section{\label{subsec:Groomed_ZvQCD}Groomed \Zp~vs. $\mathrm{QCD}$}
\begin{figure*}[h]
	\centering
	\begin{minipage}{\textwidth}
		\centering
		\subfloat[$m_{Z'}=50~\mathrm{GeV}$]
		{
			\includegraphics[width=0.33\textwidth]{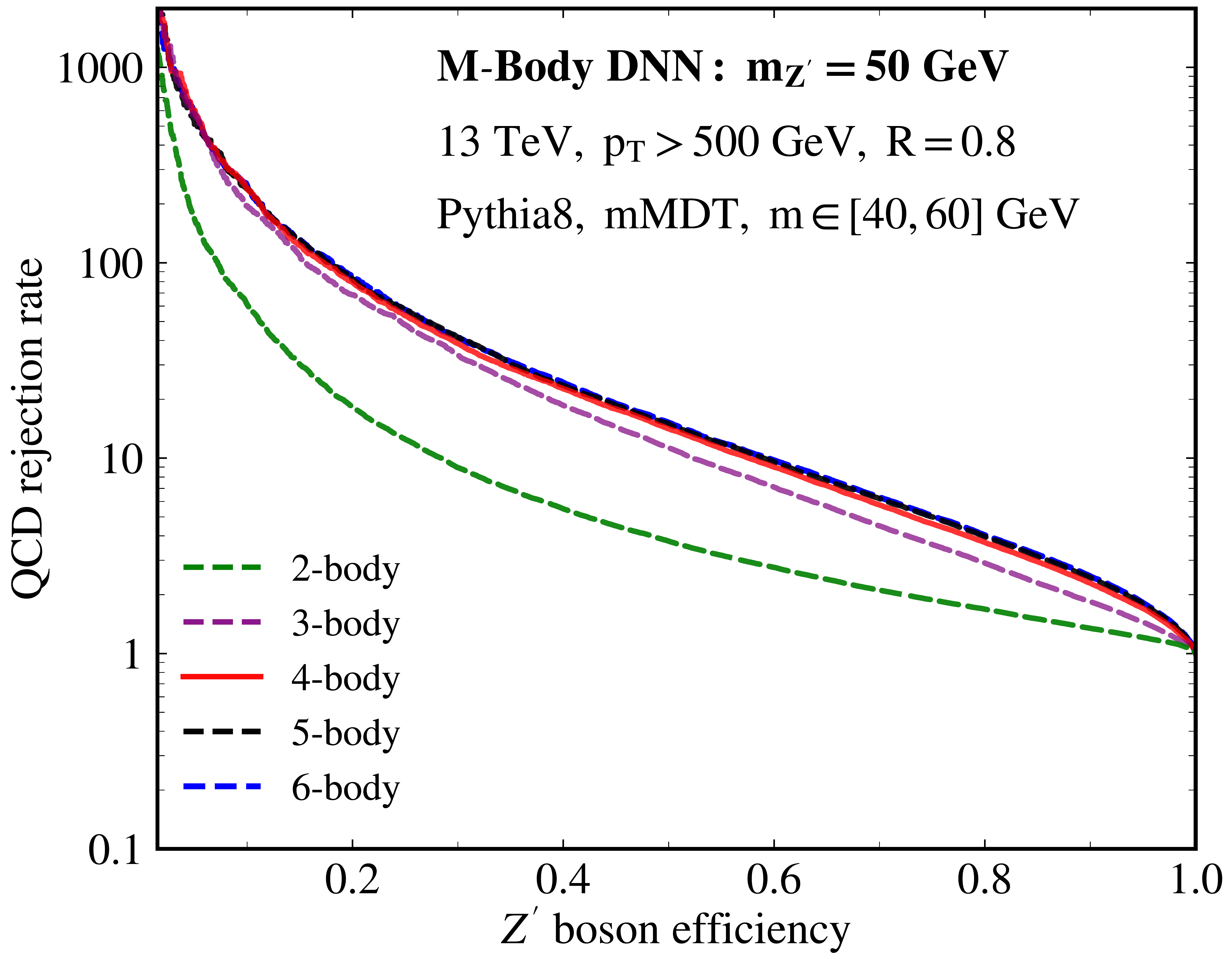}
			\label{fig:satroc50_sd}
		}
		\subfloat[$m_{Z'}=90~\mathrm{GeV}$]
		{
			\includegraphics[width=0.33\textwidth]{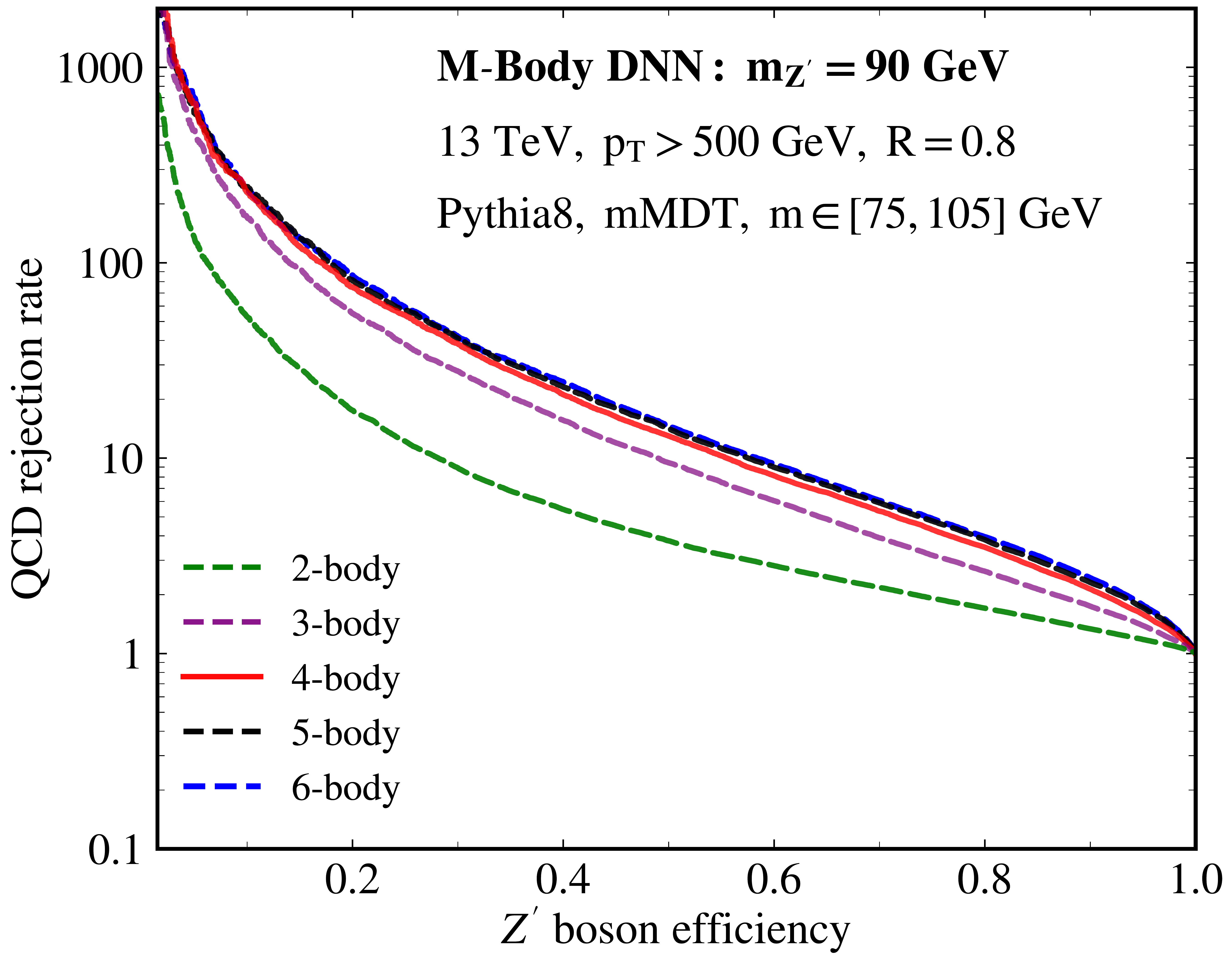}
			\label{fig:satroc90_sd}
		}
		\subfloat[$m_{Z'}=130~\mathrm{GeV}$]
		{
			\includegraphics[width=0.33\textwidth]{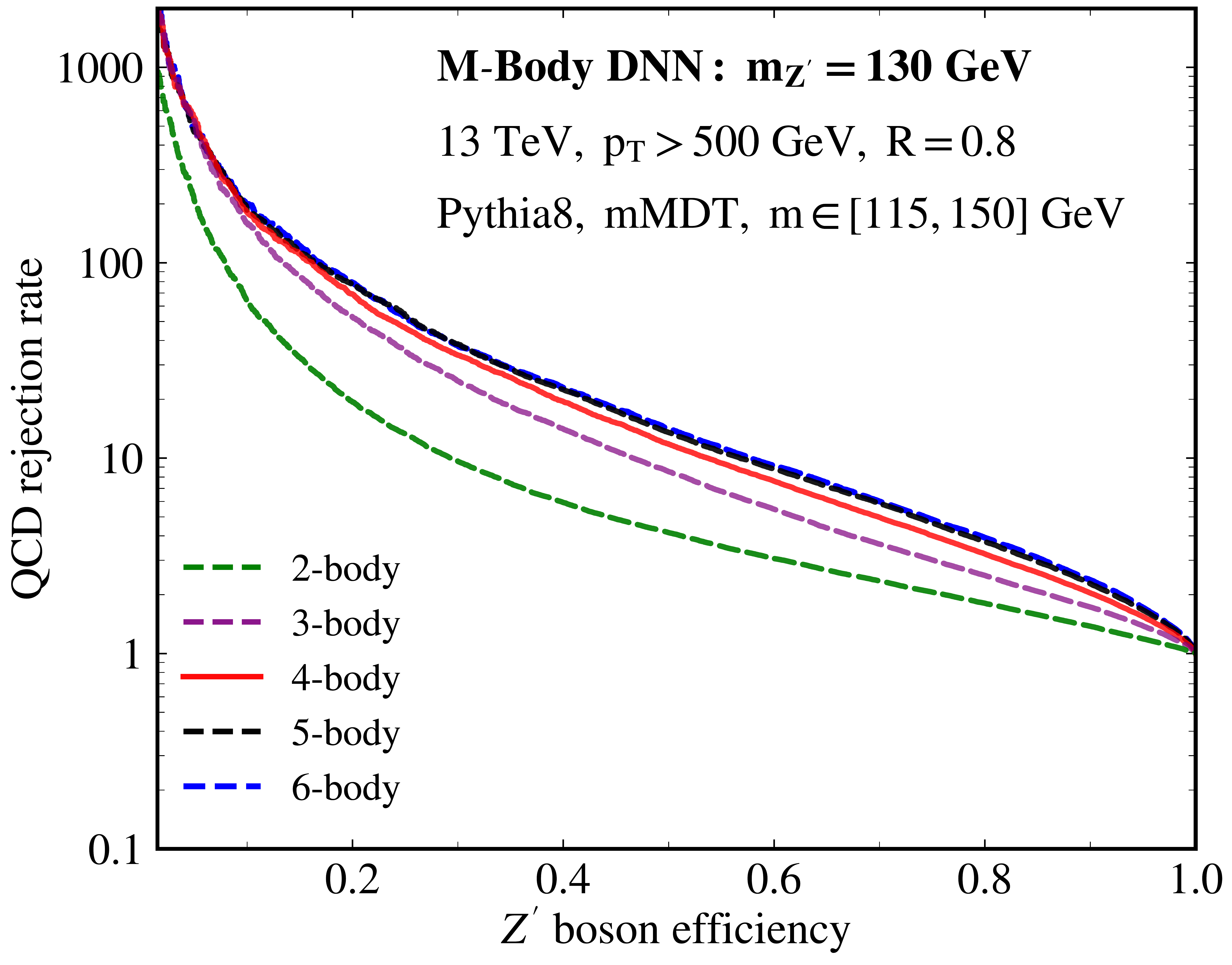}
			\label{fig:satroc130_sd}
		}
		\caption{$M$-body discrimination results of mMDT groomed  $Z^\prime$ vs. QCD jets. Here, discrimination power is again seen to effectively saturate at 4-body phase space for all considered values of $m_{Z^\prime}$.
			\label{fig:groomed_mbody_saturation}}
	\end{minipage}
\end{figure*}

In this section we carry out the same set of studies for mMDT groomed \Zp~ discrimination as for the ungroomed cases from Sec.~\ref{subsec:Ungroomed_ZvQCD}.  As in the ungroomed case, Fig.~\ref{fig:groomed_mbody_saturation} indicates that the saturation of discrimination power occurs at 4-body phase space. 

\begin{table}[h!]
	\begin{center}
		\caption{\label{tab:groomed_Z_ML}Summary of parameters for $\beta_{4}^\text{ML(g)}$ for mMDT groomed \Zp~vs. QCD discrimination at 3 mass points}
		\resizebox{\columnwidth}{!}{
			\begin{ruledtabular}
				\begin{tabular}{ccccccccc}
					$m_{Z^\prime}$[GeV]&$a$&$b$&$c$&$d$&$e$&$f$&$g$&$h$\\
					\hline	\vspace{0.05cm}	
					50 & 2.6 & -0.41 & -2.94 & -2.79 & 0.20 & 0.93 & -0.66 & 2.43 \\\vspace{0.05cm}	
					90 & 2.3 & -1.35 & -2.05 & -1.64 & -0.81 & 0.89 & 2.03 & -0.44\\
					130 & 0.80 & -1.74 & -0.28 & -1.01 & -0.38 & 0.56 & 0.82  &  0.69 \\ 
				\end{tabular}
		\end{ruledtabular}}
	\end{center}
\end{table}

\begin{table}[h!]
	\begin{center}
		\caption{\label{tab:groomed_Z_MSE}Summary of parameters for $\hat{\beta}_{4}^\text{ML(g)}$ for mMDT groomed \Zp~vs. QCD discrimination at 3 mass points}
		\resizebox{\columnwidth}{!}{
			\begin{ruledtabular}
				\begin{tabular}{ccccccccc}
					$m_{Z^\prime}$[GeV]&$a$&$b$&$c$&$d$&$e$&$f$&$g$&$h$\\
					\hline	\vspace{0.05cm}	
					
					50 & -0.35 & 0.35 & 0.56 & 1.05 & -0.17 & -0.24 & -0.34 & 0.51 \\\vspace{0.05cm}	
					90 & 0.26 & -0.41 & -0.39 & -0.68 & -0.15 & 0.11 & 0.25 & 0.42\\
					130 & 1.28 & 0.54 & 0.35 & 1.09 & 0.09 & -0.38 & -1.06  & -0.48 \\ 
				\end{tabular}
		\end{ruledtabular}}
	\end{center}
\end{table}

\begin{table}[h!]
	\begin{center}
		\caption{\label{tab:groomed_Z_AUC} Area under the ROC curve (AUC), from Fig.~\ref{fig:ROC_comp_groomed}, of standard observables and the $\beta_{4}^{\mathrm{ML(g)}}$ observables, optimized for the corresponding signal, for mMDT groomed \Zp~ vs. QCD discrimination at 3 $m_{Z^\prime}$ points. The ROC curves are calculated using the full datasets, with $\sim$300,000 events passing the mass cut for each value of $m_{Z^\prime}$.}
		\resizebox{\columnwidth}{!}{
			\begin{ruledtabular}
				\begin{tabular}{cccccc}
					$m_{Z^\prime}$[GeV]&$\hat{\beta}_{4}^\text{ML(g)}$&$\beta_{4}^\text{ML(g)}$&$\mathrm{N_2^{(2)}}$&$\mathrm{D_2^{(2)}}$&$\mathrm{\tau_{2,1}^{(2)}}$\\
					\hline				
					50 & 0.830 &0.826 & 0.796 & 0.803 & 0.780 \\
					90 & 0.822 &0.821  & 0.780 & 0.796 & 0.763\\ 
					130 & 0.814 &0.811 & 0.769 & 0.791 & 0.751
				\end{tabular}
		\end{ruledtabular}}
	\end{center}
\end{table}

\begin{figure*}[h!]
	\centering
	\begin{minipage}{\textwidth}
		\centering
		
		\subfloat[$m_{Z'}=50~\mathrm{GeV}$]
		{
			\includegraphics[width=0.29\textwidth]{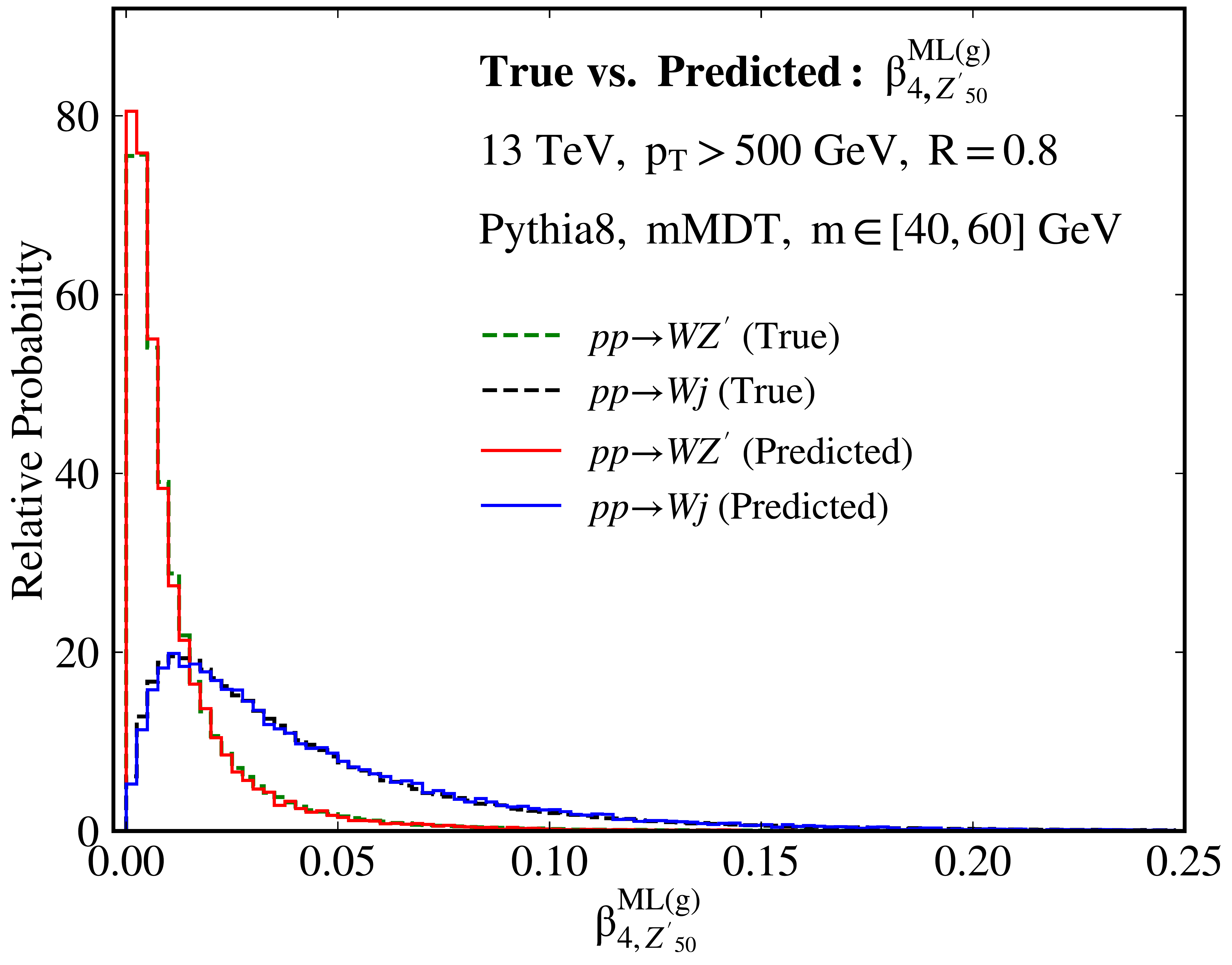}
			\label{fig:obsdist_DNN_50_sd}
		}\hspace{0.7cm}
		\subfloat[$m_{Z'}=90~\mathrm{GeV}$]
		{
			\includegraphics[width=0.29\textwidth]{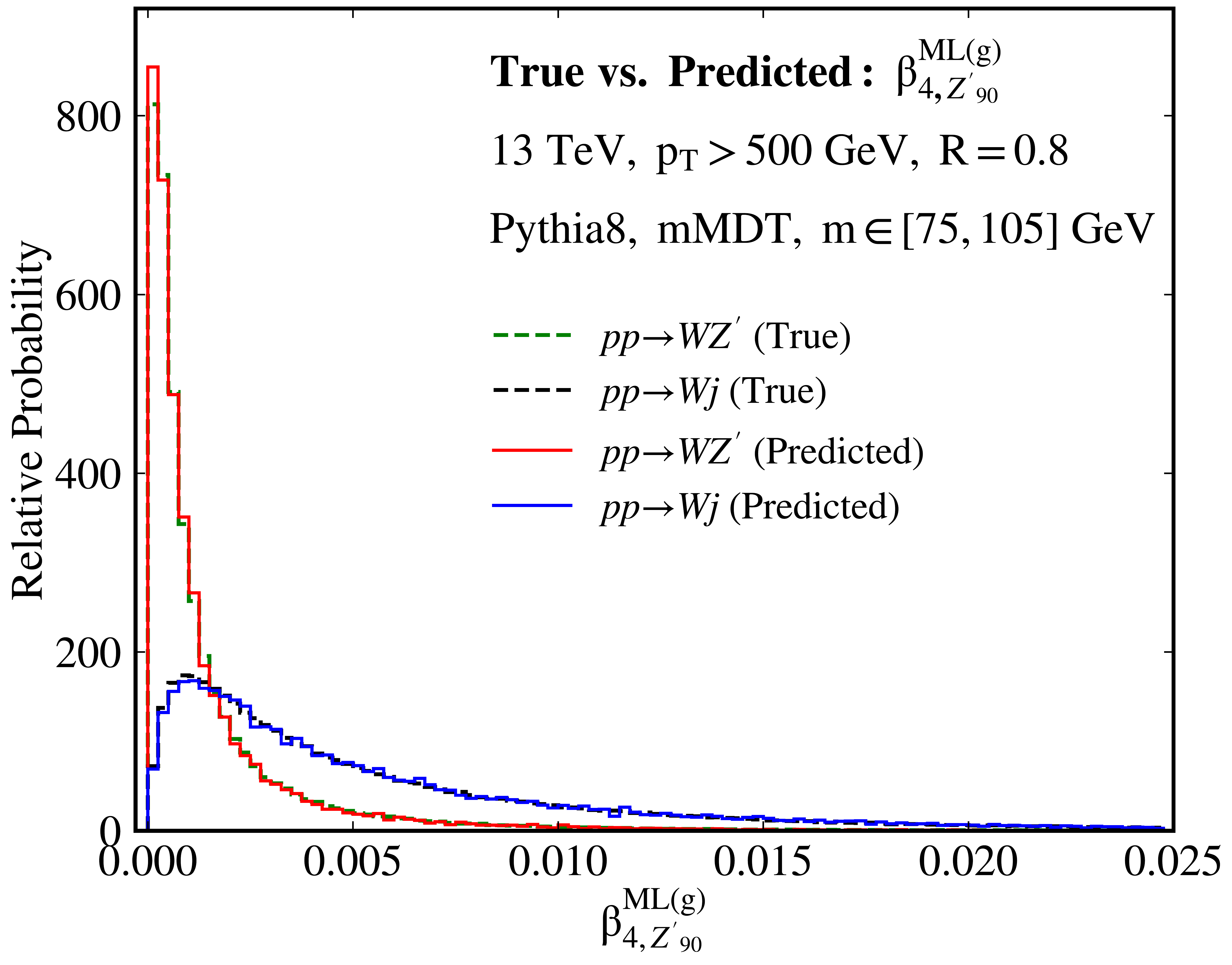}
			\label{fig:obsdist_DNN_90_sd}
		}\hspace{0.7cm}
		\subfloat[$m_{Z'}=130~\mathrm{GeV}$]
		{
			\includegraphics[width=0.29\textwidth]{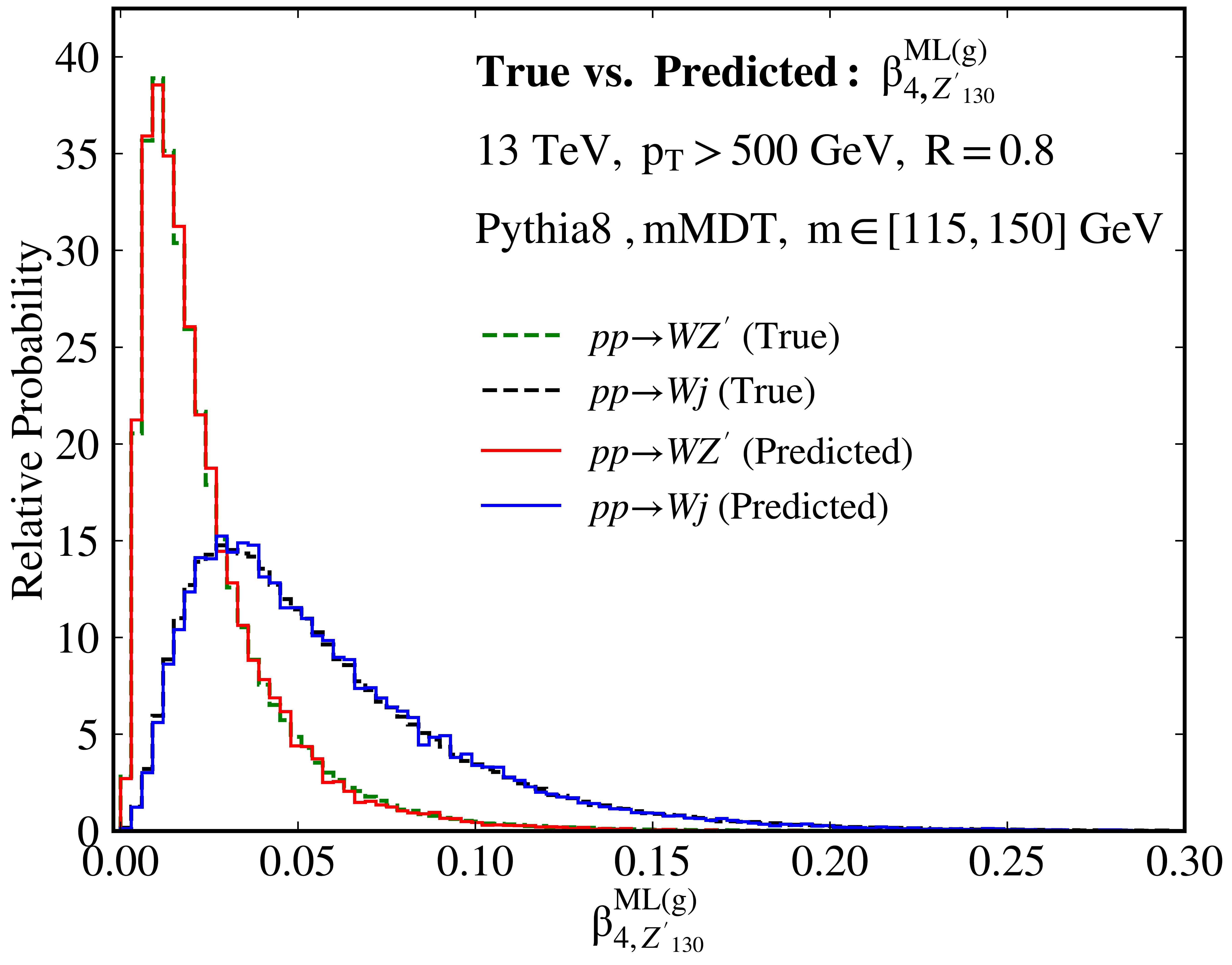}
			\label{fig:obsdist_DNN_130_sd}
		}\\
		
		\subfloat[$m_{Z'}=50~\mathrm{GeV}$]
		{
			\includegraphics[width=0.29\textwidth]{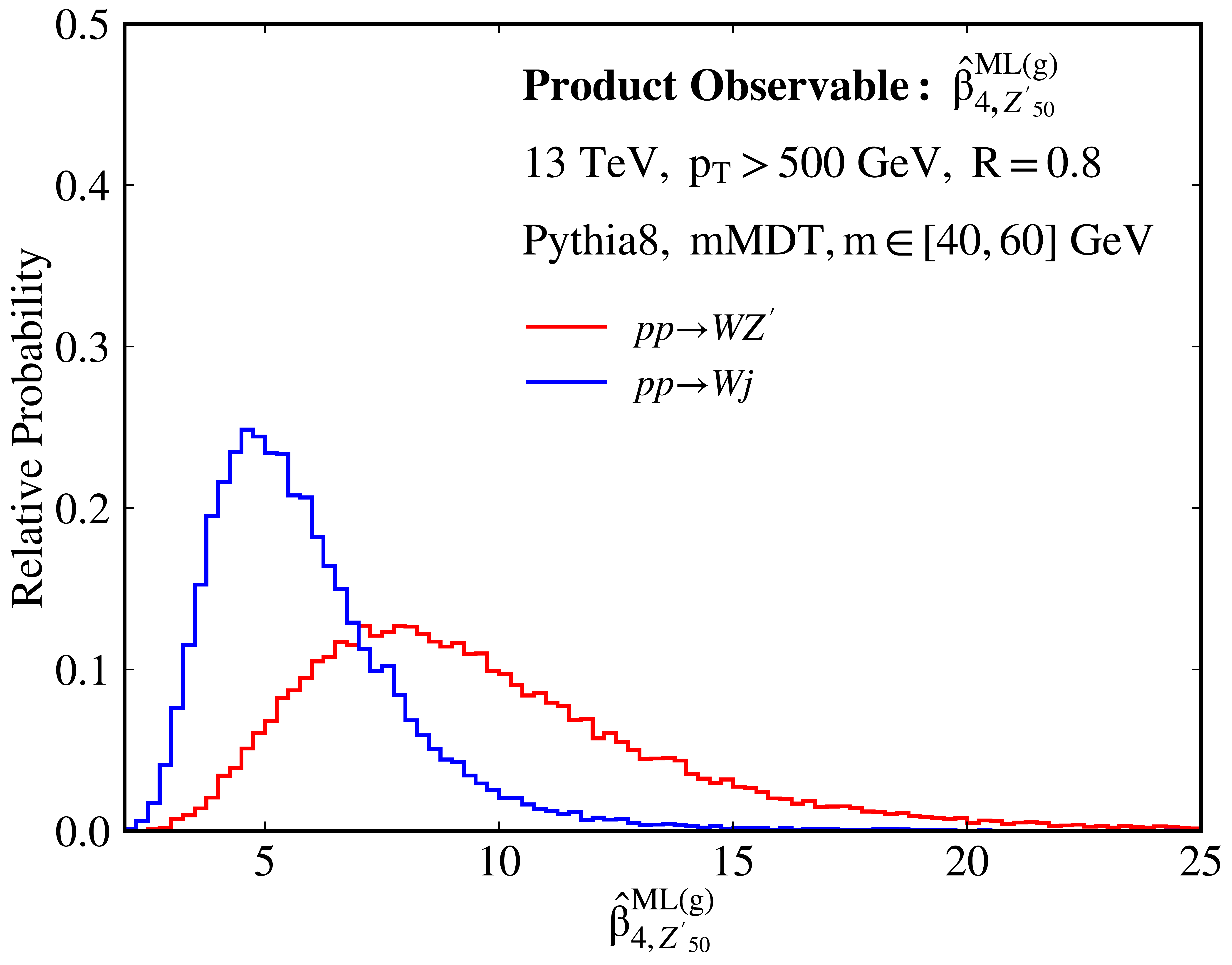}
			\label{fig:obsdist_MSE_DNN_50_sd}
		}\hspace{0.7cm}
		\subfloat[$m_{Z'}=90~\mathrm{GeV}$]
		{
			\includegraphics[width=0.29\textwidth]{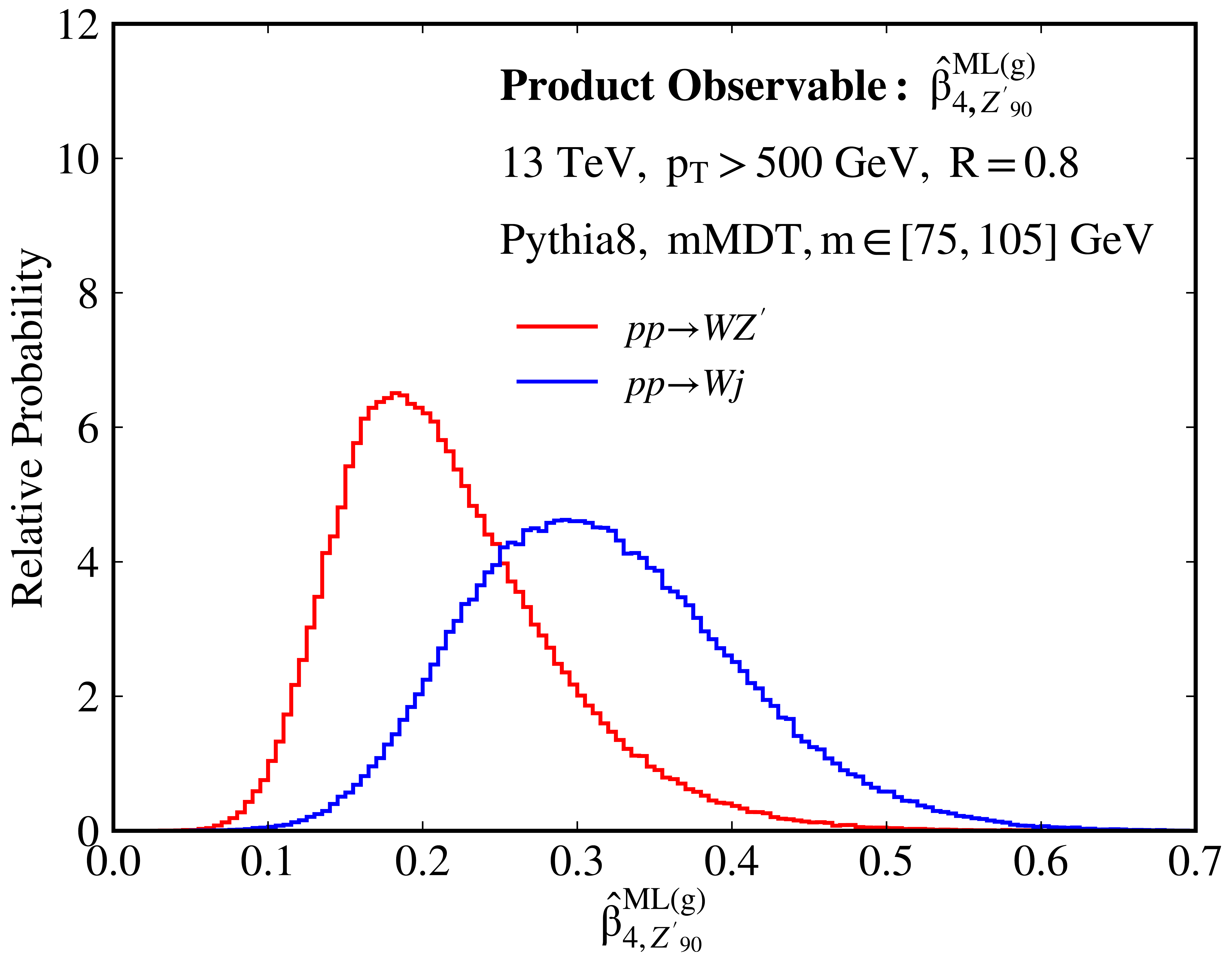}
			\label{fig:obsdist_MSE_DNN_90_sd}
		}\hspace{0.7cm}
		\subfloat[$m_{Z'}=130~\mathrm{GeV}$]
		{
			\includegraphics[width=0.29\textwidth]{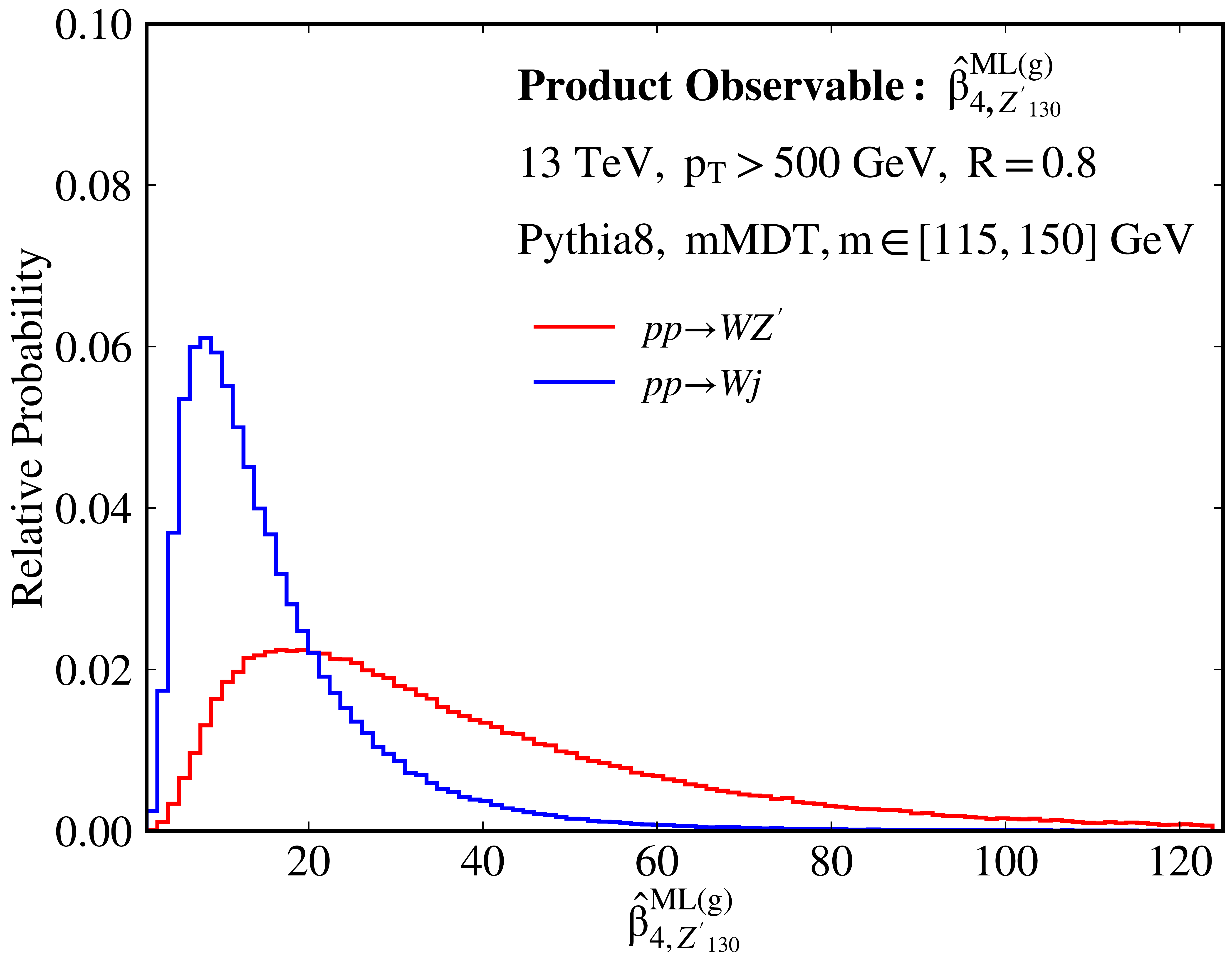}
			\label{fig:obsdist_MSE_DNN_130_sd}
		}\\
		\caption{Top panel [a-c]: Comparison of the probability density function of the new $\beta_{4}^\text{ML(g)}$ observables for mMDT groomed \Zp~discrimination, calculated for $\sim 300,000$ signal and background samples, and the distribution of the regression DNN predictions of 25,000 observable values. The  distributions are rescaled for the sake of visual comparison.  Bottom panel [d-f]: Distributions of the $\hat{\beta}_{4}^\text{ML(g)}$ observables for ungroomed \Zp~discrimination that were obtained via linear regression.} 
		\label{fig:distributions_groomed}
	\end{minipage}
	
	\begin{minipage}{\textwidth}
		\centering
		\subfloat[$m_{Z'}=50~\mathrm{GeV}$]
		{
			\includegraphics[width=0.33\textwidth]{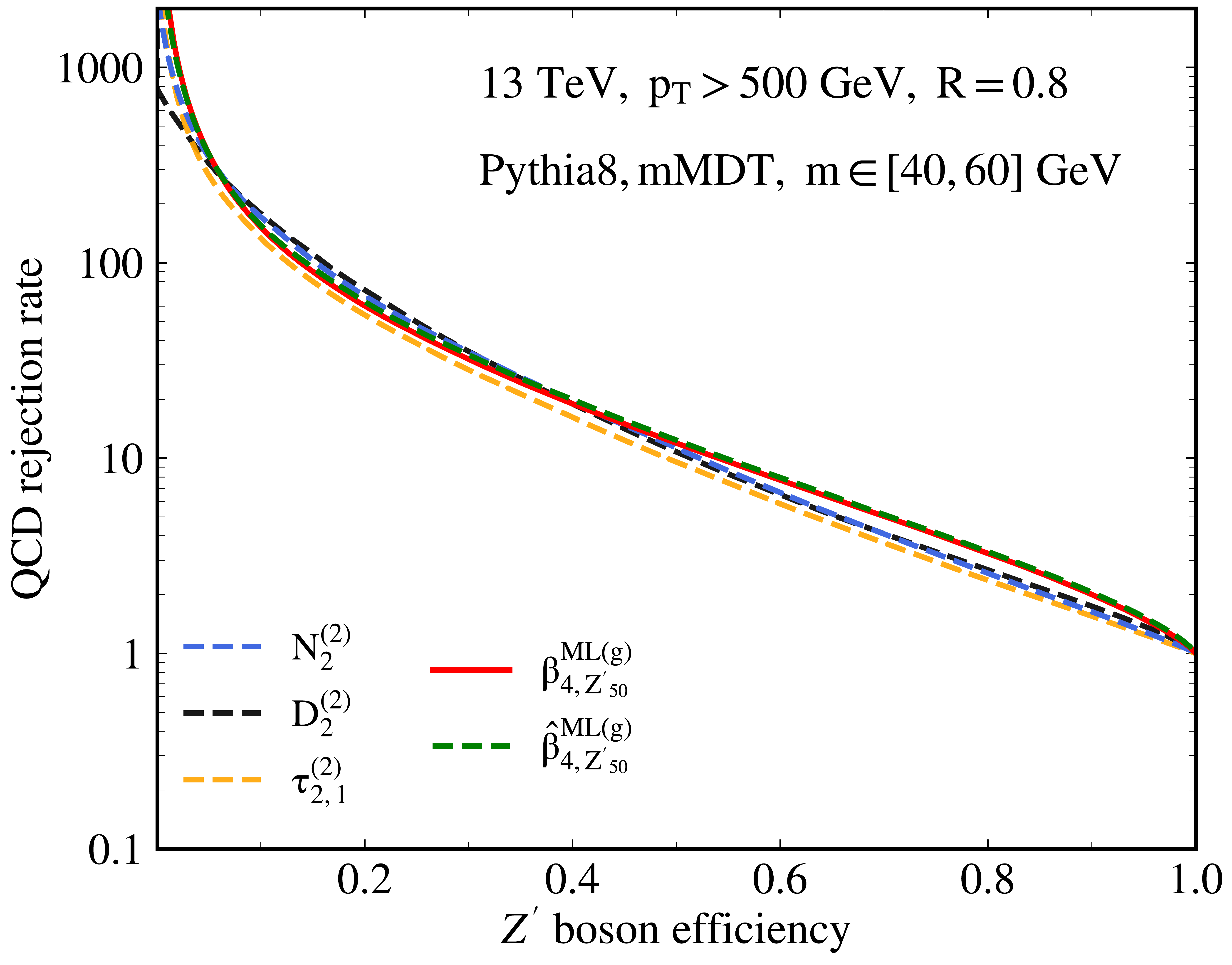}
			\label{fig:obscomp50_sd}
		}
		\subfloat[$m_{Z'}=90~\mathrm{GeV}$]
		{
			\includegraphics[width=0.33\textwidth]{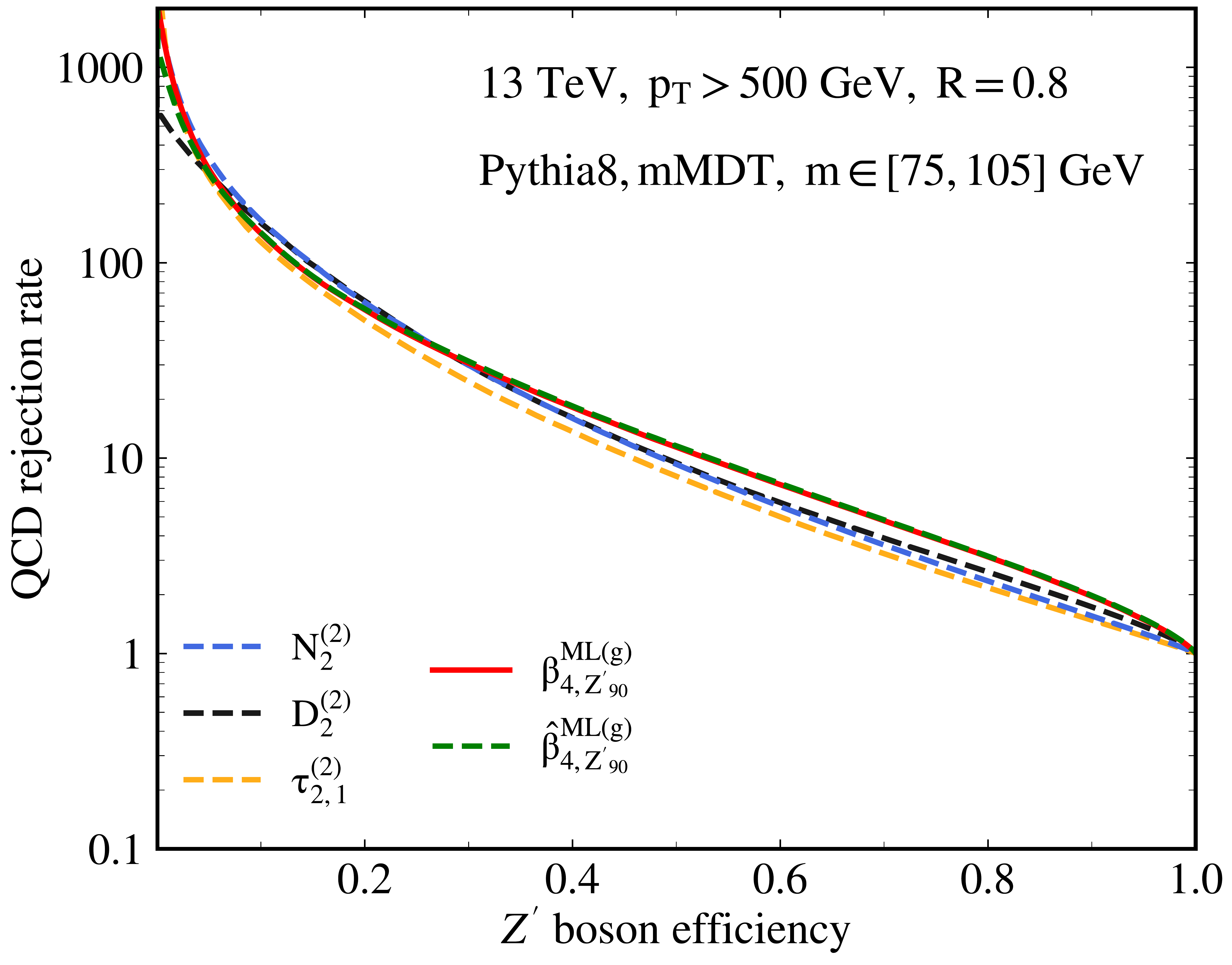}
			\label{fig:obscomp90_sd}
		}		
		\subfloat[$m_{Z'}=130~\mathrm{GeV}$]
		{
			\includegraphics[width=0.33\textwidth]{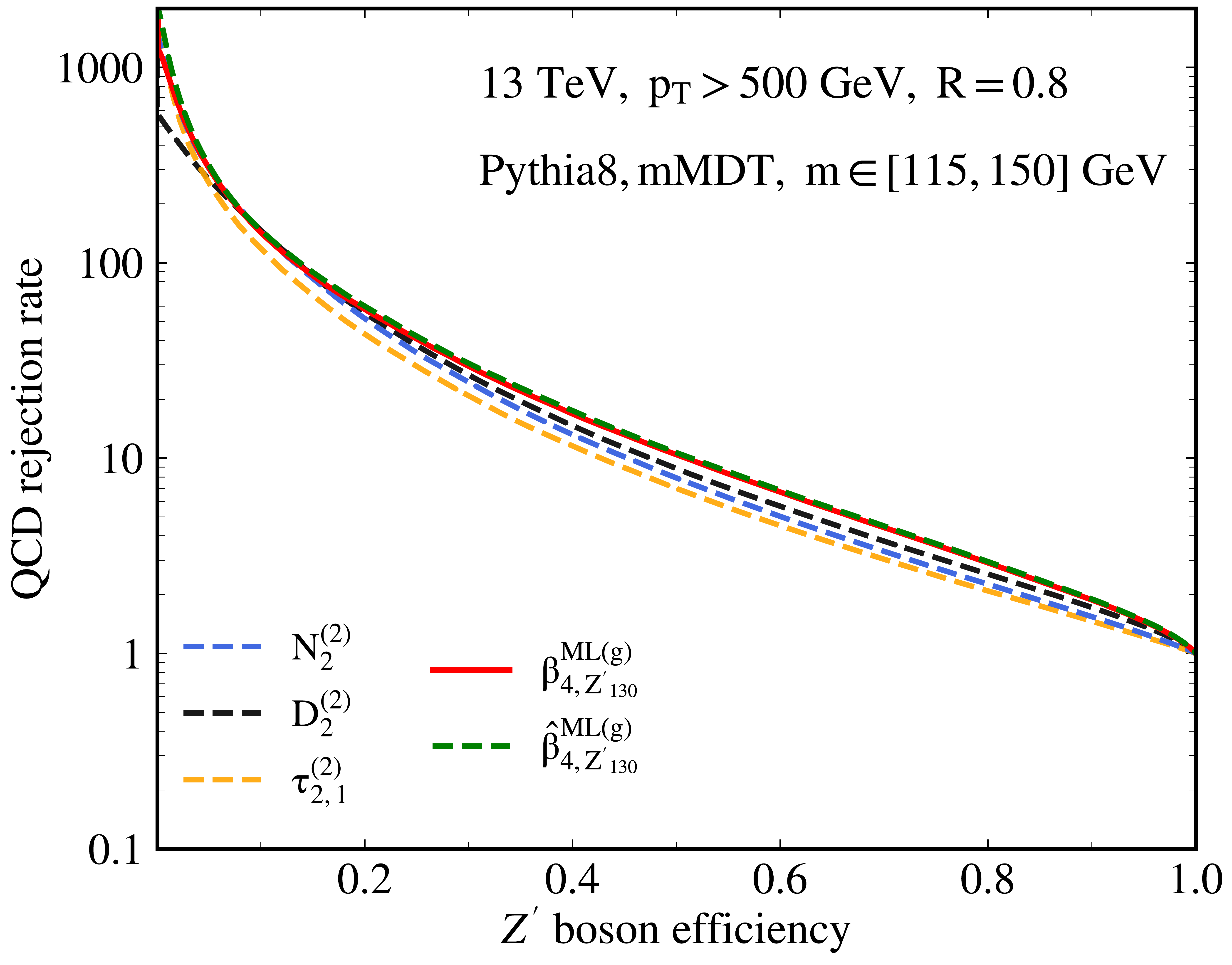}
			\label{fig:obscomp130_sd}
		}\\
	    
		\subfloat[$m_{Z'}=50~\mathrm{GeV}$]
		{
			\includegraphics[width=0.33\textwidth]{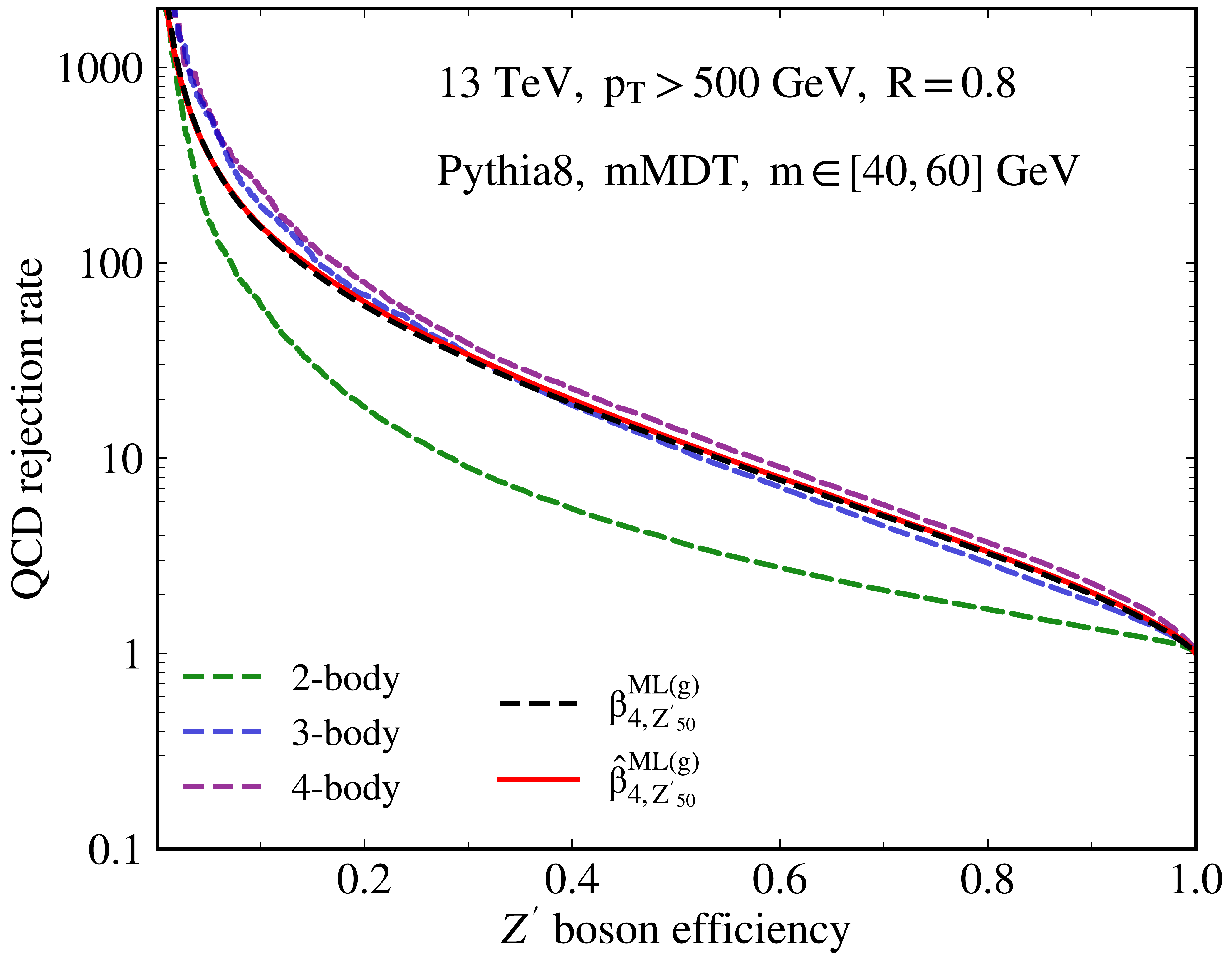}
			\label{fig:dnnobscomp50_sd}
		}
		\subfloat[$m_{Z'}=90~\mathrm{GeV}$]
		{
			\includegraphics[width=0.33\textwidth]{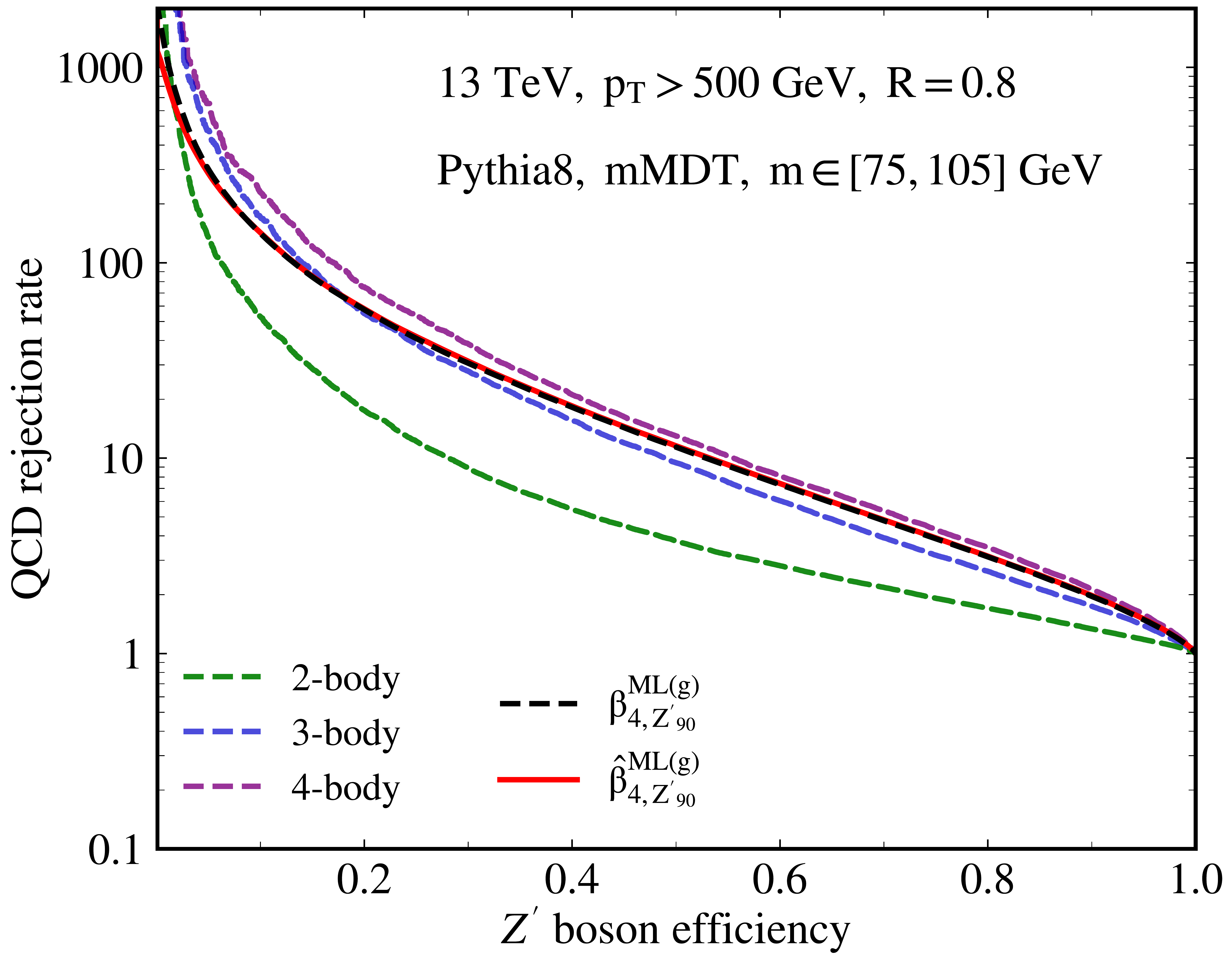}
			\label{fig:dnnobscomp90_sd}
		}		
		\subfloat[$m_{Z'}=130~\mathrm{GeV}$]
		{
			\includegraphics[width=0.33\textwidth]{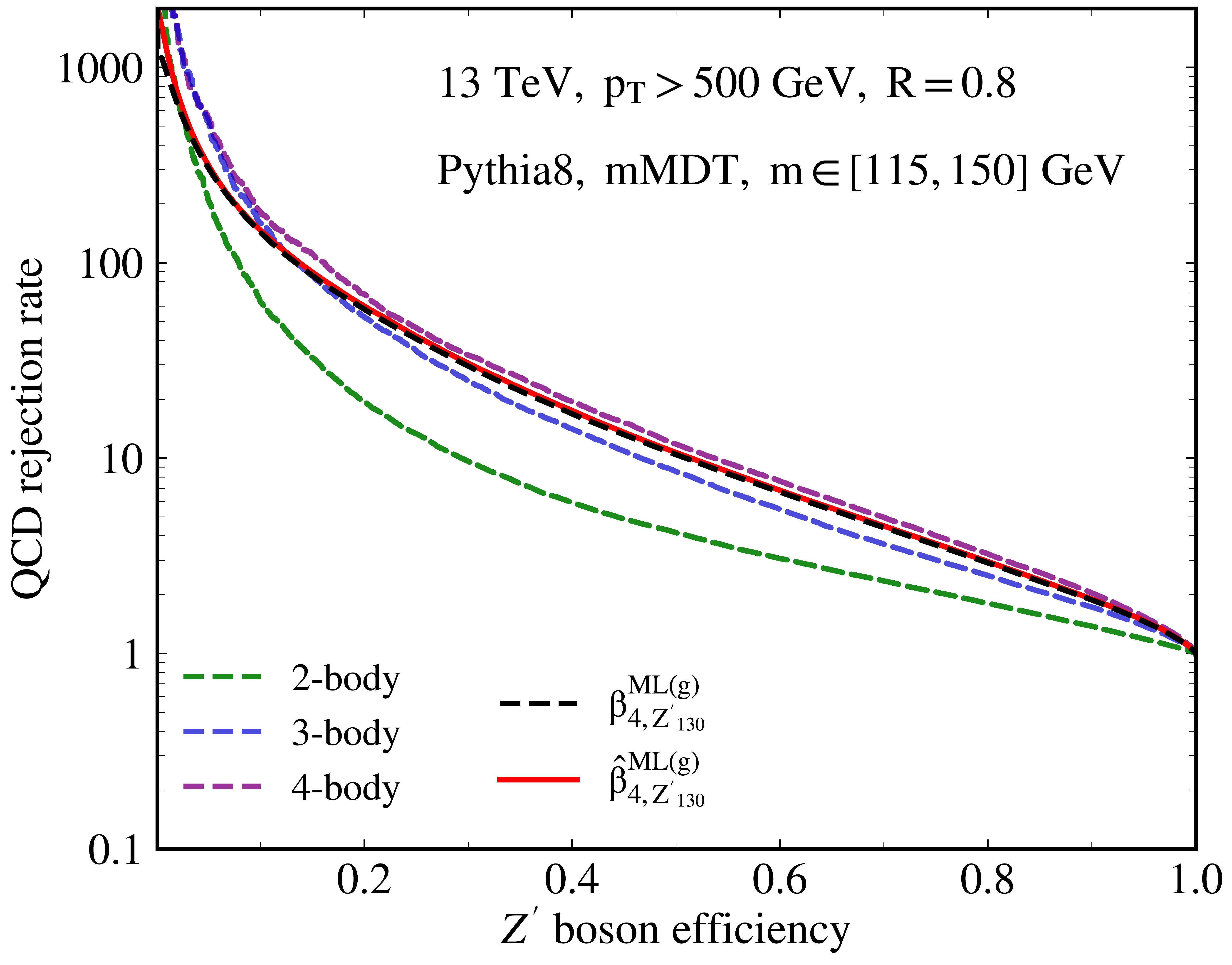}
			\label{fig:dnnobscomp130_sd}
		}
		\caption{Top panel [a-c]: Comparison of discrimination power of $\beta_{4}^\text{ML(g)}$ observables to standard observables; the latter are computed with an angular exponent of 2, for which they were observed to perform best for mMDT groomed samples. Bottom panel [d-f]: Comparison of $\beta_{4}^\text{ML(g)}$ to discrimination power of neural networks trained on the $M$-body observable bases; the observables capture almost all of the discrimination power of the 4-body neural networks.}
		\label{fig:ROC_comp_groomed}
	\end{minipage}
\end{figure*}

The results for the final observables for the three $m_{Z^\prime}$ points are presented in tables~\ref{tab:groomed_Z_ML} and~\ref{tab:groomed_Z_MSE}, and the observable distributions are plotted in Fig.~\ref{fig:distributions_groomed}. The performance of the new observables are compared to standard ones and $M$-body DNN's in Fig.~\ref{fig:ROC_comp_groomed} and the corresponding AUCs are shown in Table~\ref{tab:groomed_Z_AUC} for different mass points.  The conclusions from this section are qualitatively the same as from Sec.~\ref{subsec:Ungroomed_ZvQCD}, with a slightly lower AUC from both the product observable and the physics-motivated observables.  Importantly, the product observables for the groomed case appear to saturate the bounds from the $M$-body phase space better than in the ungroomed case.

\section{\label{app:beta4ML_comp} Mass dependence of $\beta_M^\text{ML}$}

\begin{figure*}[htb]
	\centering
	\begin{minipage}{\textwidth}
		\centering
		
		\subfloat[$m_{Z'}=50~\mathrm{GeV}$]
		{
			\includegraphics[width=0.33\textwidth]{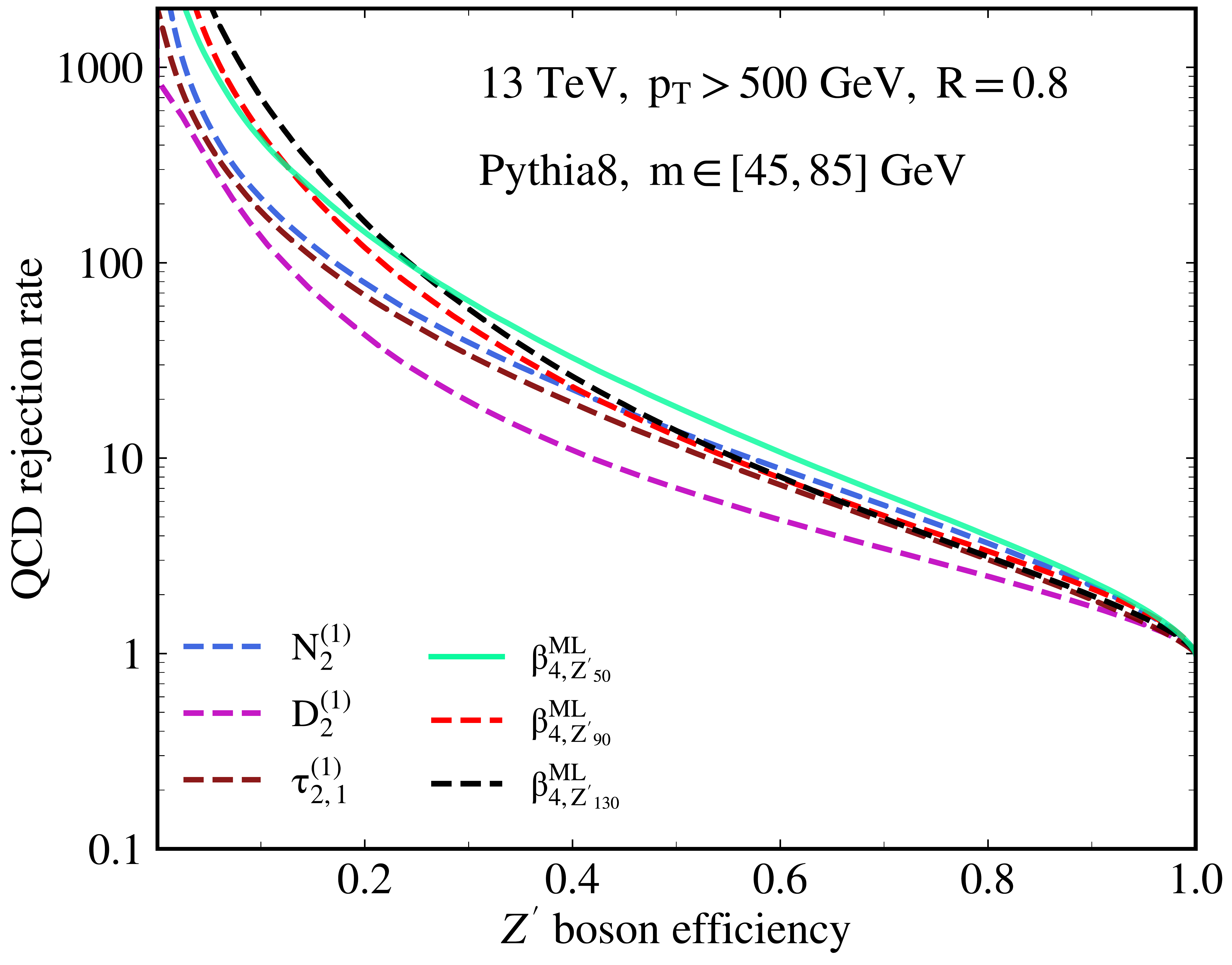}
			\label{fig:mass_roc_comp_50}
		}
		\subfloat[$m_{Z'}=90~\mathrm{GeV}$]
		{
			\includegraphics[width=0.33\textwidth]{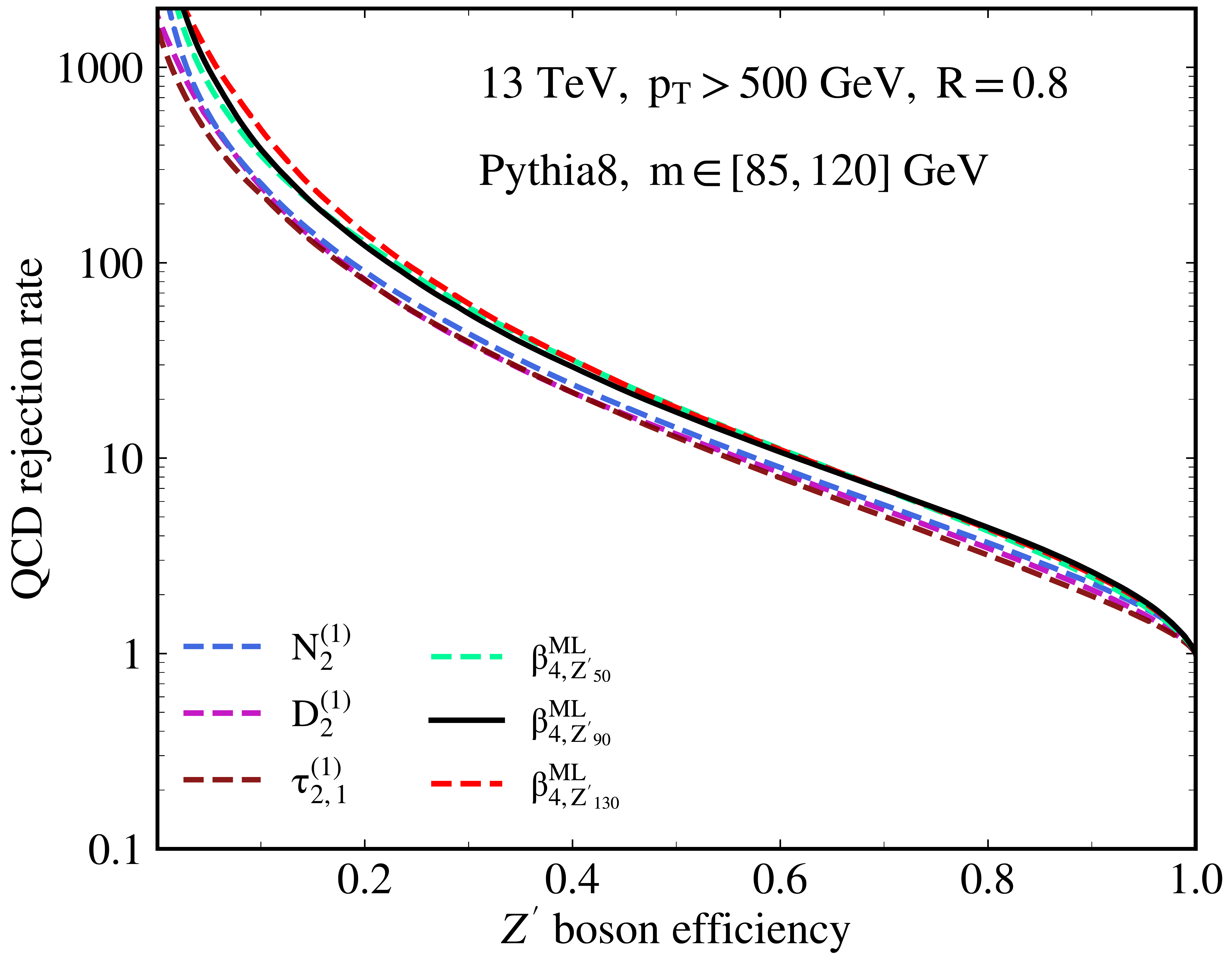}
			\label{fig:mass_roc_comp_90}
		}
		\subfloat[$m_{Z'}=130~\mathrm{GeV}$]
		{
			\includegraphics[width=0.33\textwidth]{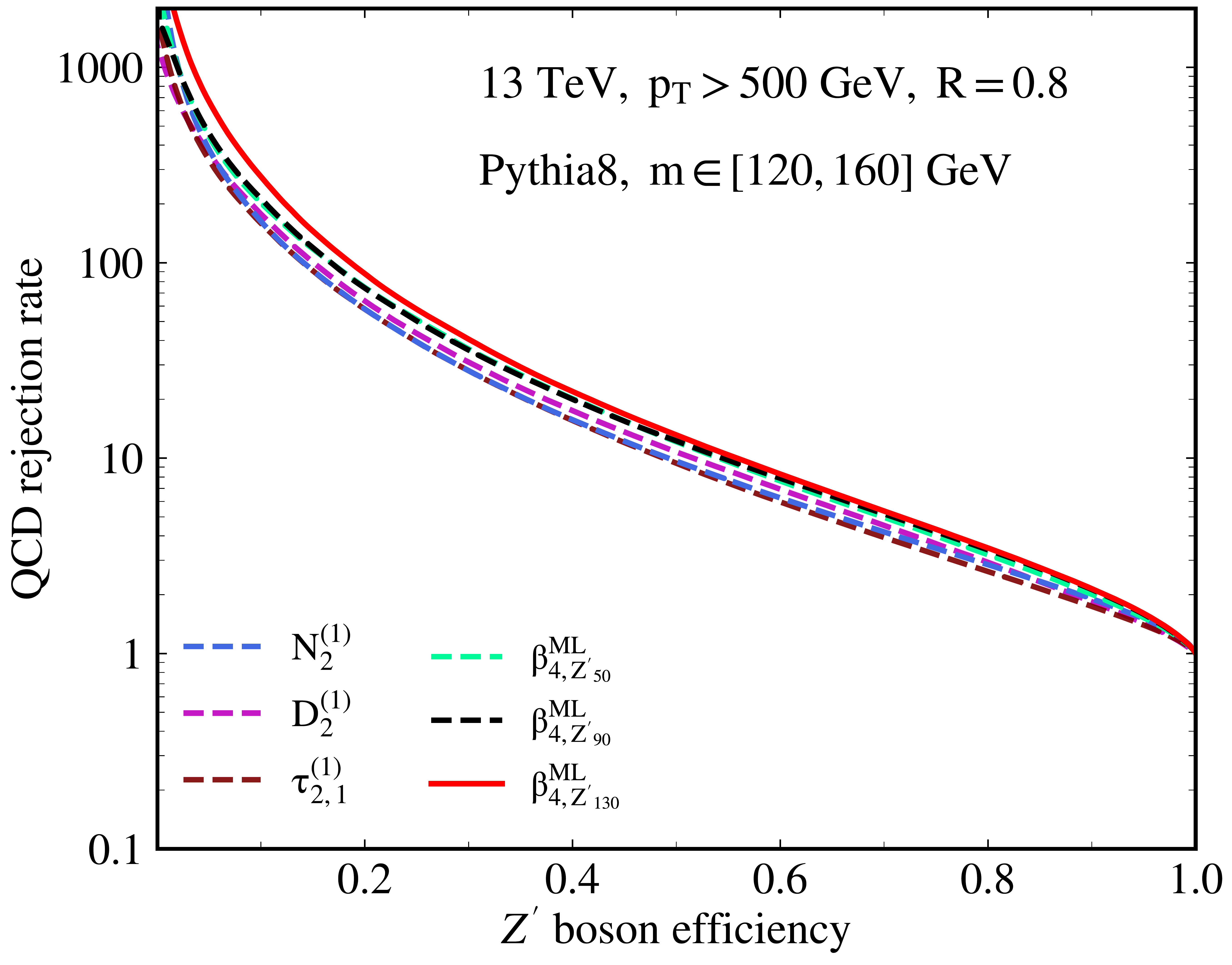}
			\label{fig:mass_roc_comp_130}
		}\\
		\subfloat[$m_{Z'}=50~\mathrm{GeV}$]
		{
			\includegraphics[width=0.33\textwidth]{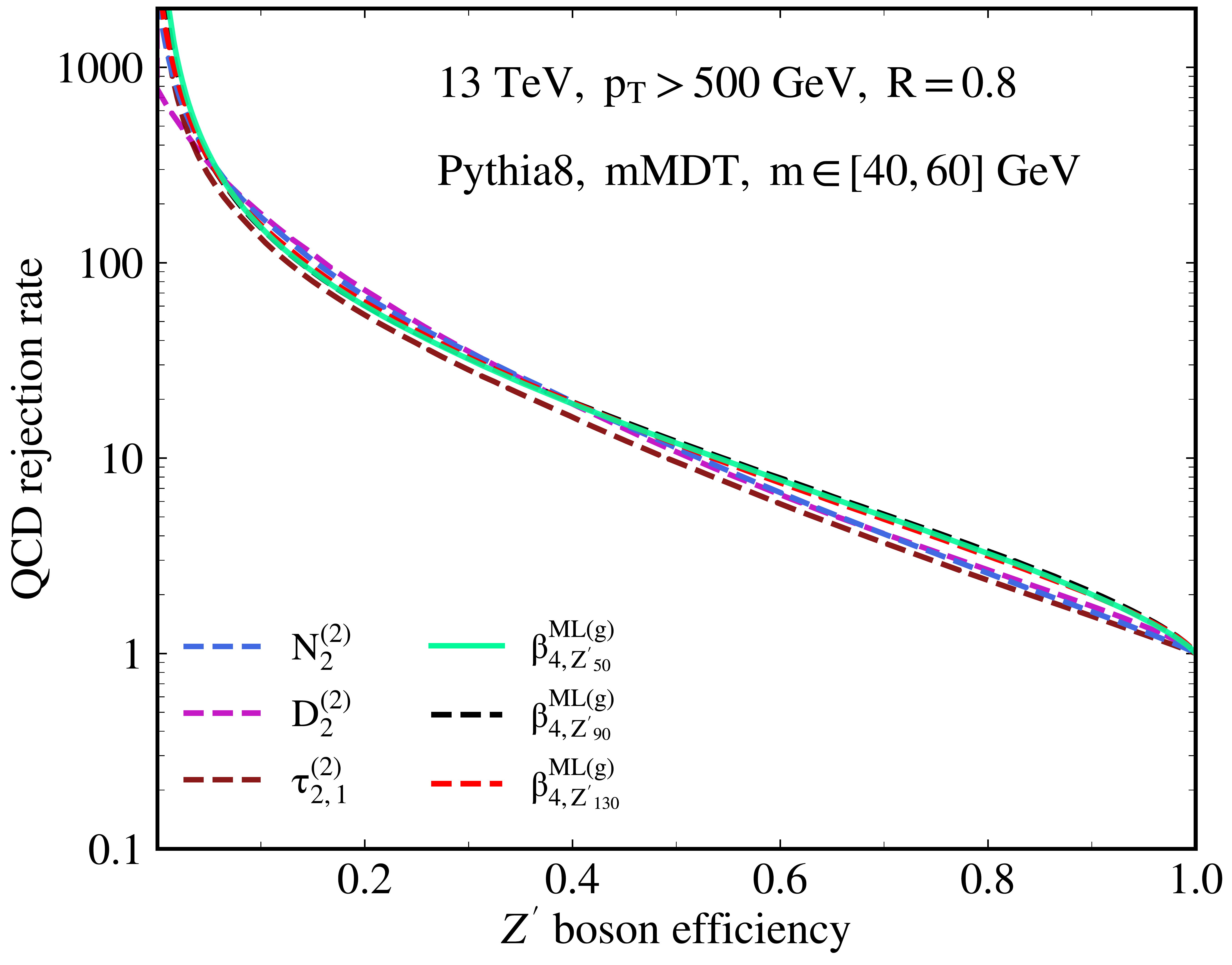}
			\label{fig:mass_roc_comp_50_sd}
		}
		\subfloat[$m_{Z'}=90~\mathrm{GeV}$]
		{
			\includegraphics[width=0.33\textwidth]{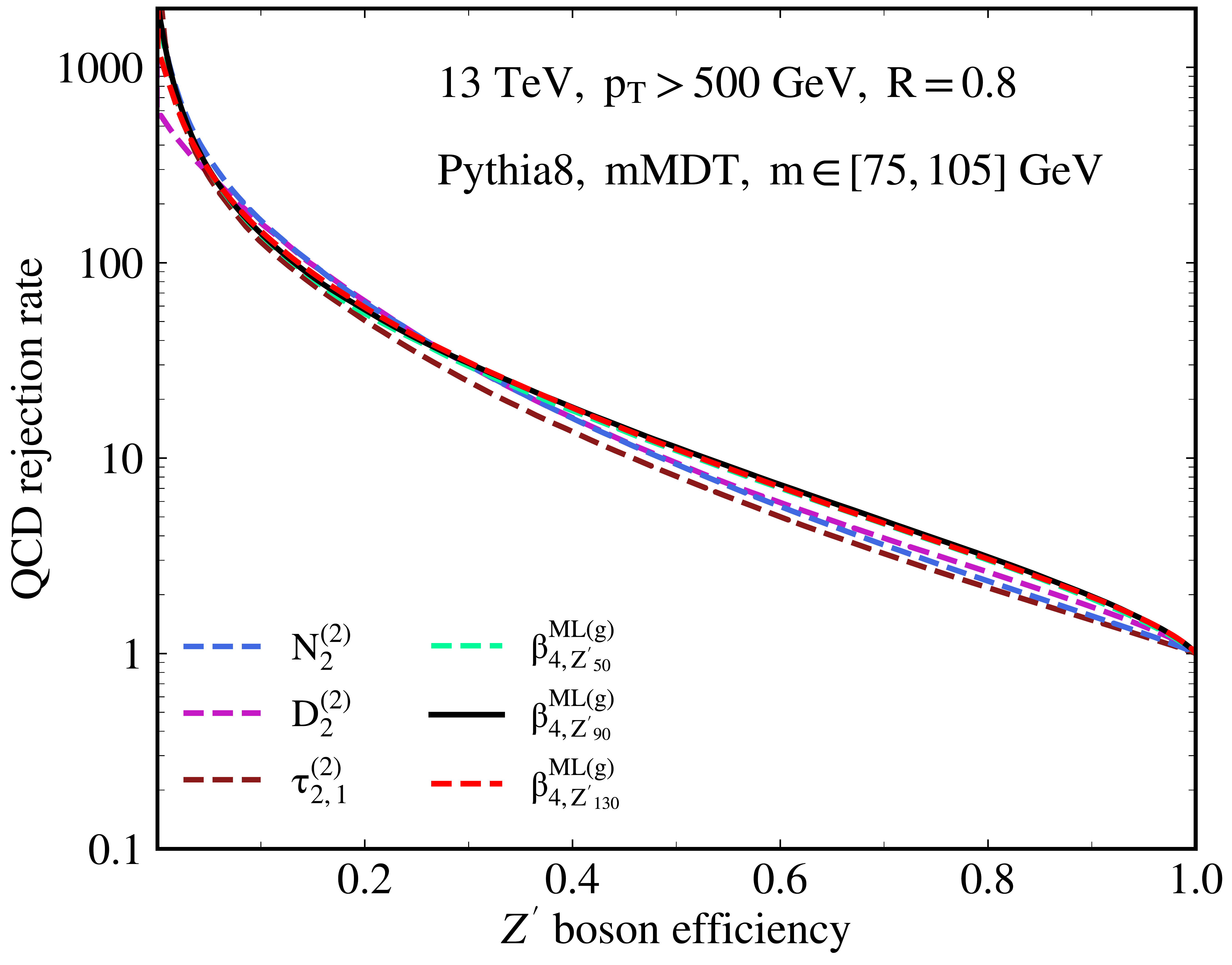}
			\label{fig:mass_roc_comp_90_sd}
		}
		\subfloat[$m_{Z'}=130~\mathrm{GeV}$]
		{
			\includegraphics[width=0.33\textwidth]{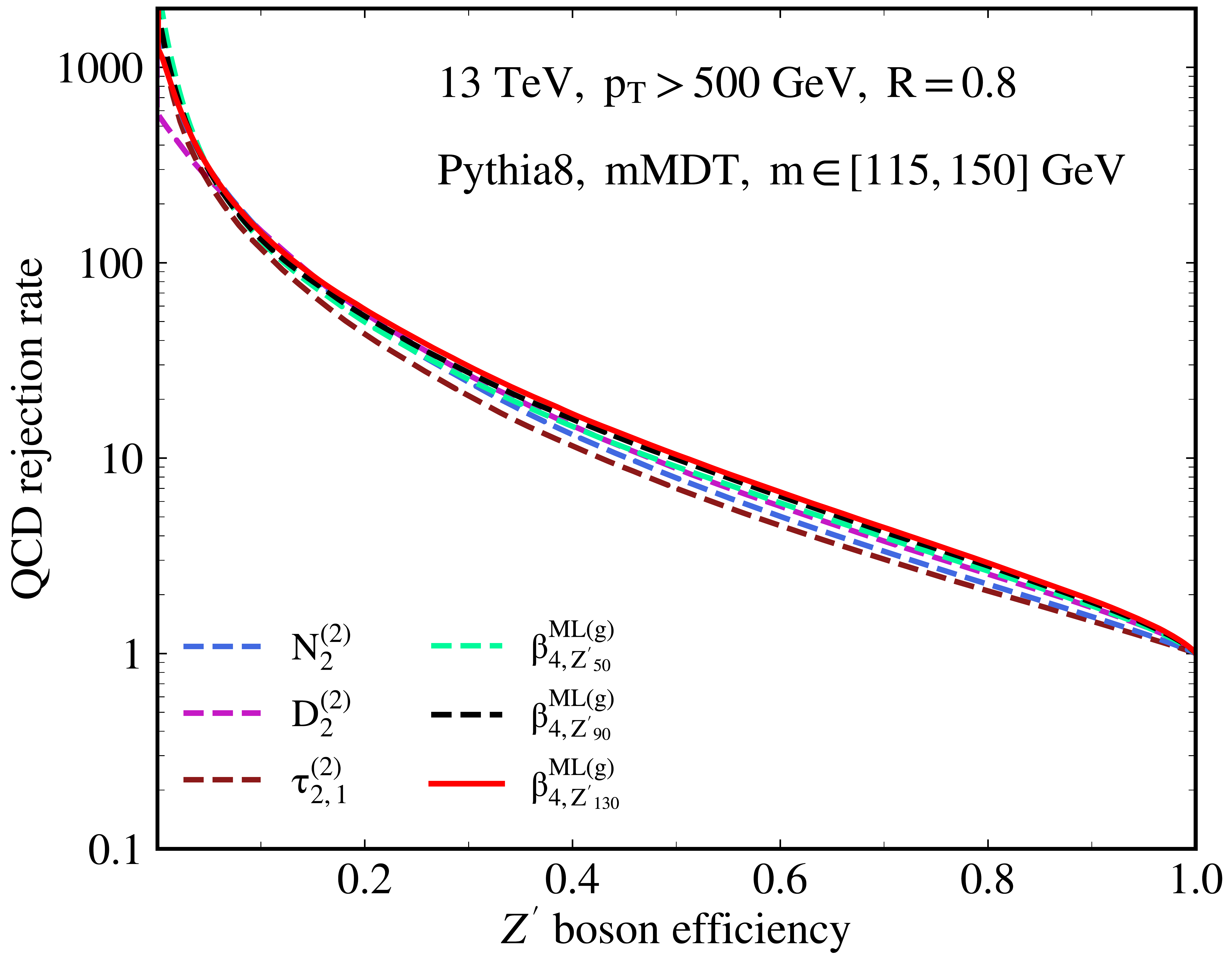}
			\label{fig:mass_roc_comp_130_sd}
		}
		
		\caption{Here we plot results for the new observables on \Zp~samples with a different mass point to that which they were optimized on, within the mass windows appropriate for the corresponding signal. We note that for all cases, all the new observables demonstrate very similar discrimination power, and outperform standard observables. 
			\label{fig:mass_comp}}
	\end{minipage}
\end{figure*}

Here, we briefly study the performance of the new observables presented in Sec.~\ref{subsec:Ungroomed_ZvQCD}. They are tested on a different combination of signal and background samples from the ones they were optimized on; for example, we calculate the new observable for $m_{Z^\prime}=130$~GeV on signal samples for $m_{Z^\prime}=90$~GeV, and background, that pass the mass window on which the 90 GeV observable was optimized. The results for this study are presented in Fig.~\ref{fig:mass_comp}, and indicate that while these observables are optimized on samples from a specific mass point, they can be applied to other classification tasks and still provide better discrimination performance than standard observables. This also suggests that the different parameter sets in tables~\ref{tab:ungroomed_Z_ML} and~\ref{tab:ungroomed_Z_MSE} may represent observables with very similar physical information even though the $N$-subjettiness variables are not invariant under transverse boosts.

\vspace{0.5cm}
\section{\label{app:betaML_comp} Saturating the discrimination power of $\hat{\beta}_M^\text{ML}$}

\begin{figure*}[h]
	\centering
	\includegraphics[width=0.42\textwidth]{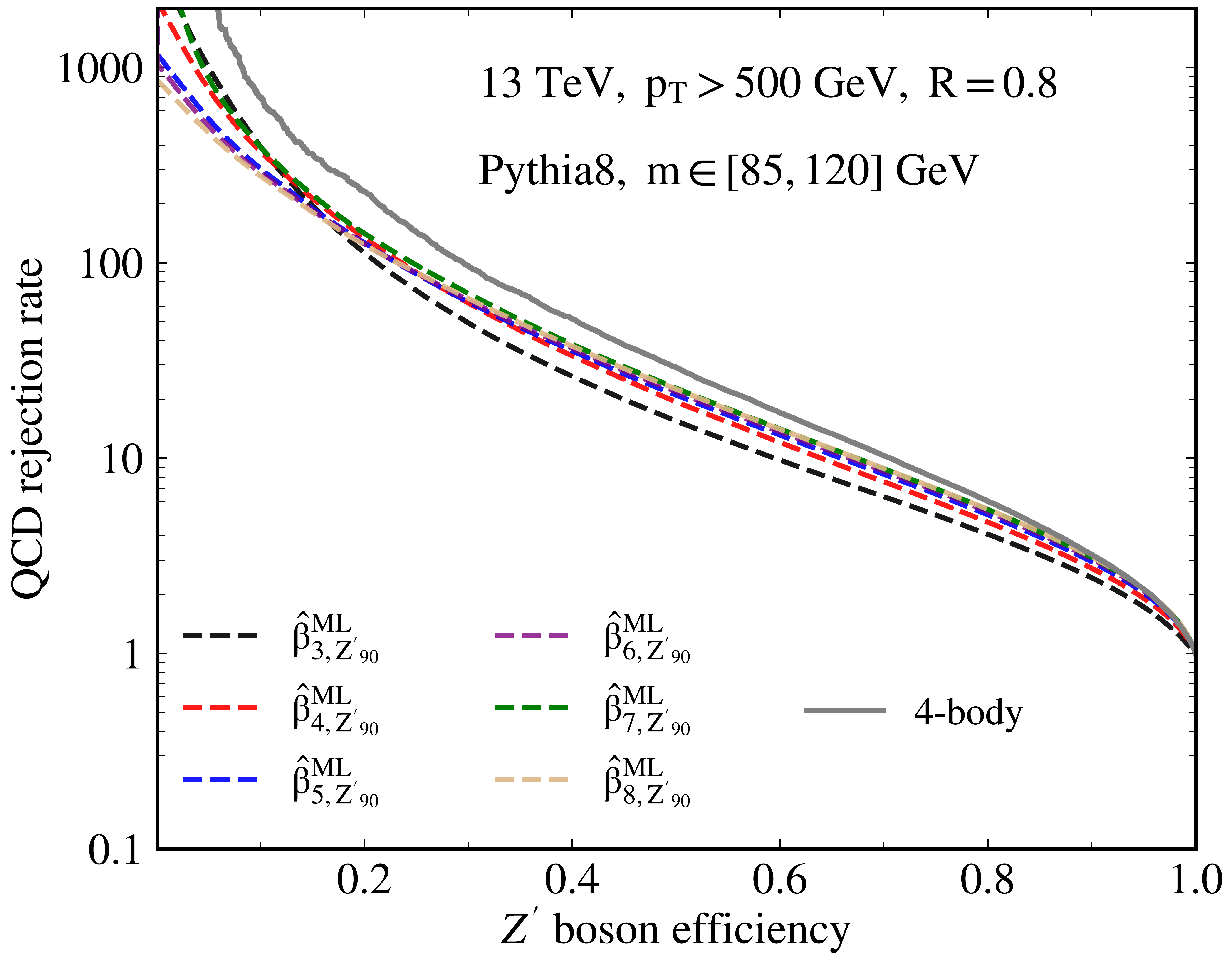}
			\label{fig:beta_comp_90}

		\caption{Here we compare the discrimination power of $\hat{\beta}_{M,Z^\prime_{90}}^{ML}$ for ungroomed $Z^\prime$ discrimination for values of $M=3,...,8$. We note that while discrimination power of the product form does increase with higher M (until the inclusion of 7- or 8-body phase space variables), it can only capture a limited amount of useful discriminating information from inclusion of variables from beyond the basis of the point of saturation of a DNN classifier (dark grey). 
			\label{fig:betacomp_90}}

\end{figure*}
In this section we briefly study the flexibility of the product form ansatz using the $\hat{\beta}_{M}^{ML}$ observables obtained via the linear regression procedure. For concreteness, we look at the $m_{Z^\prime}=90$ GeV case, and plot ROC curves for the product observables upto $M=8$ in Fig.~\ref{fig:betacomp_90}.

We observe that discrimination power gradually increases up to the inclusion of 7- or 8-body phase space variables. Compared to the ROC curve at the point of saturation, from the 4-body DNN classifier, these results suggest that while a DNN can adjust thresholds on the $M$-body inputs such that there is effectively only redundant discriminating information in higher $M$-body bases, as is also expected from the physics study in Ref.~\cite{Datta:2017rhs}, the product observables do still benefit from including $N$-subjettiness variables from beyond the point of saturation. 

Depending on the classification task, the product observables may even come very close to matching the performance of a saturated ML classifier (Fig.~\ref{fig:ROC_comp_groomed}). However, ultimately it cannot not capture all available information, due to lack of further flexibility of the product form ansatz. These observations will of course vary based on the objects being studied. We leave further physics studies of the product form or other equivalent ansatz to future work.

\end{document}